%% file: ZpT.tex
\documentclass[11pt]{article}
\usepackage{jheppub}
\usepackage{graphicx}
\usepackage{url}
\usepackage{cancel}
\usepackage{caption}
\usepackage{placeins}
\usepackage{bm}
\usepackage{hyperref}
\usepackage{amsmath}
\usepackage{amssymb}
\usepackage{listings}
\usepackage[toc, page]{appendix}
\usepackage{color}

\newcommand{\pt}{p_T^Z}
\newcommand{\y}{y_Z}
\newcommand{\Mll}{M_{ll}}

\title{The impact of the LHC $Z$-boson transverse momentum data on PDF determinations}
\author[a]{Radja Boughezal,} 
\author[b]{Alberto Guffanti,}
\author[c]{Frank Petriello,}
\author[d]{Maria Ubiali}
\affiliation[a]{High Energy Physics Division, Argonne National Laboratory, Argonne, Illinois 60439, USA}
\affiliation[b]{Dipartimento di Fisica, Universit\`{a} di Torino and INFN, Sezione di Torino,
Via P. Giuria 1, I-10125, Turin, Italy}
\affiliation[c]{High Energy Physics Division, Argonne National Laboratory, Argonne, Illinois 60439, USA and 
		   Department of Physics and Astronomy, Northwestern University,  Evanston, Illinois 60208, USA}
\affiliation[d]{Cavendish Laboratory (HEP), JJ Thomson Avenue,
Cambridge CB3 0HE, United Kingdom}
\emailAdd{rboughezal@anl.gov}
\emailAdd{alberto.guffanti@unito.it}
\emailAdd{f-petriello@northwestern.edu}
\emailAdd{ubiali@hep.phy.cam.ac.uk}

\abstract{
The LHC has recently released precise measurements of the transverse momentum distribution of the $Z$-boson that provide a 
unique constraint on the structure of the  proton.  Theoretical developments now allow the prediction of these observables through 
next-to-next-to-leading order (NNLO) in perturbative QCD.  In this work we study the impact of incorporating these latest advances 
into a determination of parton distribution functions (PDFs) through NNLO including the recent ATLAS and CMS 7 TeV and 8 TeV $\pt$ 
data.  
We investigate the consistency of these measurements in a global fit to the available data and quantify the impact of including the $\pt$ distributions 
on the PDFs.  
The inclusion of these new data sets significantly reduces the uncertainties on select parton distributions and the corresponding 
parton-parton luminosities. In particular, we find that the $\pt$ data
ultimately leads to a reduction of the PDF uncertainty on the gluon-fusion 
and vector-boson fusion Higgs production cross sections by about 30\%, while keeping the central values nearly unchanged.}

\arxivnumber{1705.00343}
\subheader{\small CAVENDISH-HEP-17/05}
\begin{document}

\maketitle
\flushbottom

\section{Introduction}
The production of a $Z$-boson that subsequently decays into a pair of leptons is a benchmark Standard Model (SM) 
process at the Large Hadron Collider (LHC).  Thanks to its large production rate and clean experimental signature, it can be be measured very accurately 
by the LHC experiments.
It can also be calculated to high accuracy within the Standard Model, with the first prediction to next-to-next-to-leading 
order (NNLO) in the strong coupling constant appearing more than two decades ago~\cite{Hamberg:1990np}, and predictions for 
differential cross sections appearing over one decade ago~\cite{Anastasiou:2003ds,Melnikov:2006kv,Catani:2009sm,Gavin:2010az}.  
This combination of precise experimental data and highly-developed theory allows this process to be used to determine quantities of 
fundamental importance to our understanding of high-energy phenomena, such as parton distribution functions (PDFs).  

Among the many distributions in $Z$-boson production that have been measured, the transverse momentum ($p_T$) distribution 
stands out as an especially interesting one.
First of all, the $Z$-boson $p_T$ spectrum is sensitive to the gluon and the light-quark PDFs in the not-so-well 
constrained intermediate Bjorken-$x$ region, which makes it a promising observable for constraining these distributions. The fact that 
the Higgs production cross section at the LHC is also sensitive to the same PDF combinations in the same region of Bjorken-$x$, 
makes the measurement of this process of direct importance to the search for beyond-the-SM phenomena in the Higgs sector. 
Second, the transverse momentum spectrum of the $Z$-boson is sensitive to both soft QCD radiation (at small $p_T$) and to large 
electroweak (EW) Sudakov logarithms (at large $p_T$). Given that PDF fits typically rely on fixed-order perturbative QCD, it is interesting 
to test how well fixed-order QCD predictions can describe this data.  This has direct impact on which range of data can be included into 
PDF fits.  

The potential for $\pt$ measurements to provide valuable constraints on PDF determinations has been considered previously, both on 
general grounds~\cite{Malik:2013kba,Rolph:2015hoa}, and when considering a recent measurement performed by the CMS collaboration~\cite{Chatterjee:2016sxt}.
Both of these studies, which are based on NLO QCD, show the potential of these measurements.  At the same time, they also stress the 
importance of including the full NNLO QCD corrections to the $Z$-boson transverse momentum distribution in order to fully exploit the 
constraining power of the data.

In present global PDF determinations, the gluon distribution at medium and large $x$ is primarily constrained by the inclusive-jet $p_T$ 
spectrum measurements. The full NNLO prediction for this observable has been recently calculated in the leading-color
approximation~\cite{Currie:2016bfm}, but results have not yet been
made available for all jet data sets included in PDF fits.  
This deficiency motivates the study of other cross sections known to NNLO for this purpose, such as the $Z$-boson $p_T$ spectrum, or 
top-pair production. For the latter, studies have appeared that
explored in great detail the possibility of making use of the total cross section~\cite{Beneke:2012wb,Czakon:2013tha} and more recently 
of the differential distribution~\cite{Czakon:2016olj} measurements. In particular, it was shown that differential distributions from top-pair 
production provide significant constraints on the large-$x$ gluon that are comparable to those obtained from inclusive jet production data.

The importance of including NNLO corrections is especially clear in the case of the $Z$-boson transverse momentum distribution given 
the recent experimental progress in measuring this observable. The data sets from the 7 and 8 TeV LHC runs from both ATLAS and CMS 
feature percent-level experimental errors, clearly requiring predictions beyond NLO in order to achieve a comparable theoretical precision.

It is our intent in this manuscript to investigate the inclusion of the $\pt$ data from the LHC into a global PDF fit.  
We perform this study in a framework based on the NNPDF3.0 global analysis~\cite{Ball:2014uwa}. The data sets we consider in our work 
are the 7 TeV measurement of the $Z$-boson $p_T$ by the ATLAS collaboration~\cite{Aad:2014xaa}, and the 8 TeV measurements from 
both ATLAS and CMS~\cite{Aad:2015auj,Khachatryan:2015oaa}.  These data sets include doubly-differential distributions in both the rapidity 
and invariant mass of the lepton pair coming from the $Z$-boson decay.  
Our theoretical predictions are based on the NNLO QCD calculation of Ref.~\cite{Boughezal:2015ded}.  We also study the impact of 
including approximate NLO electroweak corrections, as described later in the text.  The major findings of our study are summarized below.  

\begin{itemize}

\item The inclusion of the NNLO QCD corrections generally improves the agreement of theory with the experimental data.  This conclusion is
consistent with previous observations~\cite{Ridder:2016nkl,Gehrmann-DeRidder:2016jns}.  The simultaneous inclusion of the NLO electroweak contributions
together with NNLO QCD, done here for the first time, further improves the data/theory agreement at high $p_T$. 

\item The experimental errors, particularly in the higher-luminosity 8 TeV measurements from ATLAS and CMS, have dropped to the percent 
level.  With the data becoming so precise, a very careful accounting of both experimental and theoretical errors is needed.  We observe 
difficulties in fitting the data without the introduction of an additional uncorrelated error in the fit.  This can come from a combination of Monte
Carlo integration errors on the theory calculation, residual theoretical uncertainties in the prediction, or from underestimated experimental errors.
We expect this issue to become increasingly prevalent in future PDF fits as data becomes more precise.

\item We observe difficulties when attempting to simultaneously fit the 7 TeV and 8 TeV LHC data. The ATLAS 7 TeV data is provided only in terms of normalized distributions, while 
the 8~TeV measurements are also provided as absolute, unnormalized distributions. The normalization to the fiducial cross section performed for the ATLAS 7 TeV data introduces 
correlations between the low-$\pt$ bins and the $\pt>30$ GeV region to which we must restrict our fit due to the appearance of large 
logarithms in the low-$\pt$ region that require resummation.  
The covariance matrix provided for the whole data set then turns out to be incorrect when used for fitting a subset of the data. This prevents
us from consistently including the ATLAS 7 TeV data in the fit.
To validate this hypothesis, in Sec.~\ref{sec:norm} we perform a fit including the normalized ATLAS 8 TeV data rather than the unnormalized 
ones but, in analogy to what is done for the 7 TeV data, using the covariance matrix provided for the whole data set, and explore the
differences in the fit results.  
It would be interesting to revisit this issue if the unnormalized data for the 7 TeV measurement were released or if the experimental
covariance matrix for the $\pt>30$ GeV region was available.  
Attempting to include resummed predictions for the low-$\pt$ region is also possible, although this would introduce additional theoretical
uncertainties. 

\item When adding the 8 TeV LHC $Z$-boson $p_T$ data to the global NNPDF3.0-like fit, we observe a significant decrease of the gluon PDF 
uncertainty in the Bjorken-$x$ region $10^{-3}$ to $10^{-1}$ as well as a reduction of the uncertainty for light quarks.  This leads to a reduction of the PDF uncertainty on the 
gluon-fusion  and Vector Boson Fusion (VBF) Higgs boson cross section of 30\%, while the central value prediction for both 
processes increases by roughly 1\%.

\end{itemize}

Our manuscript is organized as follows.  In Section~\ref{sec:exp} we describe the experimental measurements of $\pt$ that we include in our 
fit. We also present the baseline fits that do not include these data that we use to assess their impact.  In Section~\ref{sec:theorycalc} we 
discuss the details of the theoretical calculation and settings that we use in the fit.  We give a comparison of theory with the $\pt$ data in 
Section~\ref{sec:data-theory}. 
We discuss the agreement observed upon using NLO QCD, NNLO QCD or a combined NNLO QCD + NLO EW prediction, and also consider 
several different global PDF sets.  Our fit to the $\pt$ data and several baseline fits is described in Section~\ref{sec:fits}.  We briefly discuss the
phenomenological impact of the new fits on the Higgs cross section in Section~\ref{sec:pheno}.  Finally, we conclude in Section~\ref{sec:conc}.

\section{Descriptions of the experimental data and fit settings}
\label{sec:exp}

In this Section we first discuss the features of the available experimental measurements. We then describe the methodology and settings of 
our fit to the parton distribution functions including these data.

\subsection{$\pt$ measurements from the LHC}
In this work we consider the most recent differential cross section measurements  of the $Z$-boson transverse momentum spectrum from 
ATLAS~\cite{Aad:2014xaa,Aad:2015auj} and CMS~\cite{Khachatryan:2015oaa}, both with$\sqrt{s}=7$ TeV and $\sqrt{s}=8$ TeV . 

The ATLAS measurement of the $Z$-boson
transverse momentum spectrum at the centre-of-mass energy of  $\sqrt{s}$=7 TeV~\cite{Aad:2014xaa} is performed in the $Z\to e^+e^-$ 
and $Z\to \mu^+\mu^-$ channels, using data based on an integrated luminosity of 4.7 ${\rm fb}^{-1}$. 
The results from each channel are combined for transverse momenta up to 800 GeV. 
The measurement is provided both inclusive in the $Z$-boson rapidity up to 2.4, and  separated into three rapidity bins: $0.0<|\y|<1.0$, 
$1.0<|\y|<2.0$ and  $2.0<|\y|<2.4$.
In order to maximize the constraints on PDFs, we include the data in the three exclusive rapidity bins in our analysis.
In the experimental paper only the normalized distributions are provided.
The measurement is very accurate, with statistical and systematical uncertainties below 1\% in all $\pt$ bins up to 150 GeV and for central
rapidities ($|\y|<2.0$), and about 3\% for the largest rapidity bin. 

In the ATLAS measurement at $\sqrt{s}=8$ TeV~\cite{Aad:2015auj}, the transverse momentum distribution is based on the full 8 TeV data set,
with 20.3 fb$^{-1}$ of integrated luminosity. Measurements are performed in the electron-pair and muon-pair channels and then combined.
Compared to the 7 TeV measurement~\cite{Aad:2014xaa}, this measurement has higher statistics and an improved control of experimental
systematics. 
Measurements are performed in six invariant mass bins: four bins at low invariant mass below the $Z$-peak, one on-peak invariant mass bin, 
and one bin at high invariant mass above the $Z$-peak, reaching up to $\Mll=$ 150 GeV. 
Results for the off-peak bins are provided in one inclusive rapidity bin ($0.0<|\y|<2.4$), while the $Z$-peak measurement results are given both
inclusive over the whole rapidity range $0.0<|\y|<2.4$ and separated in six rapidity bins $0.0<|\y|<0.4$, $0.4<|\y|<0.8$, $0.8<|\y|<1.2$, 
$1.2<|\y|<1.6$, $1.6<|\y|<2.0$ and $2.0<|\y|<2.4$. Again, in order to
maximize the constraints on PDF, we include the on-peak
exclusive rapidity bins in our analysis.

The measurement by the CMS collaboration at the center-of-mass energy  $\sqrt{s}=8$ TeV~\cite{Khachatryan:2015oaa} is performed
differentially in five rapidity bins: $0.0<|\y|<0.4$, $0.4<|\y|<0.8$, $0.8<|\y|<1.2$, $1.2<|\y|<1.6$ and $1.6<|\y|<2.0$.
The analysis uses the data sample of $pp$ collisions collected with the CMS detector at the LHC in 2012, which corresponds to an integrated
luminosity of 19.7 fb$^{-1}$. The $Z$-boson is identified via its decay to a pair of muons. 
We only include the measurements exclusive in the muon rapidities up to $|\y|=1.6$, given that the data in the highest rapidity bin display a
significant incompatibility with respect to the corresponding ATLAS measurement.  We leave this issue to further investigation 
by the experimental collaborations.

\subsection{Settings for the PDF analysis}
\label{sec:fitsettings}
The PDF fits presented in this work are based on the NNPDF3.0 global analysis~\cite{Ball:2014uwa} framework. As in the NNPDF3.0 fit, 
both PDF evolution and DIS structure functions are evaluated in the fit using  the public APFEL library~\cite{Bertone:2013vaa,
Carrazza:2014gfa, Bertone:2016lga}, with heavy-quark structure  functions computed in the FONLL-C general-mass variable-flavor-number
scheme~\cite{Forte:2010ta} with pole masses and with up to $n_f$=5 active flavors. 
The DGLAP evolution equations are solved up to NNLO using a truncated solution, and the input parametrization scale is taken to be 
$Q_0=1$ GeV. The strong coupling $\alpha_s$ is set to $\alpha_s(M_Z)=0.118$, in accordance with the PDG average~\cite{Agashe:2014kda}.
The charm and bottom PDFs are generated perturbatively from light quarks and gluons and the value of the heavy-quark masses are set to
$m_c=1.51$ GeV and $m_b=4.92$ GeV, corresponding to the values recommended by the Higgs Cross Section Working 
Group~\cite{Dittmaier:2011ti}.
Note that these values are different from the ones used in NNPDF3.0, which were instead set to the PDG value of the $\overline{\text{MS}}$
masses. These values will be used in the forthcoming NNPDF3.1 release~\cite{nnpdf31}.  
The dependence of the fit on the values of the heavy quark masses is moderate, and in particular is negligible for the observables under
consideration.

In the analysis performed in this work, we consider two baseline data sets. 
One consists of all available HERA deep inelastic scattering (DIS) data. An important difference with respect to 
the NNPDF3.0 HERA-only baseline is that the HERA inclusive structure function 
data, which in NNPDF3.0 were separated into the HERA-II measurements from H1 and ZEUS~\cite{Aaron:2012qi,Collaboration:2010ry,Abramowicz:2012bx}, have been replaced by the HERA legacy combination~\cite{Abramowicz:2015mha} 
that has become available recently.
This data is supplemented by the 
combined measurements of the charm production
cross section $\sigma_{cc}^{\rm red}$~\cite{Abramowicz:1900rp},
and the  H1 and ZEUS measurement of the bottom
structure function $F_2^b(x,Q^2)$~\cite{Aaron:2009af,Abramowicz:2014zub}. 
The other baseline, a global one, contains all data mentioned in the paragraph above along
with the other data analyzed in the NNPDF3.0 global fit: 
fixed-target neutral-current DIS structure functions
from NMC~\cite{Arneodo:1996kd,Arneodo:1996qe}, BCDMS~\cite{bcdms1,bcdms2},
and SLAC~\cite{Whitlow:1991uw};
charged-current structure
functions from CHORUS inclusive neutrino DIS~\cite{Onengut:2005kv} and from
NuTeV dimuon production data~\cite{Goncharov:2001qe,MasonPhD};
fixed-target E605~\cite{Moreno:1990sf} and
E866~\cite{Webb:2003ps,Webb:2003bj,Towell:2001nh} DY production
data;
Tevatron collider data including
the CDF~\cite{Aaltonen:2010zza} and D0~\cite{Abazov:2007jy} $Z$ rapidity
distributions; and LHC collider data including
ATLAS~\cite{Aad:2011dm,Aad:2013iua,Aad:2011fp},
CMS~\cite{Chatrchyan:2012xt,Chatrchyan:2013mza,Chatrchyan:2013uja,CMSDY}
and LHCb~\cite{Aaij:2012vn,Aaij:2012mda} vector boson production
measurements, adding up to a total of $N_{\rm dat}=3530$ data points.
A further difference from the global baseline (on top of the use of the HERA combined measurements) 
is that in order to ensure a consistent treatment of NNLO corrections, 
we exclude jet production measurements~\cite{Abulencia:2007ez,Aad:2011fc,Chatrchyan:2012bja,Aad:2013lpa} 
from the global baseline data set.  Only the leading color approximation has been
made available at NNLO for this process~\cite{Currie:2016bfm} and $K$-factors are not yet
available for all data sets included in global PDF determinations.

\section{Description of the theoretical calculation}
\label{sec:theorycalc}

For our study we have calculated the $Z$-boson transverse momentum distribution through 
next-to-next-to-leading order in perturbative QCD. 
This computation uses a recent result for the related process of $Z$-boson in association with a jet~\cite{Boughezal:2015ded,Boughezal:2016isb} 
based on the $N$-jettiness subtraction scheme for NNLO calculations~\cite{Boughezal:2015dva,Boughezal:2015eha,Gaunt:2015pea}.  As the $Z$-boson obtains its 
transverse momentum through recoil against jets, these two processes are identical in perturbation theory as long as the cuts on the final-state 
jets are relaxed sufficiently so that the entire hadronic phase space is integrated over for the $Z$-boson $p_T$ values under consideration.  
Since at most three jets can recoil against the $Z$-boson at NNLO, we take the lower cut on the leading-jet $p_T$ to be less than $1/3$ times 
the lowest $Z$-boson $p_T$ included in our study.   We have confirmed that our predictions are not sensitive to the exact choice of this jet cut.
We furthermore remove completely any constraints on the pseudorapidities of final-state jets. 
We note that the low transverse momentum region of $Z$-boson production requires the resummation of large logarithmic corrections of the form 
$(\alpha_s \text{ln}^2(M_Z/\pt))^n$ to all orders in perturbation theory for a proper theoretical description.  This resummation is not present 
in our fixed-order calculation.  We consequently restrict our attention to the region $\pt>30$ GeV when comparing our predictions to the 
experimental data.
In Sect.~\ref{sec:norm} we study the effect of raising the cut on $\pt$ to 50 GeV and observe that results are stable upon the choice of the $\pt$ cut.

We compare the theoretical predictions against both the unnormalized $p_T$ spectra provided by the 8 TeV ATLAS and CMS measurements, 
and also to the distributions normalized to the fiducial $Z$-boson production cross section provided by the 7 TeV ATLAS measurement.  For the
normalized distributions we compute the fiducial $Z$-boson production cross section using the $N$-jettiness subtraction scheme as implemented
in MCVM v8.0~\cite{Boughezal:2016wmq}.  We cross-check this result against FEWZ~\cite{Melnikov:2006kv,Gavin:2010az}.  For the normalized
distributions we do not expand the ratio in the strong coupling constant; {\it i.e.}, we compute both the numerator and denominator through 
relative ${\cal O}(\alpha_s^2)$.

We make the following choices for the electroweak input parameters in our calculation:
\begin{equation}
\begin{split}
M_Z &= 91.1876 \, \text{GeV},\;\;\; \Gamma_Z = 2.4925 \, \text{GeV},\\ G_{\mu} &= 1.11639\times 10^{-5} \, \text{GeV}^2,\;\;\; M_W = 80.398 \, \text{GeV}.
\end{split}
\end{equation}
We use the $G_{\mu}$ electroweak renormalization scheme.  All other couplings are therefore derived using the parameters above, including the
electromagnetic couplings and the weak mixing angle.  We choose the following dynamical scale choices for both the renormalization and
factorization scales:
\begin{equation}
\mu_R=\mu_F = \sqrt{(\pt)^2+\Mll^2}.
\end{equation}
Here, $\Mll$ denotes the invariant mass of the final-state lepton pair.  We note that our calculation includes both the $Z$-boson production and
decay to lepton pairs, the contribution from virtual photons, as well as all interferences.  The residual theoretical uncertainty on the prediction as
estimated by independently varying $\mu_R$ and $\mu_F$ around this central value is at the few-percent level.

As we will see later it is also important when describing the high-$p_T$ data to include the effect of electroweak perturbative corrections.  
The exact NLO electroweak corrections to the $Z$-boson transverse momentum spectrum, including the leptonic decay of the $Z$ boson, are
known in the literature~\cite{Denner:2011vu,Hollik:2015pja}.  However, they are not publicly available in the form of a numerical code.  
To account for their effect in our calculation we instead utilize the approximate expressions presented in Refs.~\cite{Kuhn:2004em,Kuhn:2005az}.
These include all one-loop weak corrections up to terms power-suppressed by the ratio $M_Z^2/((\pt)^2+M_Z^2)$, and additionally the leading
two-loop electroweak Sudakov logarithms.  These expressions are strictly valid only after inclusive integration over the final-state lepton phase
space; we apply them also to the cross sections with fiducial cuts on the leptons.  For the $Z$-boson peak region in 8 TeV collisions we have
checked that these approximations reproduce the numerical magnitude of the exact electroweak corrections to 2\% or better in the high-$\pt$
range where the EW effects become relevant.  Since the electroweak corrections themselves do not exceed 10\% for the entire region studied,
this furnishes an approximation to the distributions we study that is good to the few-per-mille level or better, which is sufficient for our purposes
\footnote{The exact electroweak corrections were used in the ATLAS analysis of 8 TeV data; we thank A.~Denner, S.~Dittmaier, and A.~Mueck 
for providing us with these results to cross-check our approximations.}  
When we study normalized distributions, the NLO electroweak corrections to the fiducial $Z$-boson cross section are obtained from 
FEWZ~\cite{Li:2012wna}.  To combine the electroweak and QCD corrections we assume that the two effects factorize, leading to a multiplicative
combination.  
Denoting the differential cross sections at the $m$-th order in the strong coupling constant relative to the LO result and the $n$-th order in the
QED coupling constant relative to the LO result as $d\sigma^{(m,n)}$, we assume that
\begin{equation}
\frac{d\sigma^{(2,1)}}{d\pt} = \frac{d\sigma^{(2,0)}}{d\pt} \times \frac{d\sigma^{(0,1)}/d\pt }{d\sigma^{(0,0)}/d\pt}.
\end{equation}
This factorization of the electroweak and QCD corrections is supported by a calculation of the dominant mixed ${\cal O}(\alpha \alpha_s)$ corrections in the resonance region~\cite{Dittmaier:2015rxo}.

The experimental errors in the $Z$-peak region have reached an unprecedented level for a high-energy collider experiment, approaching the 
per-mille level over two orders of magnitude in transverse momentum.  Numerous effects that were previously not relevant may now come into
play, and it is worthwhile to briefly discuss the theoretical issues that arise when attempting to reach this precision.  While we can not currently
address these issues, they should be kept in mind when considering these data sets.
\begin{itemize}

\item The uncalculated N$^3$LO perturbative QCD corrections may be needed to further improve the agreement between theory and
experimental data.  As we will see in a later section the theoretical predictions are generally below the experimental measurements.  
The inclusion of the NNLO corrections greatly improves the agreement between theory and experiment, but one may expect a further increase
 from the N$^3$LO corrections.

\item The electroweak corrections become important for $\pt \sim 100$ GeV, reaching the percent level at this point and continuing to grow as 
$\pt$ is increased.  While we assume that the electroweak and QCD corrections factorize, this assumption should be addressed, particularly in 
the high-$\pt$ region.  Non-factorizing ${\cal O}(\alpha \alpha_s)$ effects could conceivably affect the cross section at the percent level.

\item Finally, at this level of precision non-perturbative QCD effects that shift the $\pt$ distribution must be considered.
\footnote{We thank G.~Salam for raising this issue, and for discussions.}.  Since the $Z$-boson transverse momentum distribution is generated
by recoil against a final-state jet, there may be linear non-perturbative power correction of the form $\Lambda_{\text{QCD}}/\pt$.  Simple 
Monte Carlo estimates indicate that this could reach the half-per-cent level~\cite{Gavintalk}.

\end{itemize}

\section{Comparison of theory with LHC data}
\label{sec:data-theory}

In this Section we compare the theoretical predictions for the $\pt$ spectrum to the experimental measurements described in 
Sect.~\ref{sec:exp}. 
We assess the impact of NNLO QCD and NLO electroweak corrections and quantify the agreement between data and theory by computing 
the fully-correlated $\chi^2$ for each of the experiments that we include in our analysis using as input the most recent public releases of 
four PDF determinations: ABMP16~\cite{Alekhin:2017kpj} CT14~\cite{Dulat:2015mca}, MMHT2014~\cite{Harland-Lang:2014zoa} and 
NNPDF3.0~\cite{Ball:2014uwa}.

\begin{figure}[th]
	\centering
	\includegraphics[width=0.32\textwidth]{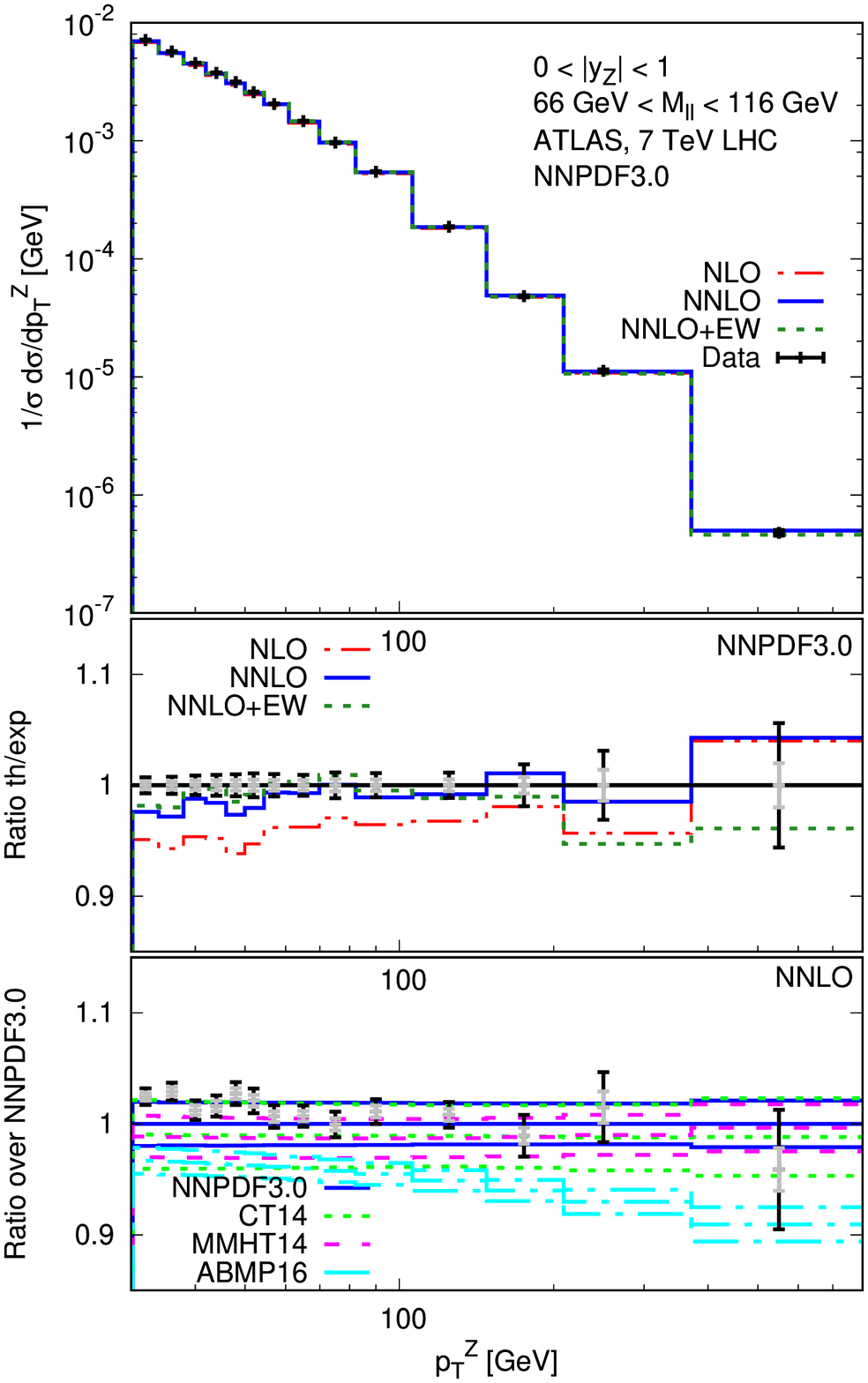}
	\includegraphics[width=0.32\textwidth]{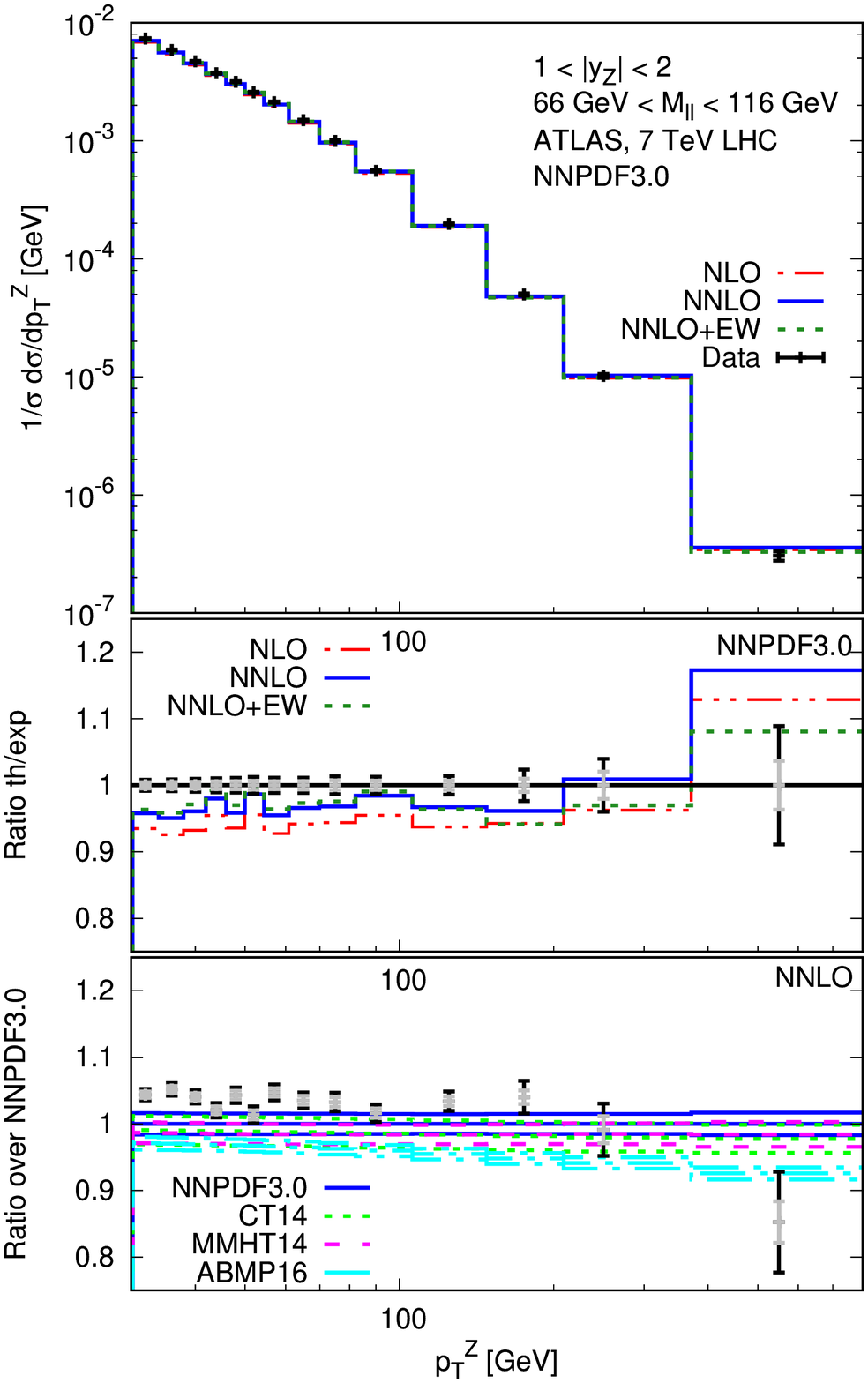}
	\includegraphics[width=0.32\textwidth]{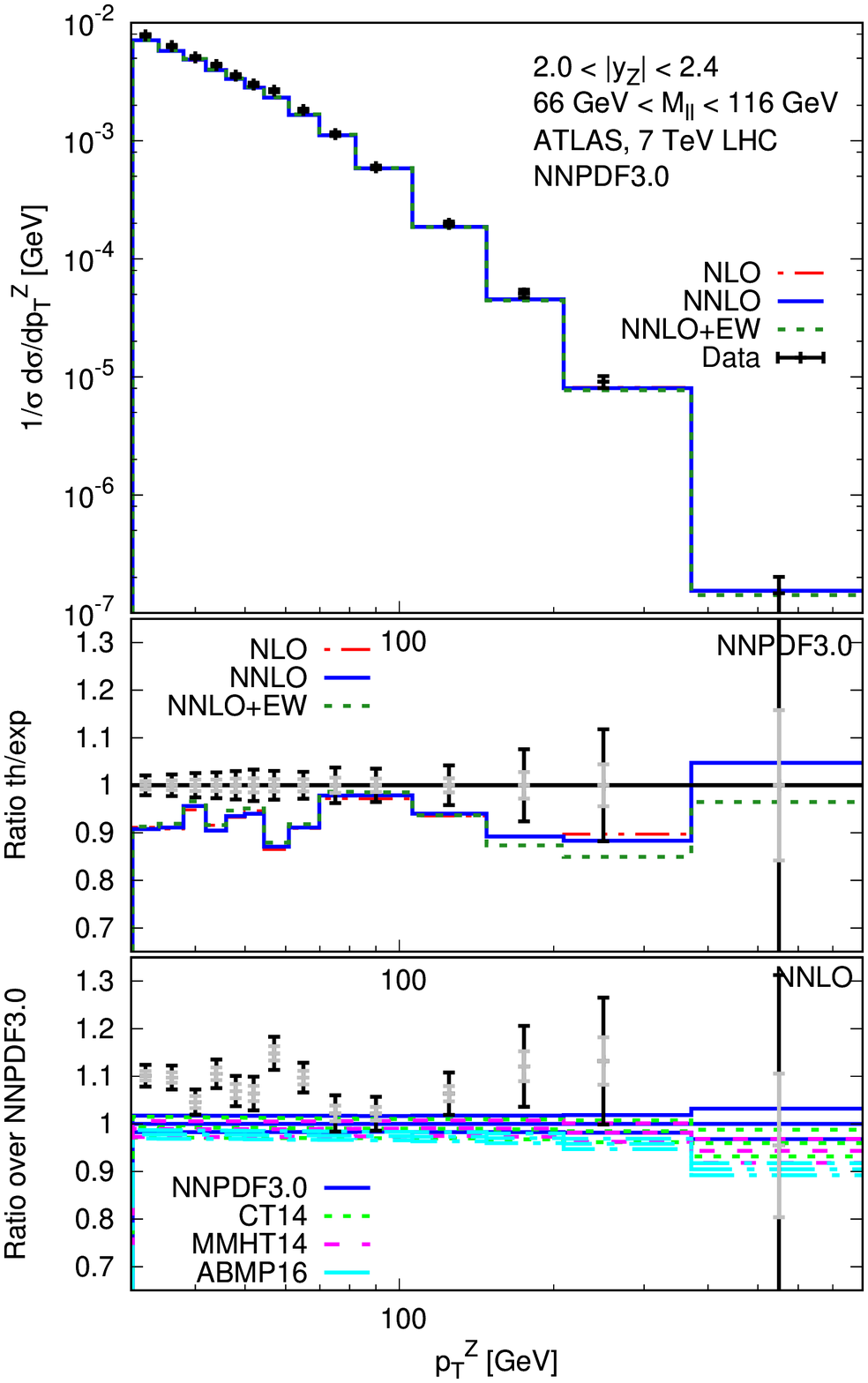}
	\caption{
         Top inset: Theory-data comparison for the ATLAS 7 TeV data~\cite{Aad:2014xaa}
          using NLO (dashed red), NNLO (solid blue) and NNLO+EW(dot-dashed green) predictions. The 
          NNPDF3.0 (N)NLO sets with $\alpha_s(M_Z) = 0.118$ are used for the 
          (N)NLO predictions. Middle inset: NLO, NNLO
        and NNLO QCD + NLO EW predictions are divided by the experimental
        central value. The outer error bar (black) of the data points is given by the total experimental
        uncertainty, while the inner error bar (grey) is given by sum in quadrature of the
        bin-by-bin statistical and uncorrelated uncertainties. Third
        inset: the NNLO predictions computed with the CT14 (green
        dotted), MMHT2014 (pink dashed),
        ABMP16 (cyan dot-dashed) NNLO PDF sets are normalized to 
        the NNLO predictions computed
        with the NNPDF3.0 (solid blue) PDF set. Error bands represent
        the 68\% C.L. PDF uncertainties.}
	\label{fig:theoryexp_atlas7tev_1}
\end{figure}

In Fig.~\ref{fig:theoryexp_atlas7tev_1} we compare the NLO and NNLO predictions 
to the experimental measurements performed by the 7 TeV ATLAS measurements, described 
in Ref.~\cite{Aad:2014xaa}, after imposing the additional cut of $\pt>30$ GeV discussed earlier. We also include
the NLO EW corrections as described in Section~\ref{sec:theorycalc}. All three rapidity bins measured by ATLAS are shown.

\begin{table}[t]
\centering
\caption{$\chi^2$ per degree of freedom for the normalized ATLAS 7 TeV $\pt$ on-peak distributions 
in the separate rapidity bins before their inclusion in the fit.
 As input PDFs we use the NNPDF3.0 set with $\alpha_S(M_Z)=0118$.  The computation is done at NLO, NNLO and NNLO QCD + NLO EW, with (N)NLO PDF set for
(N)NLO computations.  Results for CT14, MMHT2014 and ABMP16 are also shown.}
\label{tab:chi2prefit:atlas7tev}
\begin{footnotesize}
\begin{tabular}{c|c|c|c|c|c|c}
\hline
  Bin  & Order &  $N_{\rm dat}$ &$\chi^2_{\rm d.o.f.}$ \tiny{(NN30)}
  &$\chi^2_{\rm d.o.f.}$\tiny{(CT14)} &$\chi^2_{\rm d.o.f.}$ \tiny{(MMHT14)} &$\chi^2_{\rm d.o.f.}$\tiny{(ABMP16)}\\ 
\hline
 $0.0<\y<1.0$  & NLO      & 14  & 10 & 21 & 9.2 & n.a.\\
               & NNLO     & 14  & 2.2 & 3.8 & 4.3 & 11 \\
               & NNLO+EW  & 14  & 1.3 & 2.3 & 2.6 & 9.1\\
\hline
 $1.0<\y<2.0$  & NLO      & 14  & 13 & 18 & 12& n.a.\\
               & NNLO     & 14  & 5.6 & 8.2 & 9.3      &  15.\\
               & NNLO+EW  & 14  & 3.9 & 6.0 & 6.8  &  12.\\
\hline
$2.0<\y<2.4$   & NLO      & 14  & 7.0 & 7.1 & 6.0 &n.a.\\
               & NNLO     & 14  & 7.0 & 8.2 & 8.7 &11.\\
               & NNLO+EW  & 14  & 5.9 & 7.1 & 7.5 &9.5\\
\hline
All bins   & NLO      & 42  & 9.9 & 15 & 9.1 &n.a.\\
               & NNLO     & 42  & 4.9 & 6.7 & 7.4 &13.\\
               & NNLO+EW  & 42  & 3.7 & 5.2 & 5.6 &12.\\
\hline
\end{tabular}
\end{footnotesize}
\end{table}

We observe that the NNLO corrections significantly increase the NLO predictions, bringing them closer to the 
measured values of the distribution.  The NNLO corrections are approximately constant as a function of $\pt$.  The EW corrections become
significant only for the last three $\pt$ bins.
The quantitative agreement with the theory is summarized in table~\ref{tab:chi2prefit:atlas7tev}, in which the fully-correlated $\chi^2$ is provided, for each 
bin separately and for the three bins together.

For MMHT2014, CT14 and NNPDF3.0 the agreement is improved for central rapidities after the inclusion of NNLO QCD corrections, with a 
further improvement observed upon including NLO electroweak corrections.  For ABMP16 only the NNLO fit is available, so in this case we can 
only test that the agreement is improved upon adding electroweak corrections. 
In the highest rapidity bin this improvement is only observed for NNPDF3.0.  The CT14 $\chi^2_{\rm d.o.f.}$ remains unchanged after 
including NNLO QCD+NLO electroweak, while the result for MMHT2014 becomes slightly worse.  
For all PDF sets the $\chi^2_{\rm d.o.f.}$ is much larger than one, indicating a poor agreement between theory and data (before the fit) even after including
higher-order corrections.

\begin{figure}[tbp]
	\centering
	\includegraphics[width=0.45\textwidth]{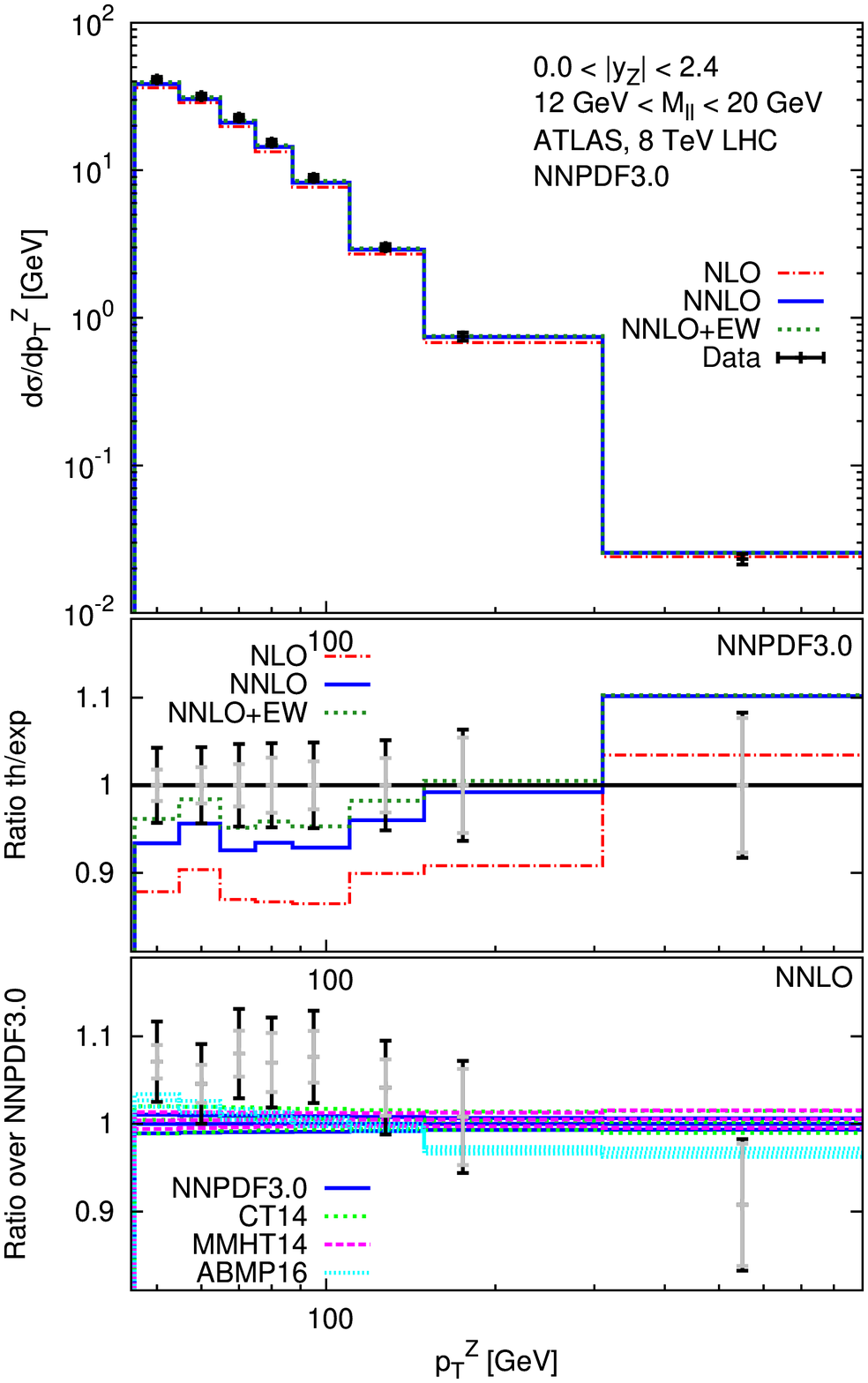}
	\includegraphics[width=0.45\textwidth]{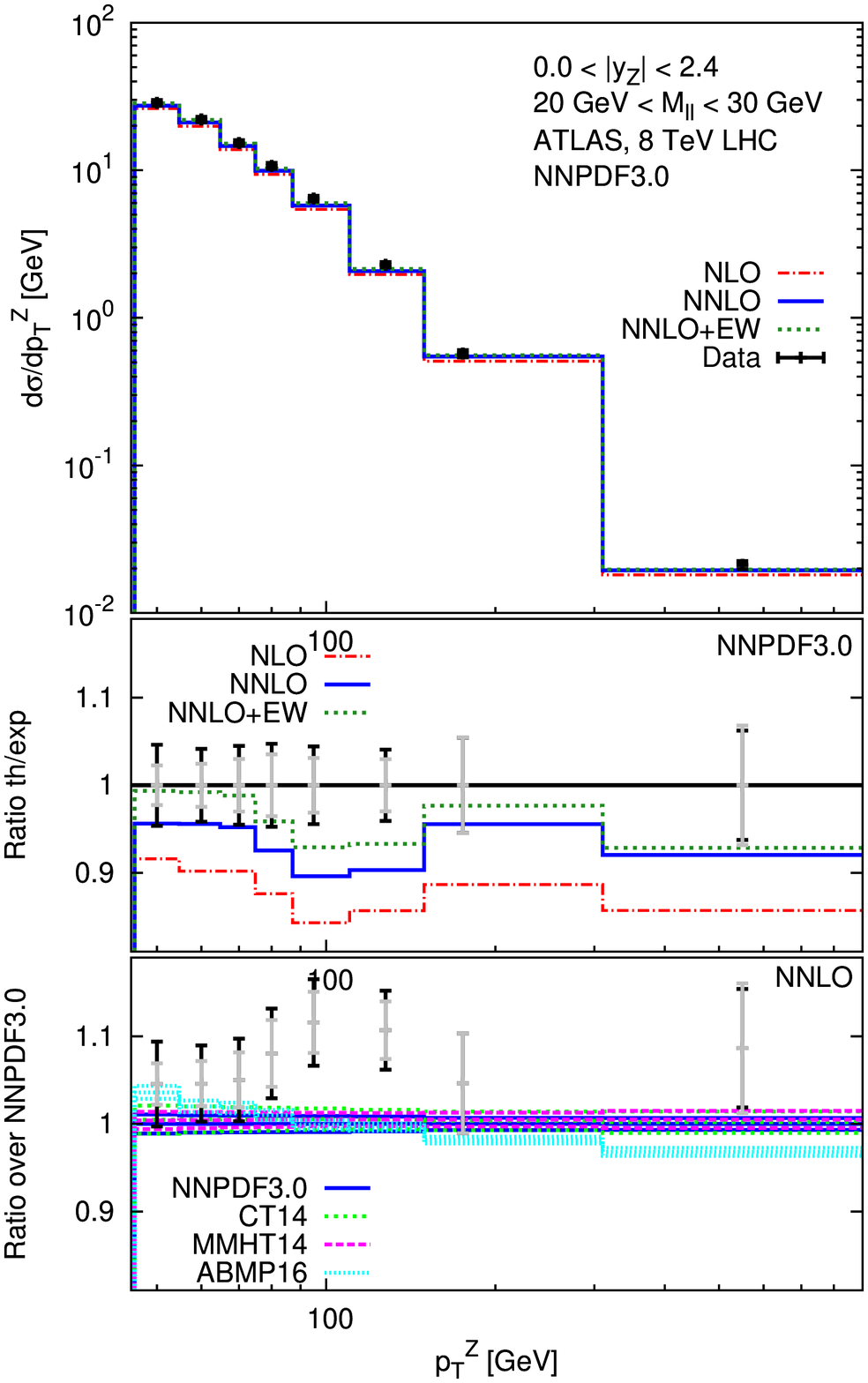}
	\caption{Same as Fig.~\ref{fig:theoryexp_atlas7tev_1} for the ATLAS 8 TeV data 
          divided into invariant mass bins~\cite{Aad:2015auj}. The two lowest invariant mass bins
	  are displayed.}
	\label{fig:theoryexp_atlas8tev_1}
\end{figure}

\begin{figure}[tbp]
	\centering
	\includegraphics[width=0.32\textwidth]{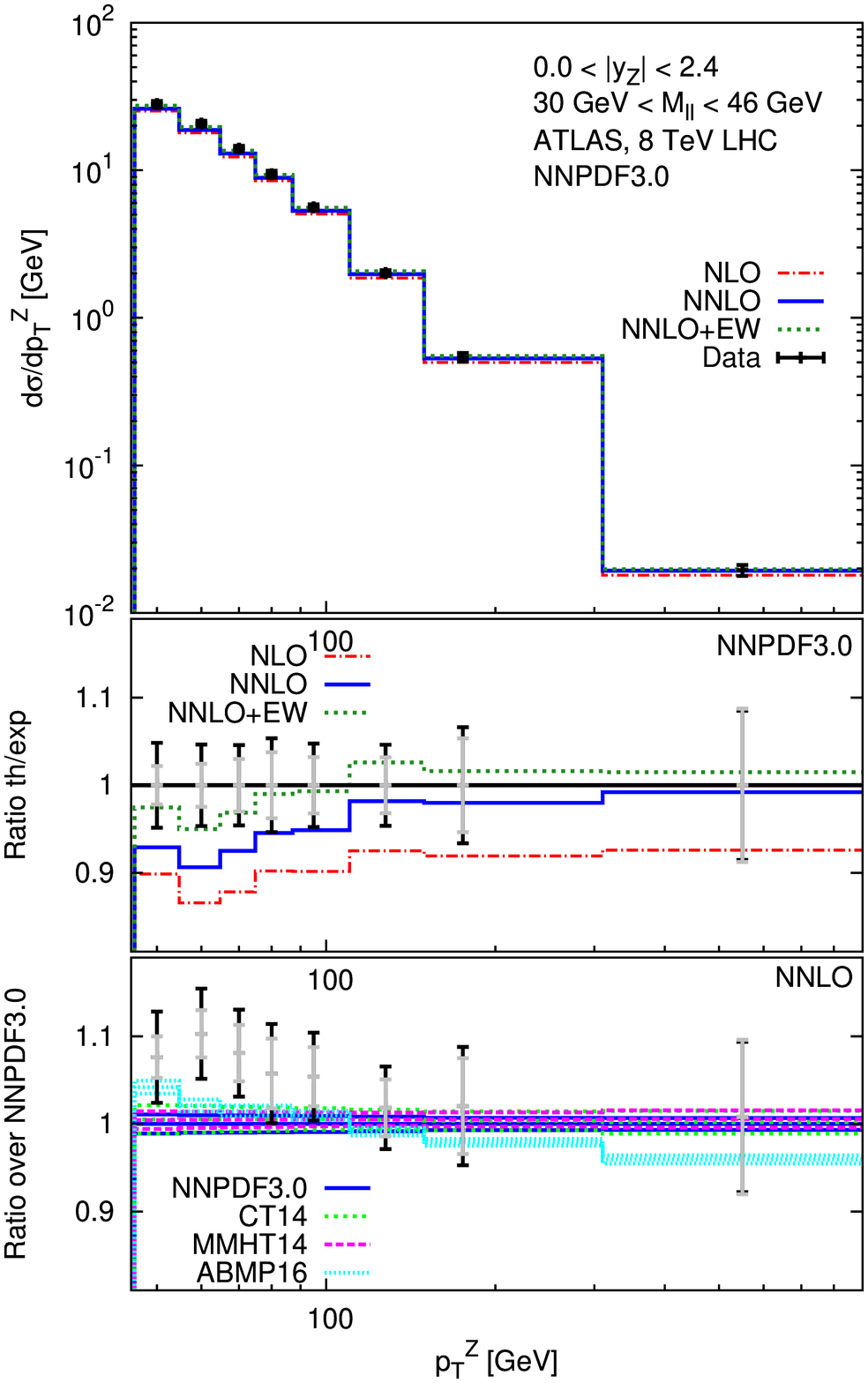}
	\includegraphics[width=0.32\textwidth]{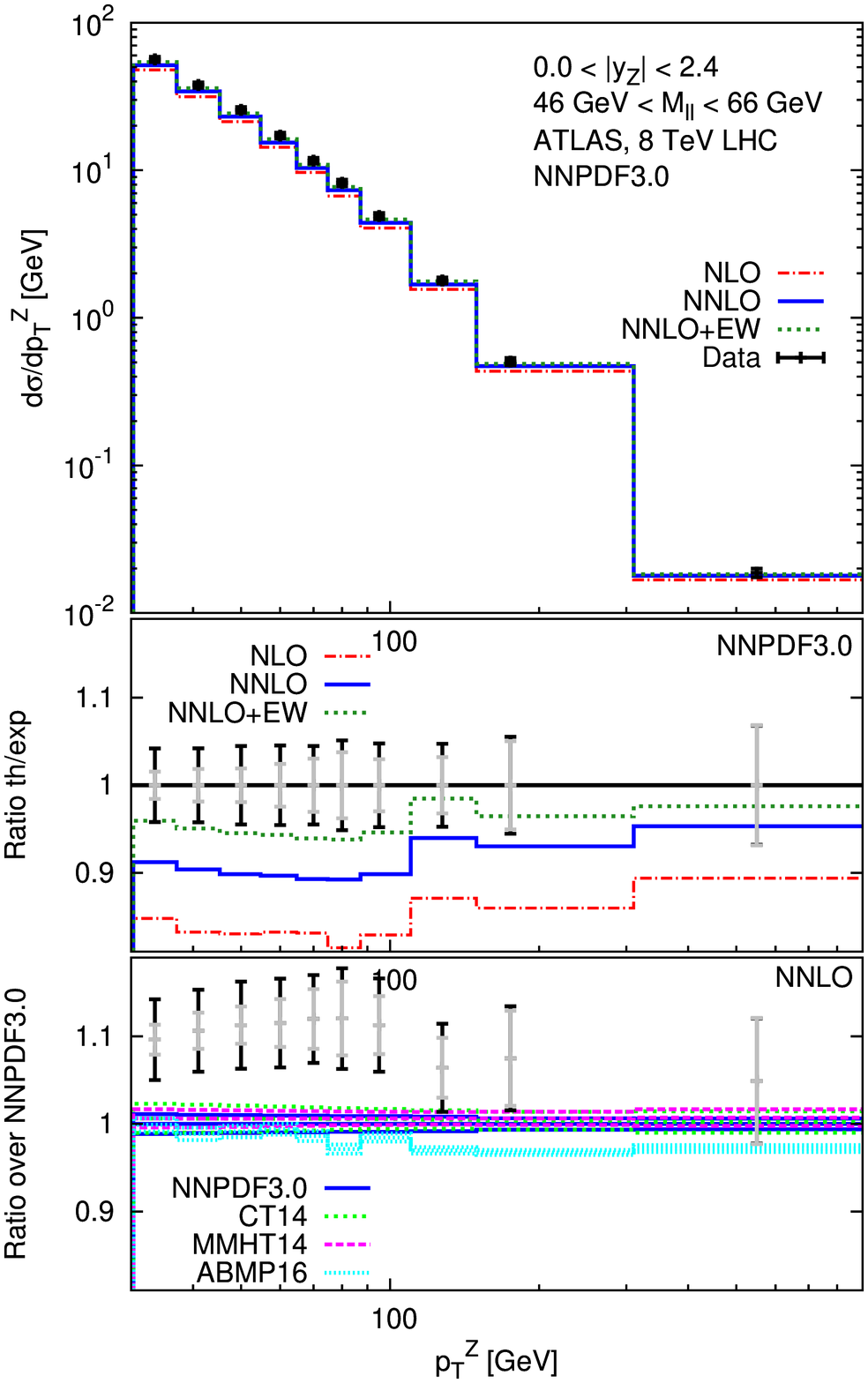}
	\includegraphics[width=0.32\textwidth]{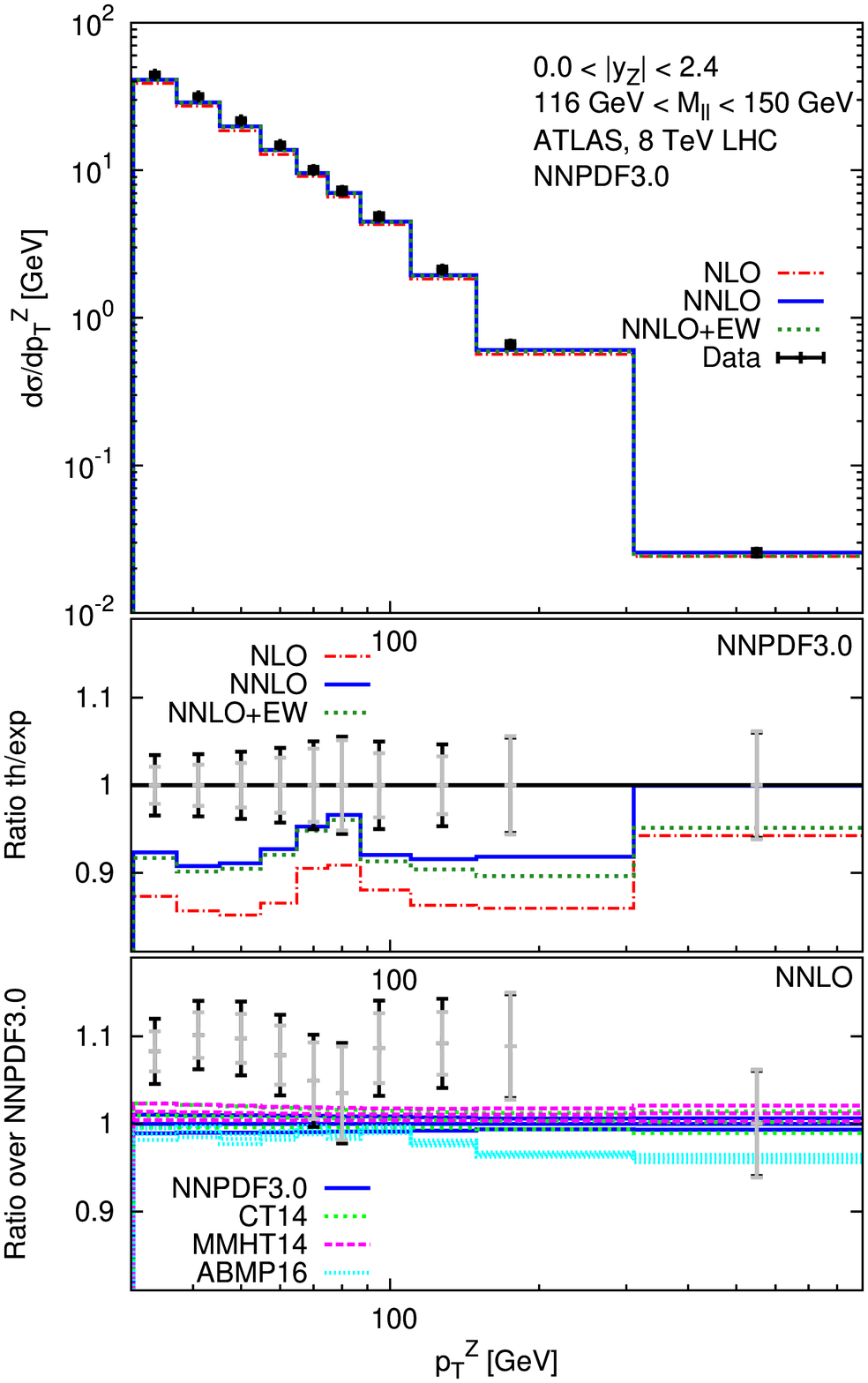}
	\caption{Same as Fig.~\ref{fig:theoryexp_atlas8tev_1} for 
          the two remaining low invariant-mass bins and the high invariant-mass bin.}
	\label{fig:theoryexp_atlas8tev_2}
\end{figure}

In Figs.~\ref{fig:theoryexp_atlas8tev_1} and~\ref{fig:theoryexp_atlas8tev_2} a similar comparison is performed for the off $Z$-peak  bins of 
the 8 TeV ATLAS measurement~\cite{Aad:2015auj}.  The NNLO QCD corrections again provide a positive shift of the NLO result that is
approximately independent of $\pt$, with NLO electroweak corrections causing a approximatively constant upwards (downwards) shift for the bins
below (above) the $Z$-peak.  
While the NNLO predictions are in better agreement with the data than the NLO ones, the data are again higher than the theoretical predictions.
The quantitative comparison of the NNPDF3.0, MMHT2014, CT14 and ABMP16 PDF sets using the $\chi^2_{\rm d.o.f.}$ defined previously is 
shown in Table~\ref{tab:chi2prefit:atlas8tev-mdist}.  In all cases an improvement is seen upon inclusion of the NNLO QCD corrections, while the
incorporation of the NLO electroweak corrections as well further improves the agreement in all individual bins below the $Z$ peak.  

\begin{table}[ht]
\centering
\caption{Same as Table~\ref{tab:chi2prefit:atlas7tev} for the ATLAS 8 TeV $\pt$ distributions 
in the low and high invariant-mass bins before their inclusion in the fit.}
\label{tab:chi2prefit:atlas8tev-mdist}
\begin{footnotesize}
\begin{tabular}{c|c|c|c|c|c|c}
\hline
  Bin  & Order &  $N_{\rm dat}$ &$\chi^2_{\rm d.o.f.}$ \tiny{(NN30)}
  &$\chi^2_{\rm d.o.f.}$ \tiny{(CT14)} &$\chi^2_{\rm d.o.f.}$ \tiny{(MMHT14)}&$\chi^2_{\rm d.o.f.}$ \tiny{(ABMP16)}\\ 
\hline
 $12<\Mll<20$ GeV  & NLO  & 8  & 2.5 & 2.2 & 1.9 & n.a.\\
               & NNLO     & 8  & 1.7 & 1.6 & 1.7 & 1.1\\
               & NNLO+EW  & 8  & 1.2 & 1.1 & 1.2 & 0.9\\
\hline
 $20<\Mll<30$ GeV & NLO   & 8  & 2.3 & 2.6 & 2.3 & n.a\\
               & NNLO     & 8  & 1.3 & 1.2 & 1.2 & 2.1\\
               & NNLO+EW  & 8  & 1.1 & 1.1 & 1.0 & 2.1\\
\hline
 $30<\Mll<46$ GeV  & NLO  & 8  & 1.4 & 1.3 & 1.0 & n.a\\
               & NNLO     & 8  & 1.1 & 1.0 & 1.0 & 0.7\\
               & NNLO+EW  & 8  & 0.8 & 0.8 & 0.8 & 0.6\\
\hline
 $46<\Mll<66$ GeV  & NLO  & 10  & 1.9 & 1.9 & 1.5 & n.a\\
               & NNLO     & 10  & 0.8 & 0.8 & 0.8 & 1.0\\
               & NNLO+EW  & 10  & 0.4 & 0.4 & 0.4 & 0.5\\
\hline
 $116<\Mll<150$ GeV  & NLO& 10  & 2.3 & 2.1 & 1.6 & n.a\\
               & NNLO     & 10  & 1.2 & 1.0 & 1.0 & 1.3\\
               & NNLO+EW  & 10  & 1.2 & 1.1 & 1.0 & 1.5\\
\hline
All bins   & NLO      & 44  & 1.3  & 1.2&1.1 & n.a\\
               & NNLO     & 44  & 1.0 &  0.9&0.9 & 1.1\\
               & NNLO+EW  & 44  &  1.0 & 1.0 & 1.0& 1.4 \\
\hline
\end{tabular}
\end{footnotesize}
\end{table}

\begin{figure}[tbp]
	\centering
	\includegraphics[width=0.32\textwidth]{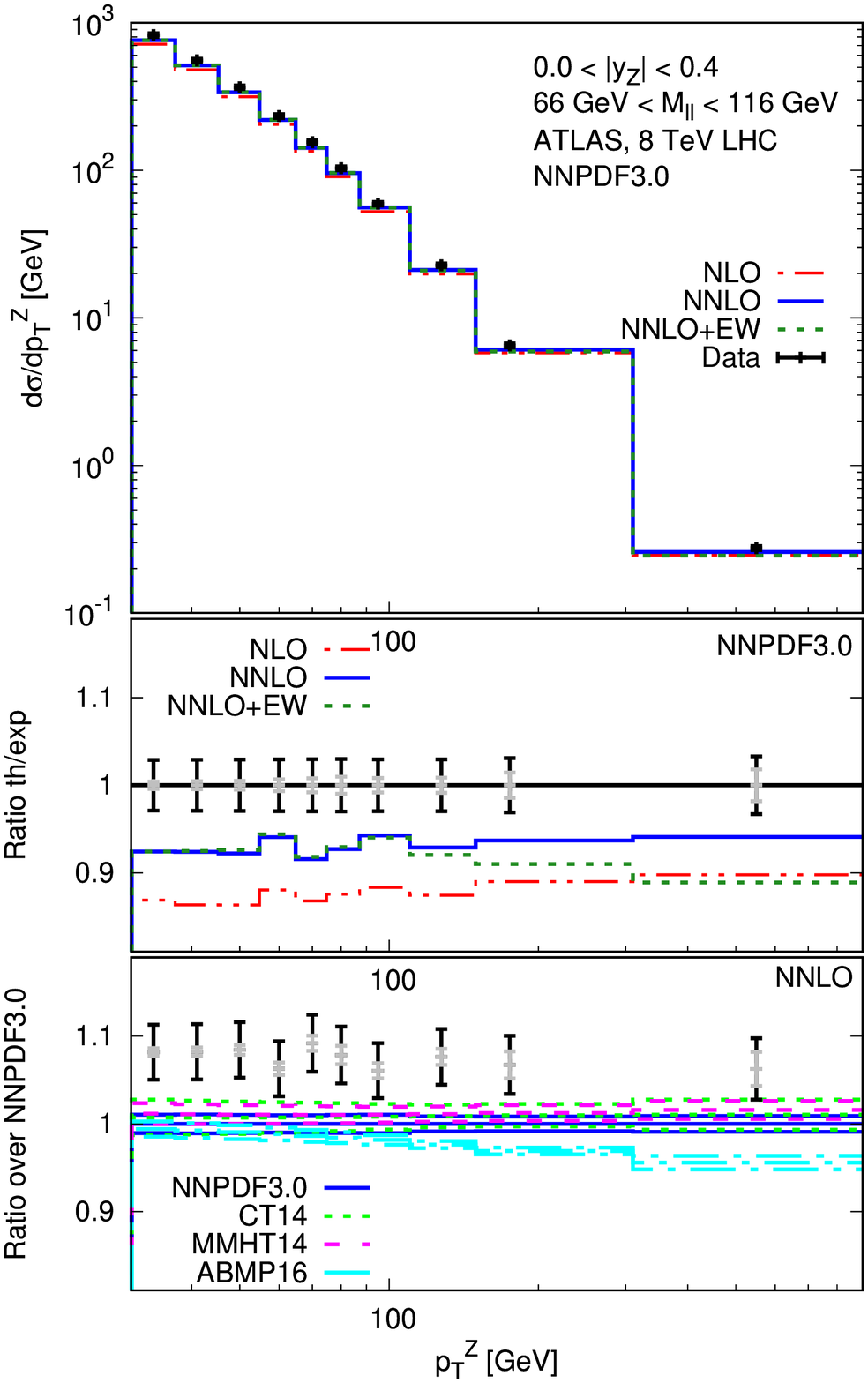}
	\includegraphics[width=0.32\textwidth]{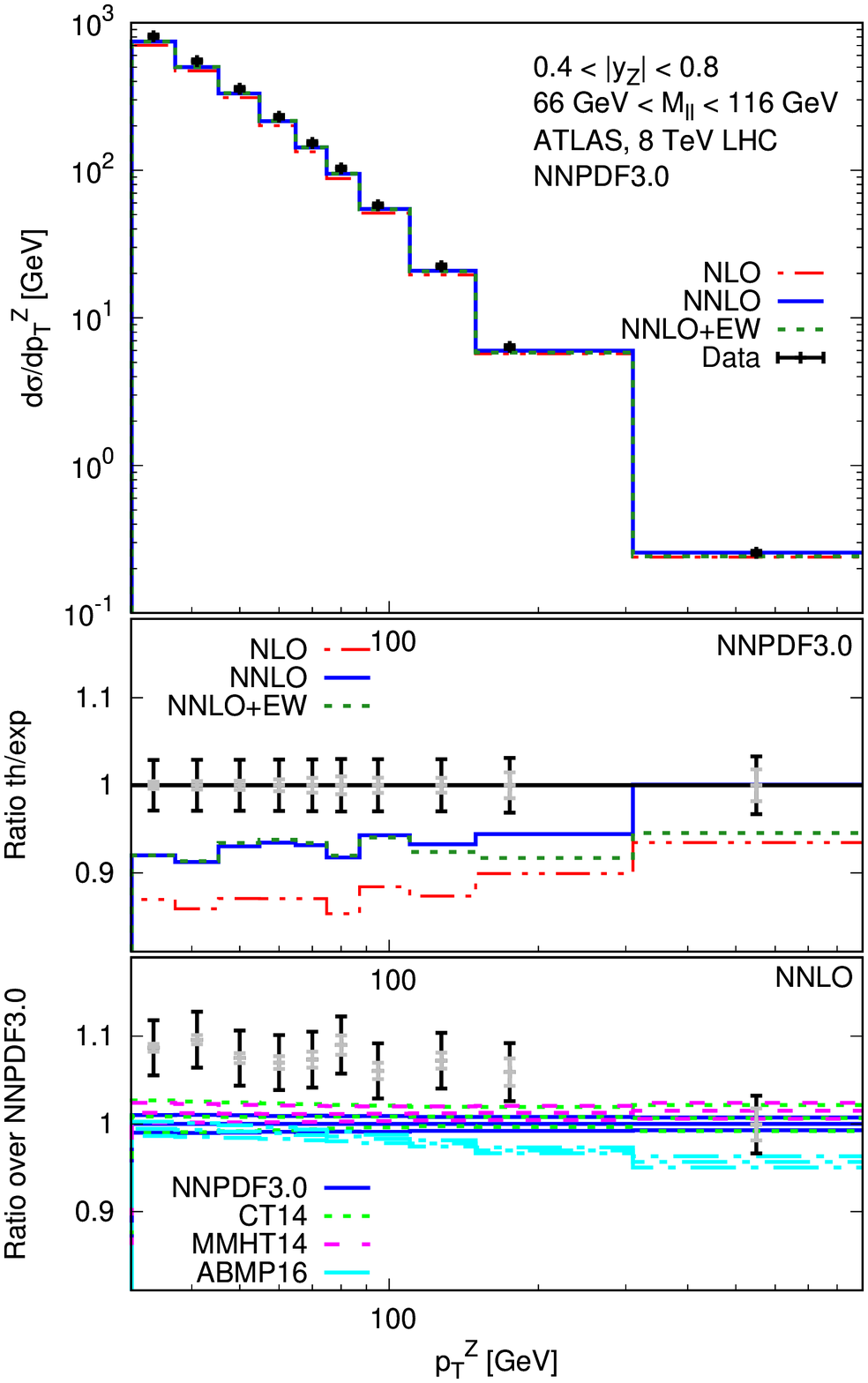}
	\includegraphics[width=0.32\textwidth]{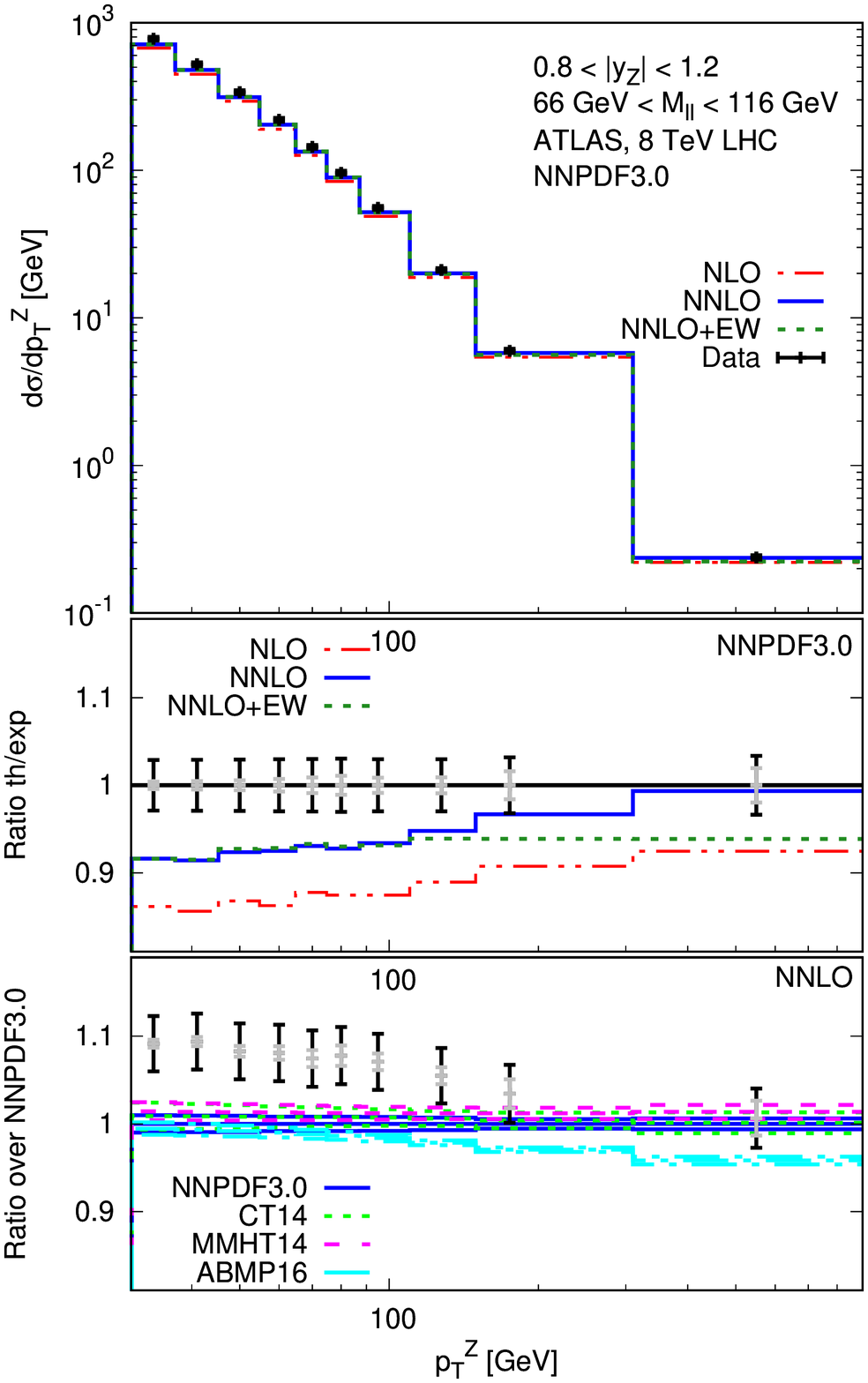}
	\caption{Same as Fig.~\ref{fig:theoryexp_atlas7tev_1} for the ATLAS 8 TeV on-peak data 
          divided into rapidity bins~\cite{Aad:2015auj}. The three lowest rapidity bins are displayed.}
	\label{fig:theoryexp_atlas8tev-a}
\end{figure}

\begin{figure}[tbp]
	\centering
	\includegraphics[width=0.32\textwidth]{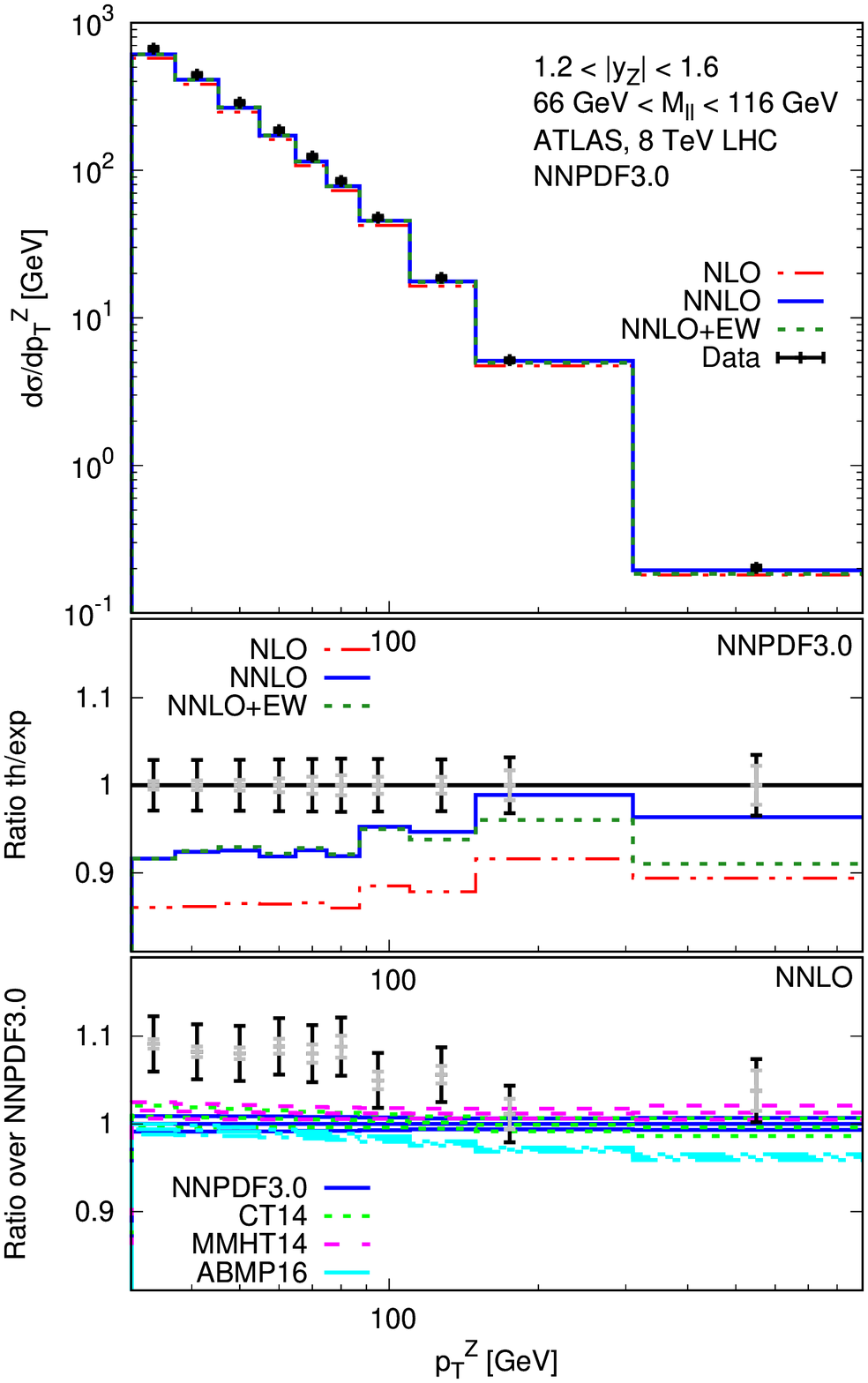}
	\includegraphics[width=0.32\textwidth]{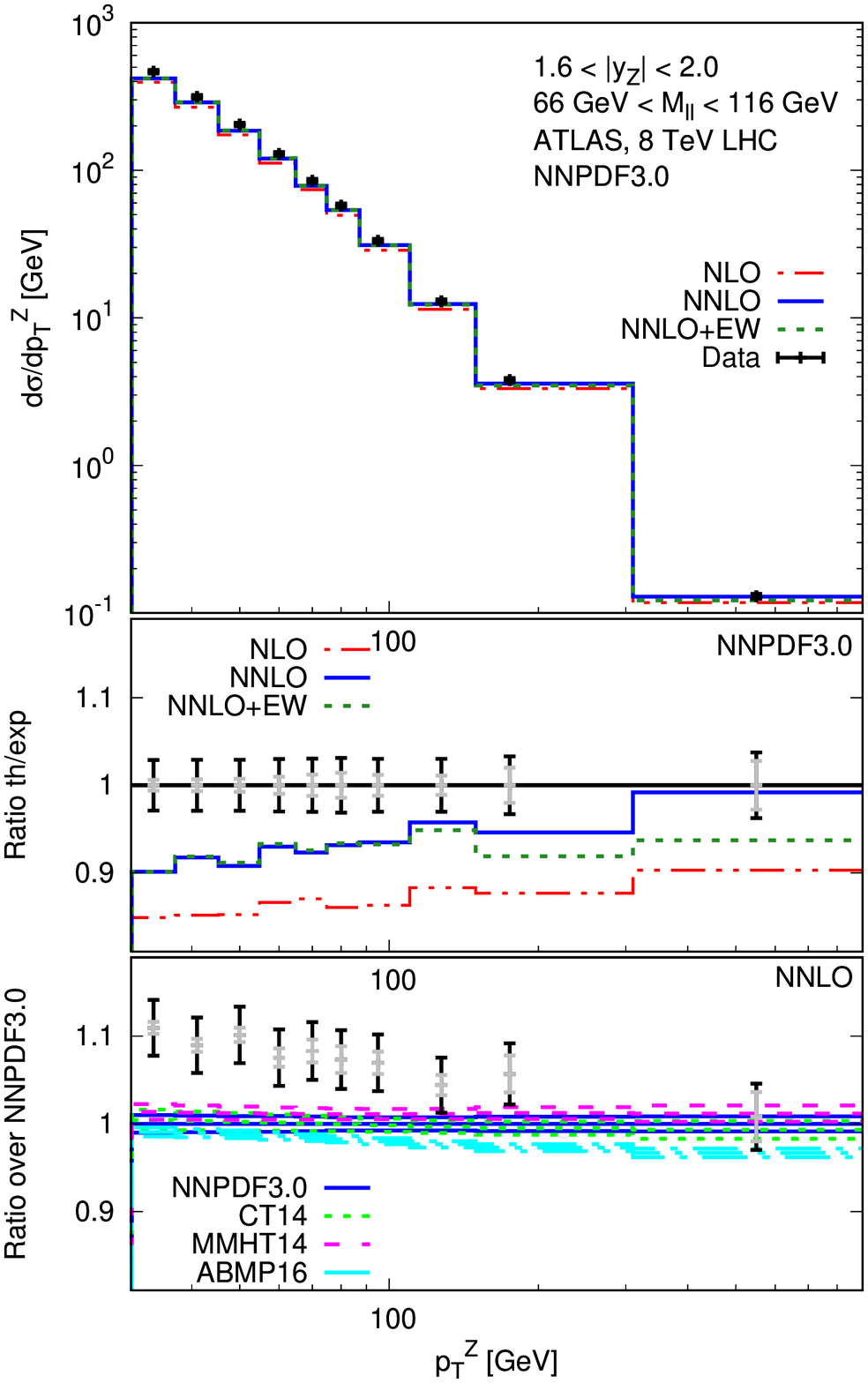}
	\includegraphics[width=0.32\textwidth]{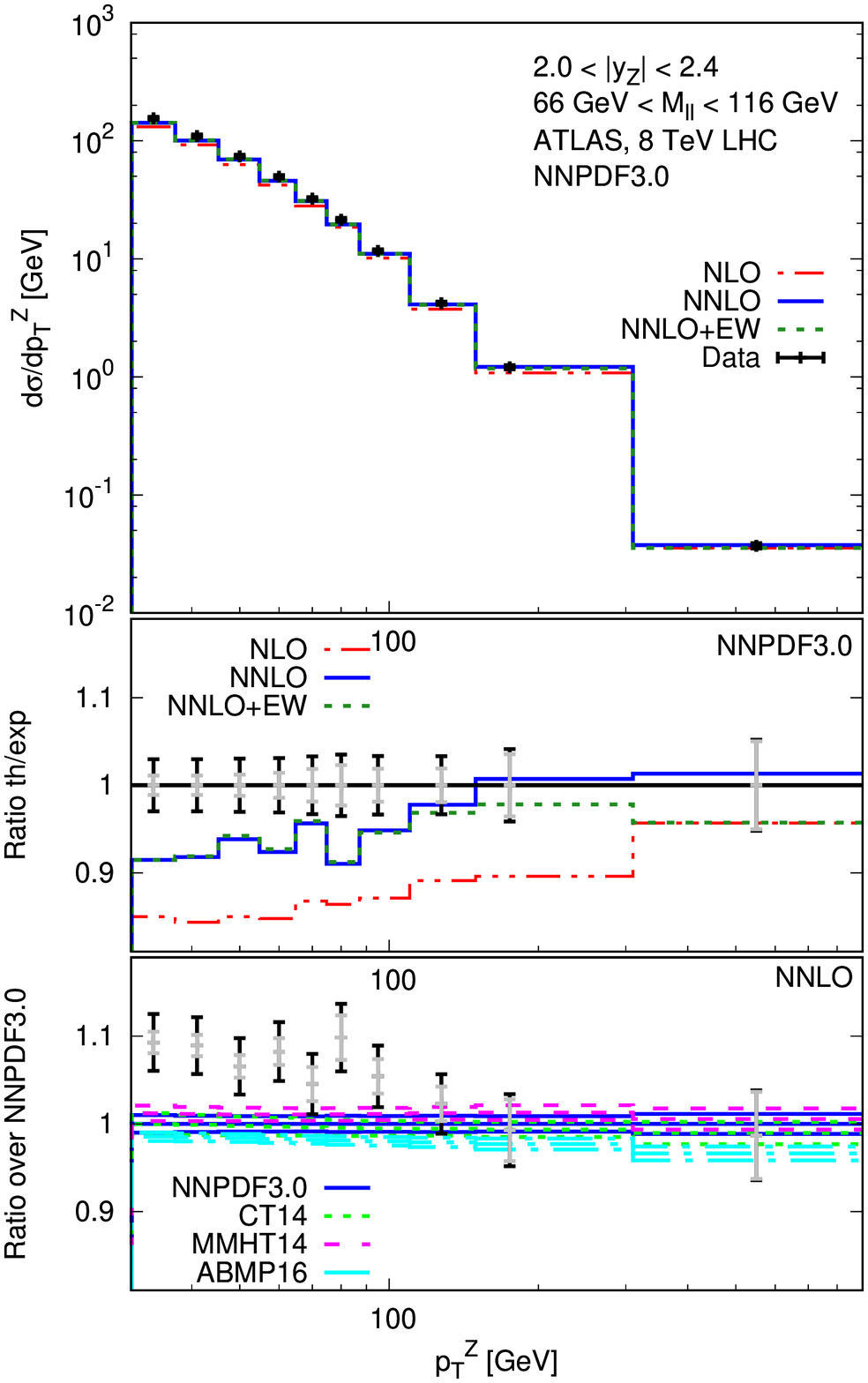}
	\caption{Same as Fig.~\ref{fig:theoryexp_atlas8tev-a} for the more forward three rapidity bins.}
	\label{fig:theoryexp_atlas8tev-b}
\end{figure}

We next consider the 8 TeV ATLAS data on the $Z$-peak divided into rapidity bins. The comparisons of NLO, NNLO and NNLO+EW theory
with data are shown in Figs.~\ref{fig:theoryexp_atlas8tev-a} and~\ref{fig:theoryexp_atlas8tev-b}, while the $\chi^2_{\rm d.o.f.}$ results for
NNPDF3.0, MMHT2014, CT14 and ABMP16 are shown in Table~\ref{tab:chi2prefit:atlas8tev-ydist}.  The general trends observed in this 
comparison are similar to those seen in the ATLAS 7 TeV comparison and in the comparison of the invariant mass binned data: the NNLO
corrections increase the NLO predictions by an amount almost independent of $\pt$, bringing theory closer to data.  
The quantitative comparison of $\chi^2_{\rm d.o.f.}$ in Table~\ref{tab:chi2prefit:atlas8tev-ydist} reveals that NNLO improves upon the NLO
description in four of the six rapidity bins for NNPDF3.0, while NNLO+EW improves upon NLO for all six bins.  For CT14 NNLO+EW improves
upon NLO for five of the six bins, while for MMHT the improvement is only observed for two bins.  One reason that the inclusion of the NNLO
corrections does not improve the theory/data agreement as significantly as in the other data sets is because the experimental error in this case 
is very small, and is dominated by the correlated systematic error.  Even if NNLO reduces the normalization difference between theory and
experiment, remaining shape differences between the predictions and data prevent a large improvement in $\chi^2_{\rm d.o.f.}$ from being
obtained.  This issue will arise again when we attempt to add this data set to the PDF fit.


\begin{table}[ht]
\centering
\caption{Same as Table~\ref{tab:chi2prefit:atlas7tev} for the ATLAS 8 TeV $\pt$ on-peak distributions 
in the separate rapidity bins before their inclusion in the fit.}
\label{tab:chi2prefit:atlas8tev-ydist}
\begin{footnotesize}
\begin{tabular}{c|c|c|c|c|c|c}
\hline
  Bin  & Order &  $N_{\rm dat}$ &$\chi^2_{\rm d.o.f.}$ \tiny{(NN30)}
  &$\chi^2_{\rm d.o.f.}$ \tiny{(CT14)} &$\chi^2_{\rm d.o.f.}$ \tiny{(MMHT14)}&$\chi^2_{\rm d.o.f.}$ \tiny{(ABMP16)}\\ 
\hline
 $0.0<\y<0.4$ & NLO      & 10  & 4.0 & 3.2 & 2.4 & n.a.\\
               & NNLO     & 10  & 2.7 & 2.7 & 2.6 & 2.7\\
               & NNLO+EW  & 10  & 3.4 & 3.2 & 3.1 & 5.4\\
\hline
 $0.4<\y<0.8$ & NLO      & 10  & 5.6 & 4.6 & 3.8& n.a.\\
               & NNLO     & 10  & 5.4 & 5.2 & 5.3& 3.3\\
               & NNLO+EW  & 10  & 4.0 & 3.9 & 3.7& 3.8\\
\hline
 $0.8<\y<1.2$  & NLO      & 10  & 5.8 & 3.8 & 3.0 & n.a.\\
               & NNLO     & 10  & 4.7 & 4.0 & 4.3& 2.1\\
               & NNLO+EW  & 10  & 2.3 & 2.0 & 1.9& 1.7\\
\hline
 $1.2<\y<1.6$  & NLO      & 10  & 4.5 & 3.2 & 2.5& n.a.\\
               & NNLO     & 10  & 5.1 & 4.0 & 4.6&3.0 \\
               & NNLO+EW  & 10  & 3.3 & 2.6 & 2.7& 2.5\\
\hline
 $1.6<\y<2.0$  & NLO      & 10  & 4.4 & 3.2 & 2.4& n.a.\\
               & NNLO     & 10  & 5.4 & 4.3& 5.0& 3.7\\
               & NNLO+EW  & 10  & 3.9 & 3.2 & 3.4& 3.0\\
\hline
 $2.0<\y<2.4$  & NLO      & 10  & 4.1 & 3.2 & 2.4& n.a.\\
               & NNLO     & 10  & 3.4 & 3.1 & 3.3& 3.2\\
               & NNLO+EW  & 10  & 2.6 & 2.3 & 2.4& 2.5\\
\hline
All bins   & NLO      & 60  & 3.4 &2.0 & 1.9& n.a.\\
               & NNLO     & 60  & 4.5& 4.0 & 4.4& 2.6\\
               & NNLO+EW  & 60  & 2.9 & 2.5 & 2.7 & 2.4\\
\hline
\end{tabular}
\end{footnotesize}
\end{table}

\begin{figure}[tbp]
	\centering
	\includegraphics[width=0.45\textwidth]{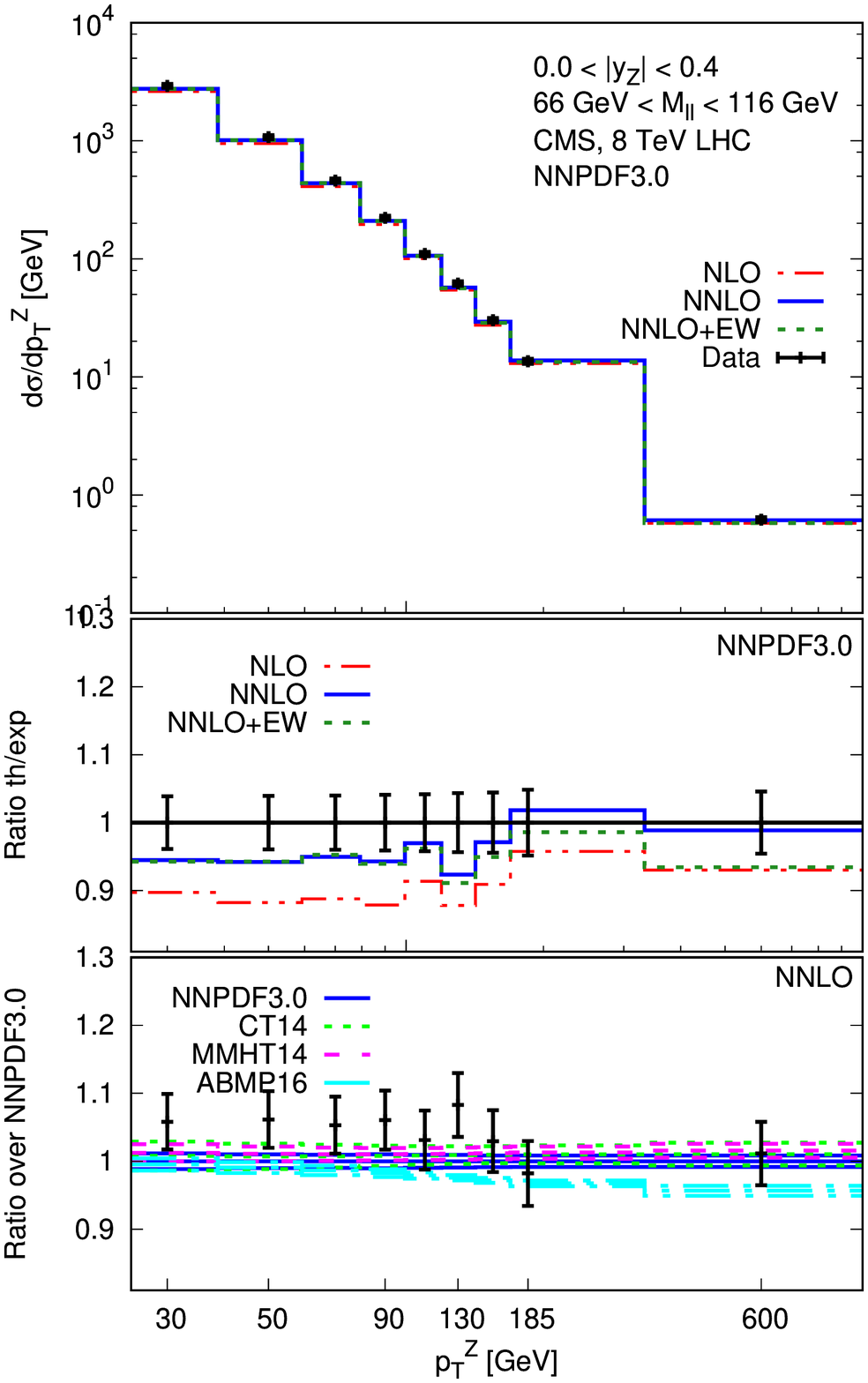}
	\includegraphics[width=0.45\textwidth]{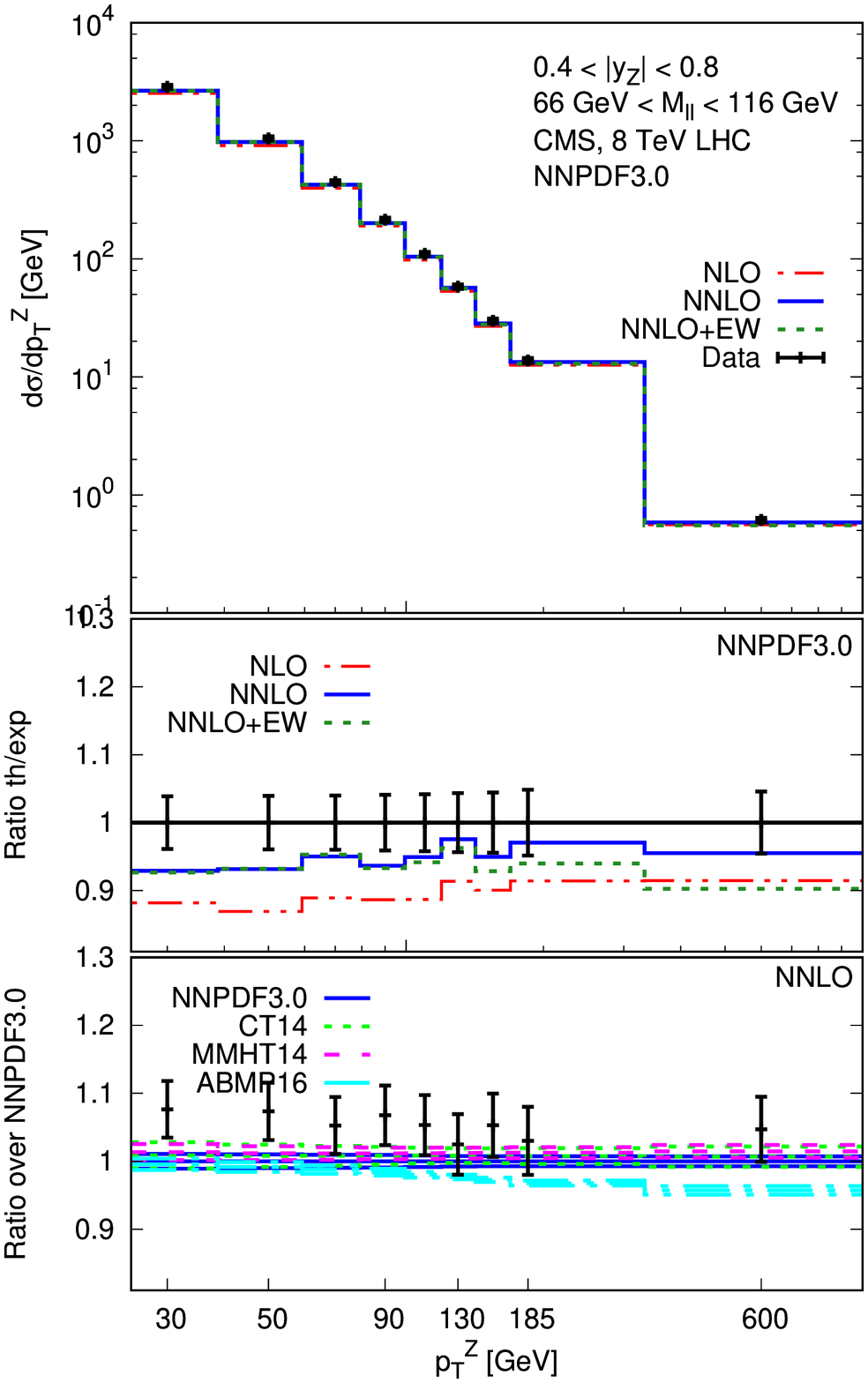}
	\caption{Same as Fig.~\ref{fig:theoryexp_atlas7tev_1} for the CMS on-peak 8 TeV data 
          divided into rapidity bins~\cite{Khachatryan:2015oaa}. Only the total uncertainty of 
          data points is displayed, given that separate statistical and uncorrelated uncertainties 
          are not available.}
	\label{fig:theoryexp_cms8tev-1}
\end{figure}

\begin{figure}[tbp]
	\centering
	\includegraphics[width=0.45\textwidth]{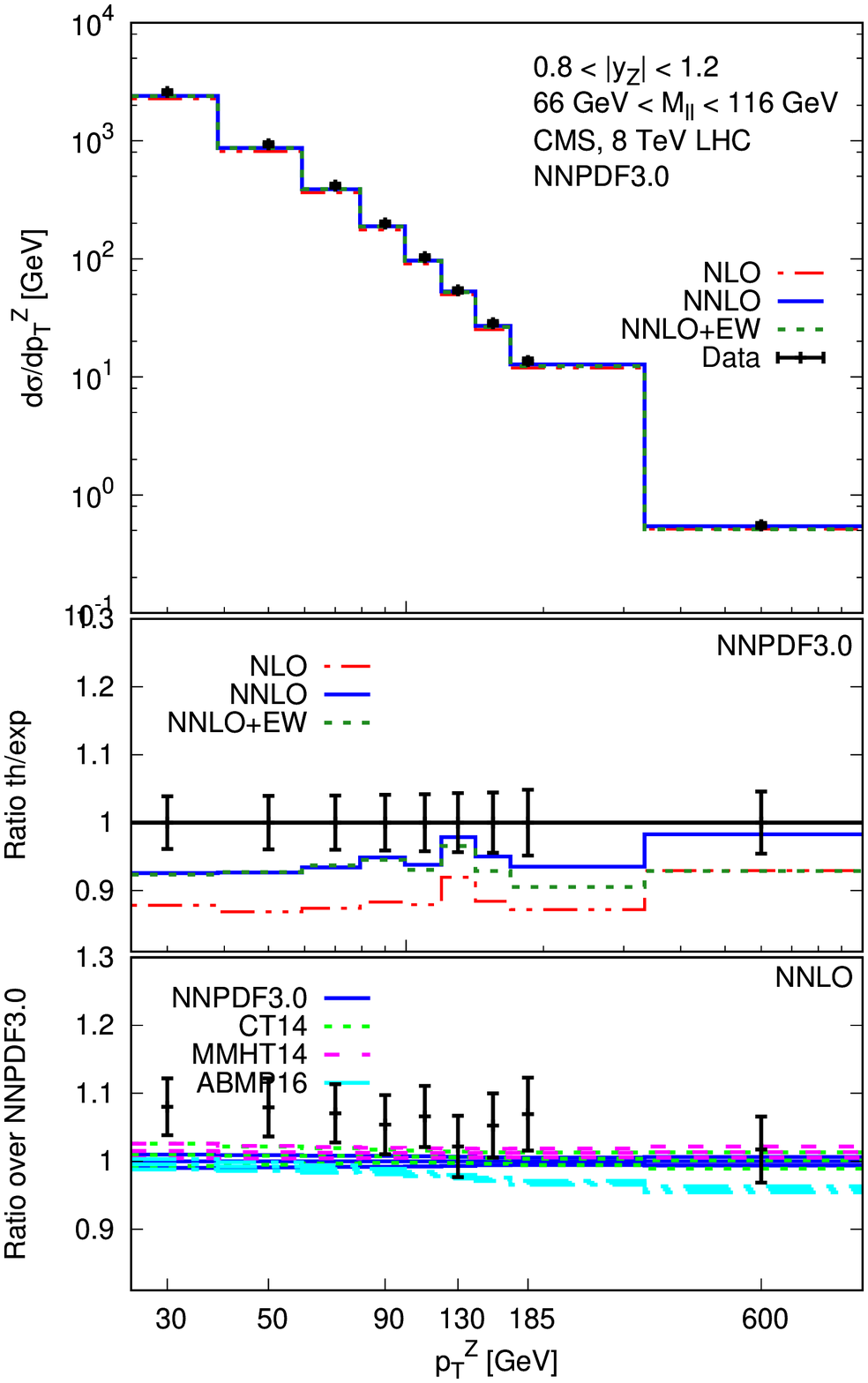}
	\includegraphics[width=0.45\textwidth]{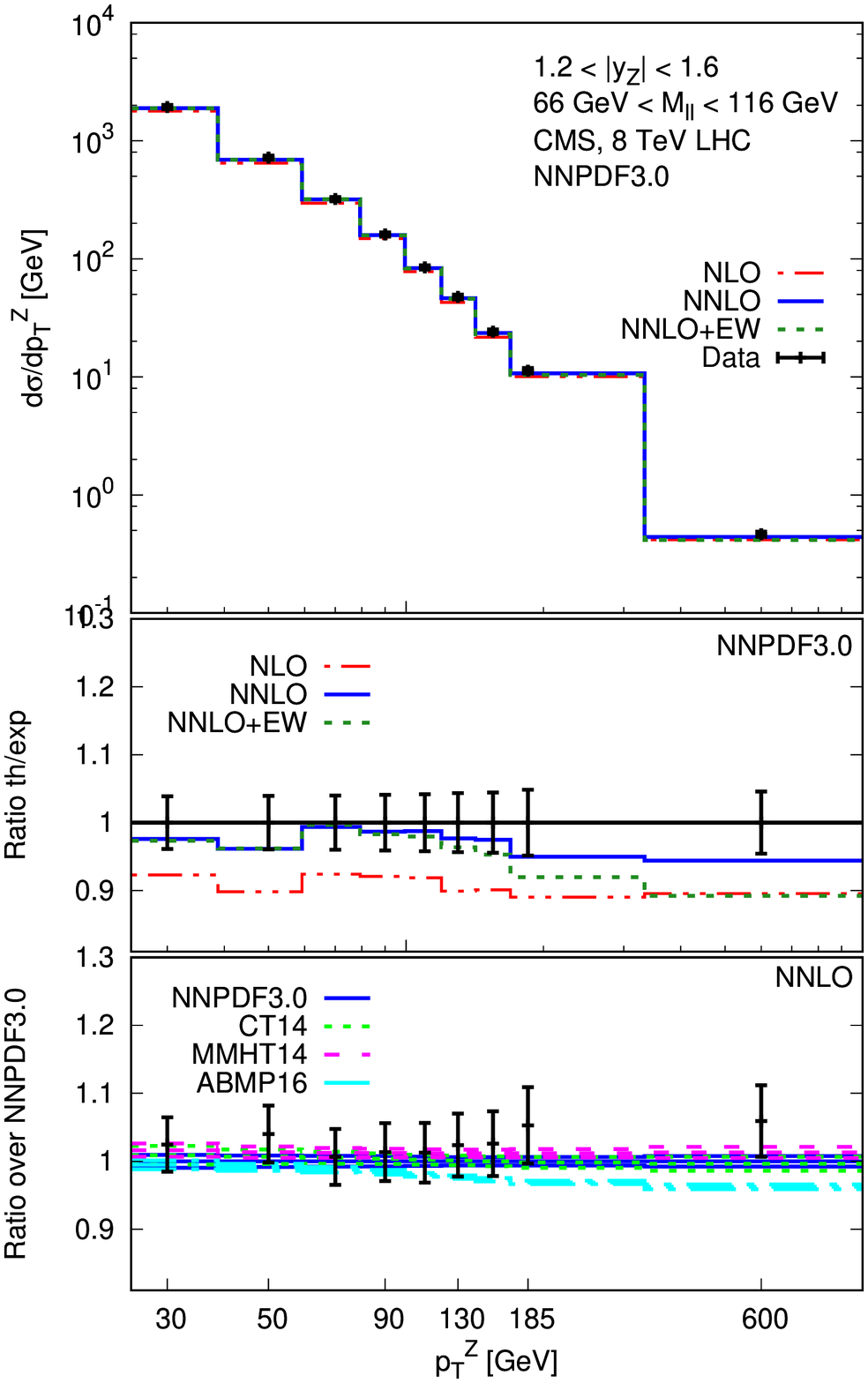}
	\caption{Same as Fig.~\ref{fig:theoryexp_cms8tev-1} for two higher rapidity bins.}
	\label{fig:theoryexp_cms8tev-2}
\end{figure}

Finally, in Figs.~\ref{fig:theoryexp_cms8tev-1} and~\ref{fig:theoryexp_cms8tev-2}, we show the comparison of the various theoretical 
predictions with the CMS 8 TeV data divided into rapidity bins~\cite{Khachatryan:2015oaa}.  The $\chi^2_{\rm d.o.f.}$ is shown in 
Table~\ref{tab:chi2prefit:cms8tev-ydist}.  As discussed when describing the data in Sect. 2, we focus on the region $|\y|<1.6$.
Including NNLO corrections improves the  agreement between theory and data in all four rapidity bins, while adding NLO EW corrections 
further improves the comparison in all but the highest rapidity bins.  We note that the CMS relative errors are larger than those found by 
ATLAS, and the issues seen in the $\chi^2_{\rm d.o.f.}$ comparison are not as pronounced as for the ATLAS 8 TeV data set.  
Interestingly, even though each individual rapidity bin is improved upon including NNLO, the $\chi^2_{\rm d.o.f.}$ combining all bins is 
slightly worsened at NNLO, again showing the impact of the correlated
uncertainties when attempting to describe these very precise data
sets.  Fitting the data modifies the PDF shape, thus significantly improving the data description. 

\begin{table}[htb]
\centering
\caption{Same as Table~\ref{tab:chi2prefit:atlas7tev} for the CMS 8 TeV $\pt$ distributions 
in the separate rapidity bins before their inclusion in the fit.}
\label{tab:chi2prefit:cms8tev-ydist}
\begin{footnotesize}
\begin{tabular}{c|c|c|c|c|c}
\hline
  Bin  & Order &  $N_{\rm dat}$ &$\chi^2_{\rm d.o.f.}$ (NN30)
  &$\chi^2_{\rm d.o.f.}$ (CT14) &$\chi^2_{\rm d.o.f.}$ (MMHT14)\\ 
\hline
 $0.0<\y<0.4$ GeV  & NLO     & 9  & 3.1 & 2.6 & 2.2\\
               & NNLO                    & 9  & 2.2 & 2.4 & 2.3\\
               & NNLO+EW             & 9  & 1.4 & 1.4 & 1.3\\
\hline
 $0.4<\y<0.8$ GeV & NLO      & 9  & 2.4 & 1.9 & 1.5\\
               & NNLO                    & 9  & 1.4 & 1.4 & 1.3\\
               & NNLO+EW             & 9  & 1.9 & 1.9 & 1.7\\
\hline
 $0.8<\y<1.2$ GeV  & NLO     & 9 & 2.3 & 2.0 & 1.6\\
               & NNLO                   & 9  & 1.6 & 1.4 & 1.4\\
               & NNLO+EW            & 9  & 1.2 & 1.2 & 1.0\\
\hline
 $1.2<\y<1.6$ GeV  & NLO    & 9  & 1.6 & 2.0 & 1.9\\
               & NNLO                   & 9  & 1.4 & 1.4 & 1.3\\
               & NNLO+EW            & 9  & 2.8 & 3.1 & 2.8\\
\hline
All bins   & NLO      & 36  &  3.3 & 3.2 & 3.3\\
               & NNLO     & 36  & 3.5 & 3.5 & 3.7 \\
               & NNLO+EW  & 36  & 3.9 & 4.0 & 4.0 \\
\hline
\end{tabular}
\end{footnotesize}
\end{table}

\section{Inclusion of the $\pt$ distribution in PDF fits} 
\label{sec:fits}

In this Section we first look at the correlation between the measured distributions and the various PDF combinations, which provides a first
intuition for what parton distributions and at what value of $x$ we should expect to observe the largest impact when including these data 
in the fit. We then add each data set separately to a DIS HERA-only fit to determine basic compatibility of different data sets and to assess the 
impact of including EW corrections. Finally, we perform a fit adding $\pt$ data to a global data set to estimate the impact of including
these data in a realistic PDF determination.

\subsection{Correlations between PDFs and $\pt$ measurements}
\label{sec:correlations}
\begin{figure}[tbp]
	\centering
        \includegraphics[width=0.98\textwidth]{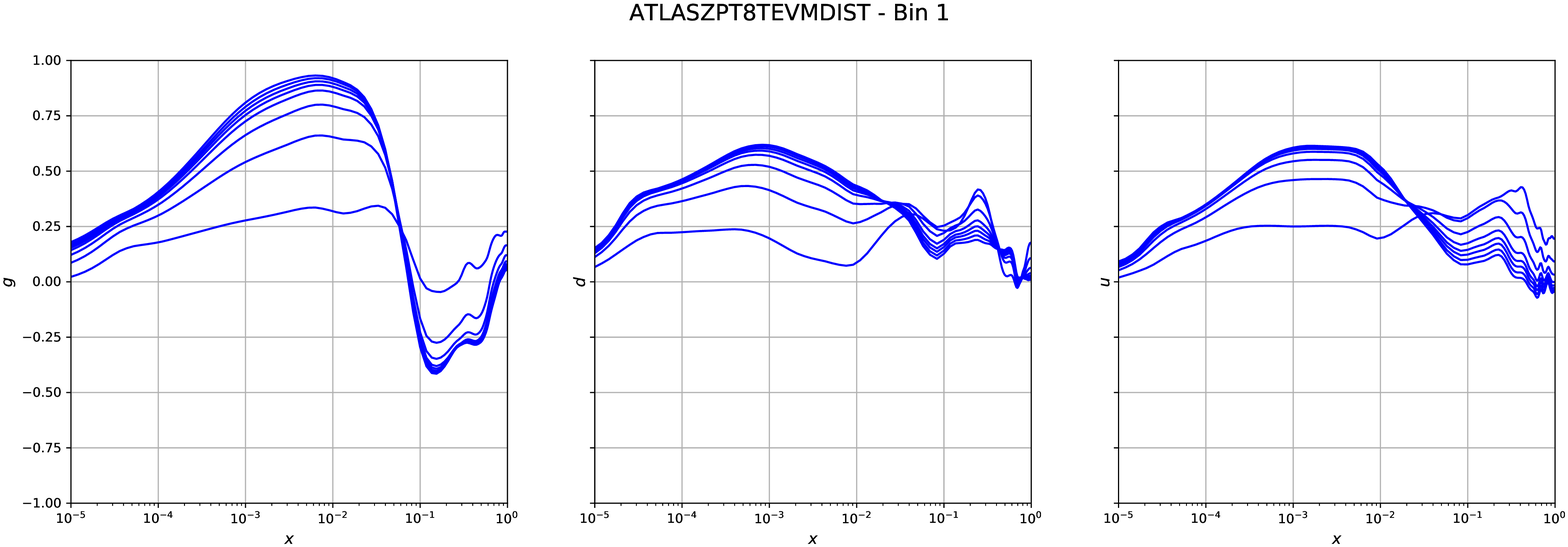}
        \includegraphics[width=0.98\textwidth]{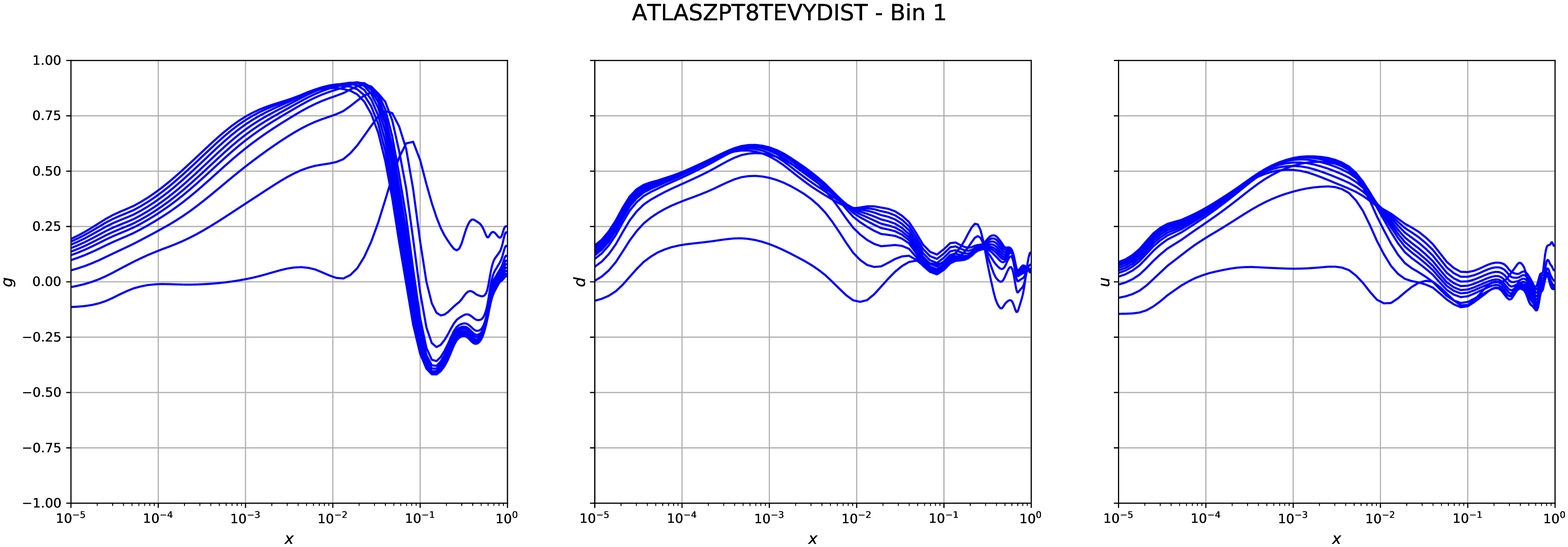}
	\caption{Top row: Correlations between the ATLAS 8 TeV $\pt$ measurement binned in the invariant mass of the lepton pair 
                 and the gluon, up- and down-quark distributions as a function of $x$ ($\pt$ bin 1; 12 GeV $<  M_{ll} < 20$ GeV;  $0.0 < |Y_Z| < 2.4$).
		Bottom row: Correlations between the ATLAS 8 TeV $\pt$ measurement in the $Z$-peak invariant mass bin, binned in
                 rapidity of the vector boson and the gluon, up- and down-quark distributions as a function of $x$ ($\pt$ bin 1;
                  66 GeV $<  M_{ll}  < 116$ GeV; $0.0 < |Y_Z| < 0.4$).}
	\label{fig:atlas8tevmdistbin1-pdfobs-corr}
\end{figure}

To determine the specific PDFs and regions in $x$ for which the $Z$-boson transverse momentum distribution measurements from ATLAS and 
CMS provide the most stringent constraints we study the correlation coefficient as a function of $x$ ($\rho(x)$), between PDFs at a 
given scale $Q$ and each bin of the measurements included in the present analysis. 
In Figure~\ref{fig:atlas8tevmdistbin1-pdfobs-corr} we plot the correlations, computed using the SMPDF code~\cite{Carrazza:2016htc}, of
the gluon, up-quark and down-quark distributions with the lowest invariant mass bin of the ATLAS 8 TeV measurement, and with the on-peak 
8 TeV measurement of ATLAS, for the lowest rapidity bin. Each line corresponds to one $\pt$ bin. 
These are representative examples, the pattern of correlations found for the other measurements is similar.  
We observe a strong correlation between the gluon distribution in the region $x \approx 10^{-3}-10^{-2}$ with the $\pt$ measurements, 
with the correlation coefficient reaching nearly 90\%.  Slightly weaker correlations of approximately 60\% are found for the up-quark and 
down-quark distributions.  These plots make it clear that these data sets have a strong potential to improve our knowledge of PDFs in 
the $10^{-3}-10^{-2}$ region. The largest $\pt$ bins are correlated to the $10^{-2}-10^{-1}$ region, thus
an increase in the experimental statistics in that region would provide a stronger constraint also in the large-$x$ region.

\clearpage

\subsection{Impact of the  $\pt$ data on a DIS HERA-only fit}

We begin by assessing the quality of a fit to the HERA DIS data upon inclusion of the available $\pt$ data at 8 TeV. The inclusion of the
normalized ATLAS 7 TeV data is problematic and we discuss it separately in Sect.~\ref{sec:norm}. 
We perform several fits that add the individual ATLAS and CMS data sets to HERA separately and in various combinations.  
As discussed in previous Sections we impose the following cuts on the $\pt$ data:
\begin{align*}
& \pt\, > 30\,{\rm GeV}\\
& |\y| < 1.6 \qquad ({\rm CMS}\,{\rm only}).
\end{align*}
These constraints leave us with 60 data points for the ATLAS 8 TeV doubly-differential distributions in rapidity and $p_T$ on the
$Z$-peak, 44 data points for the ATLAS 8 TeV doubly-differential distributions in the dilepton invariant mass and $p_T$, and 36 data 
points for the CMS  8 TeV doubly-differential distributions in rapidity and  $p_T$ on the $Z$-peak.  Additionally, we consider
fits using pure NNLO QCD theory and fits with NNLO QCD and NLO EW corrections combined.  In the pure NNLO fits we remove the 
$\pt$ bins for which the EW corrections are larger than the sum quadrature of the statistical and uncorrelated systematic uncertainty of 
that data point to avoid fitting EW effects.  This imposes the additional constraints
\begin{align*}
& \pt\, < 150\,{\rm GeV}\qquad ({\rm ATLAS}\,8\,{\rm TeV, peak\,region})\\
& \pt\, < 170\,{\rm GeV}\qquad ({\rm CMS}\,8\,{\rm TeV}).
\end{align*}
These  cuts reduce the number of data points to 48 for the ATLAS 8 TeV doubly-differential distributions in rapidity and  $p_T$ on the 
$Z$-peak, 44 data points for the ATLAS 8 TeV doubly-differential distributions in invariant mass and $p_T$, and 28 data points for the 
CMS  8 TeV doubly-differential distributions in rapidity and $p_T$ in the $Z$-peak region.

Since we have considered numerous combinations of the available data and several different settings, we begin by summarizing the 
fits in Table~\ref{tab:descrfitsHERA8}. These are labelled (a)-(j).  Our baseline fit with only HERA data is labelled (a).  
Fits (b) and (c) add individually the ATLAS 8 TeV data and CMS 8 TeV data sets. Fit (d) adds all 8 TeV data sets.  
A new feature we find necessary in our analysis is the inclusion of an additional uncorrelated uncertainty. 
This uncertainty can be due to a combination of Monte-Carlo integration uncertainties on the computationally 
expensive NNLO theoretical calculations, to residual theoretical uncertainties not accounted for in our fit, 
and to possible underestimated experimental errors.  
The need and approximate size of this contribution to the uncertainty can be inferred from an analysis based on modelling the NLO and
NNLO theoretical predictions and their fluctuations along the lines of the one described for inclusive jet production in~\cite{Carrazza:2017bjw}.
The addition of this new effect is needed to obtain a 
good $\chi^2$ in our fit, as shown later in this section.  To study the stability of our fit with respect to this 
uncertainty we consider the values 0\%, 0.5\%, and 1\%.  Fits (b)-(d) use a 1\% uncorrelated uncertainty, while fits (e)-(g) use 0.5\%. 
 This uncertainty is removed in fits (h)-(j).  We will see later that the fitted PDFs are insensitive to the value of this parameter.

 \begin{table}[ht]
 \centering
 \caption{Overview of fits run with HERA-only as a baseline. For each fit, we indicate
 which measurements from ATLAS and CMS has been included, whether an
 uncorrelated uncertainty has been added to the $\chi^2$ (in
 brackets unless it is set to 0).}
 \label{tab:descrfitsHERA8}
 \begin{small}
 \begin{tabular}{c|c|ccc|ccc|ccc}
 \hline
                      & (a) & (b)        & (c) & (d) & (e) & (f) & (g) & (h) & (i) & (j)\\
 \hline
 HERA            & y     & y         &y        &y         & y           &y           &y            &y & y & y\\ 
 ATLAS8TEV  & n     & y(1\%) &n        &y(1\%) & y(0.5\%)&n           &y(0.5\%) &y & n & y\\
 CMS8TEV     & n     & n         &y(1\%)&y(1\%) & n           &y(0.5\%)&y(0.5\%) &n & y & y\\ 
 \hline
 \end{tabular}
 \end{small}
 \end{table}

The results of fits (a)-(j) are summarized in Table~\ref{tab:overview8}.
For each fit the $\chi^2$ per degree of freedom ($\chi^2_{{\rm d.o.f.}}$) of 
the experiments included in the fit, and of the prediction for the observables not included in the fit (in brackets), are
displayed.  The additional uncorrelated uncertainty added to the fit is denoted by $\Delta$.  
We have repeated the baseline HERA-only fit (a) at the beginning of each Table section for ease of comparison.  
A few things are apparent from the table.  
\begin{itemize}

\item The addition of $\Delta$ improves the description of the ATLAS 8 TeV on-peak and CMS 8 TeV data.  
The $\chi^2_{{\rm d.o.f.}}$ decreases from 1.66 to 0.77 for the ATLAS 8 TeV set and from 2.51 to 1.21 for the CMS 8 TeV set as 
$\Delta$ is changed from 0\% to 1\% in the baseline fit.  This effect is less noticeable for the invariant-mass binned ATLAS data 
due to the slightly larger errors for this set. 

\item Comparing fit (b) (where only the ATLAS 8 TeV data is fit along with HERA) to fit (c) (where only the CMS 
8 TeV data is fit together with HERA) shows that the ATLAS 8 TeV data is slightly more consistent with HERA than CMS.  
The  $\chi^2_{{\rm d.o.f.}}$ is below one for the ATLAS sets in fit (b) after including them in the fit, while it is at 1.21 in (c) when CMS 
is combined with HERA.  

\item Fit (d) shows that it is possible to obtain a reasonably good fit of ATLAS 8 TeV data, CMS 8 TeV data, and HERA with the inclusion 
of a $\Delta=1\%$ additional uncorrelated uncertainty.  Reducing this uncertainty to 0.5\% in fit (g) leads to a noticeably worse description
of the CMS data.  Both the CMS and on-peak ATLAS 8 TeV data sets get a worse $\chi^2_{{\rm d.o.f.}}$ if $\Delta$ is removed completely, 
as in fit (j).

\item It is clear from the table that the ATLAS 7 TeV measurement is inconsistent with the other data sets.  We discuss this further in Section 5.3.

\end{itemize}

\begin{table}[ht]
\centering
\caption{Fully correlated $\chi^2_{{\rm d.o.f.}}$ for the fits described in Table~\ref{tab:descrfitsHERA8}. 
The numbers in brackets correspond to the $\chi^2$ for experiments which are not fitted.  
In particular the ATLAS 7 TeV data are not fitted in any of these fits. The total $\chi^2$ is
  computed over all baseline HERA data the included $\pt$ distributions.}
\label{tab:overview8}
\begin{small}
\begin{tabular}{c|c|ccccc}
\hline
fit id & extra $\Delta$ & $\chi^2_{\rm ATLAS7tev}$ & $\chi^2_{\rm ATLAS8tev,m}$
  &$\chi^2_{\rm ATLAS8tev,y}$ & $\chi^2_{\rm CMS8tev}$ & $\chi^2_{\rm tot}$\\
\hline
(a) & 1\% & (21.8) & (1.00) & (1.56) & (1.55) & 1.168 \\  
\hline
(b) & 1\% & (19.6)&  0.91  &  0.70  & (1.61)   & 1.146 \\
(c) & 1\% & (16.2)& (1.04) & (1.56) &  1.21   & 1.176\\ 
(d) & 1\% & (18.0)&  0.90   & 0.77  &  1.42   & 1.156\\ 
\hline
(a) & 0.5\% & (27.6) & (1.10) & (2.83) & (2.46) & 1.168 \\
\hline
(e) & 0.5\% & (23.0) &  0.99   &  1.05  & (3.01)   & 1.168 \\ 
(f)  & 0.5\% & (20.5) & (1.13)  & (3.15) &  1.91    &1.198\\ 
(g)  & 0.5\% & (21.4) &  0.99   &  1.29  &  2.44    &1.207 \\ 
\hline
(a) & no  & (30.6) & (1.15) & (4.65) & (3.46)  & 1.168 \\
\hline
(h) & no & (25.5) & 1.02   &  1.66  & (4.79) & 1.193 \\ 
(i)  & no & (19.5) & (1.28) & (5.44) & 2.51   &1.225 \\ 
(j)  & no & (24.5) & 1.03   & 2.09   & 3.59    &1.251 \\ 
\hline
\end{tabular}
\end{small}
\end{table}

\begin{figure}[htbp]
        \centering
        \includegraphics[width=0.45\textwidth]{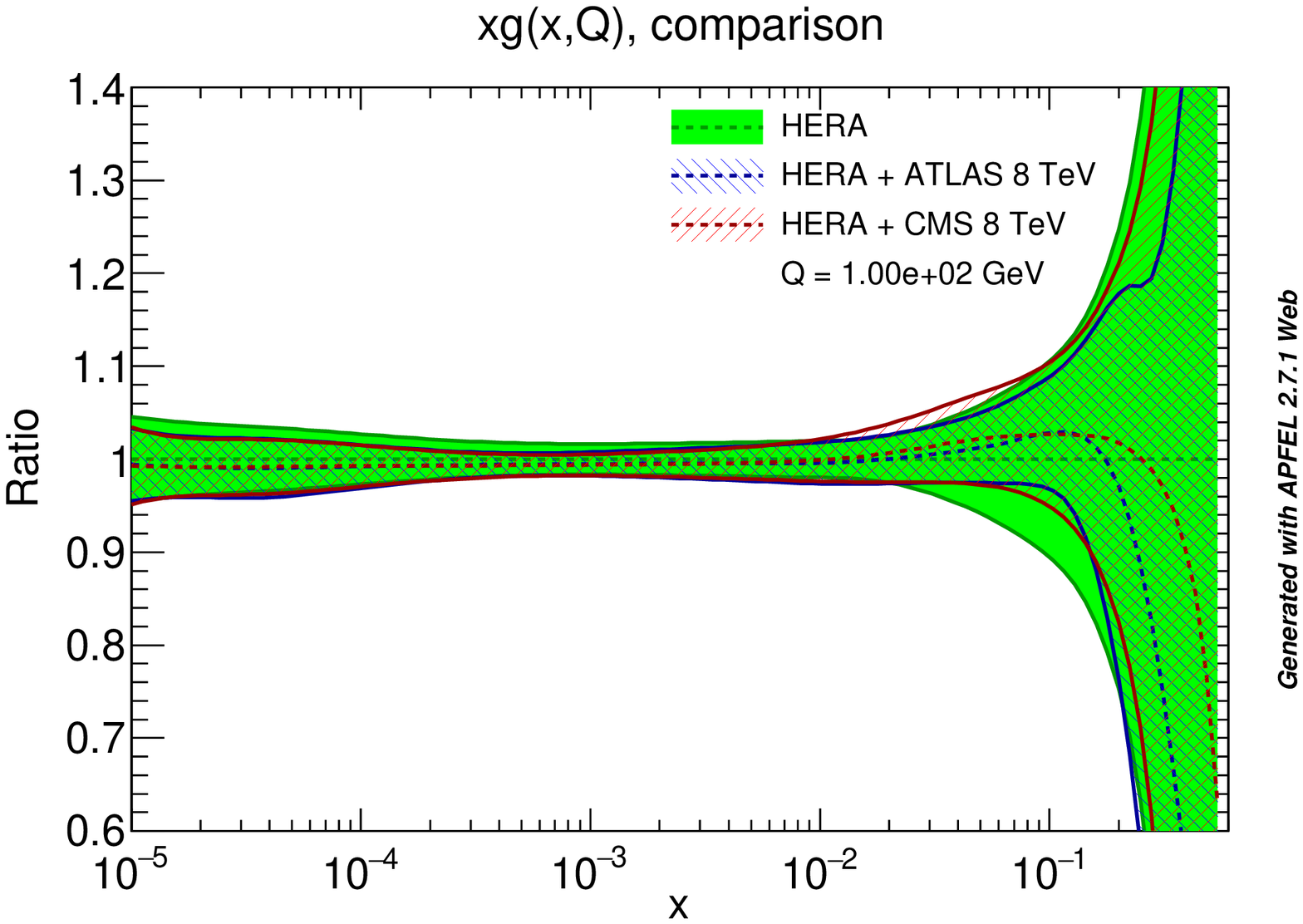}
        \includegraphics[width=0.45\textwidth]{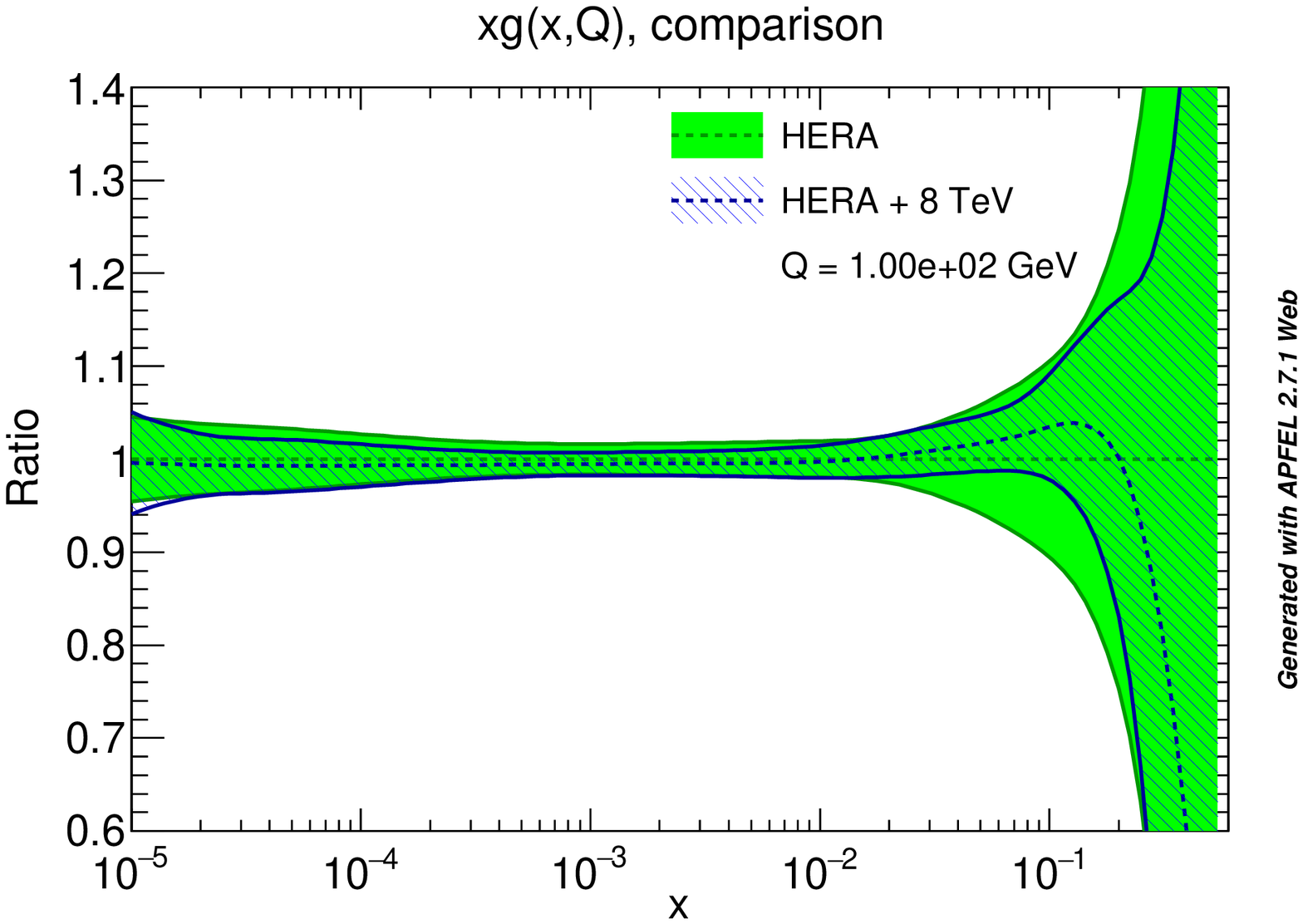}\\
        \includegraphics[width=0.45\textwidth]{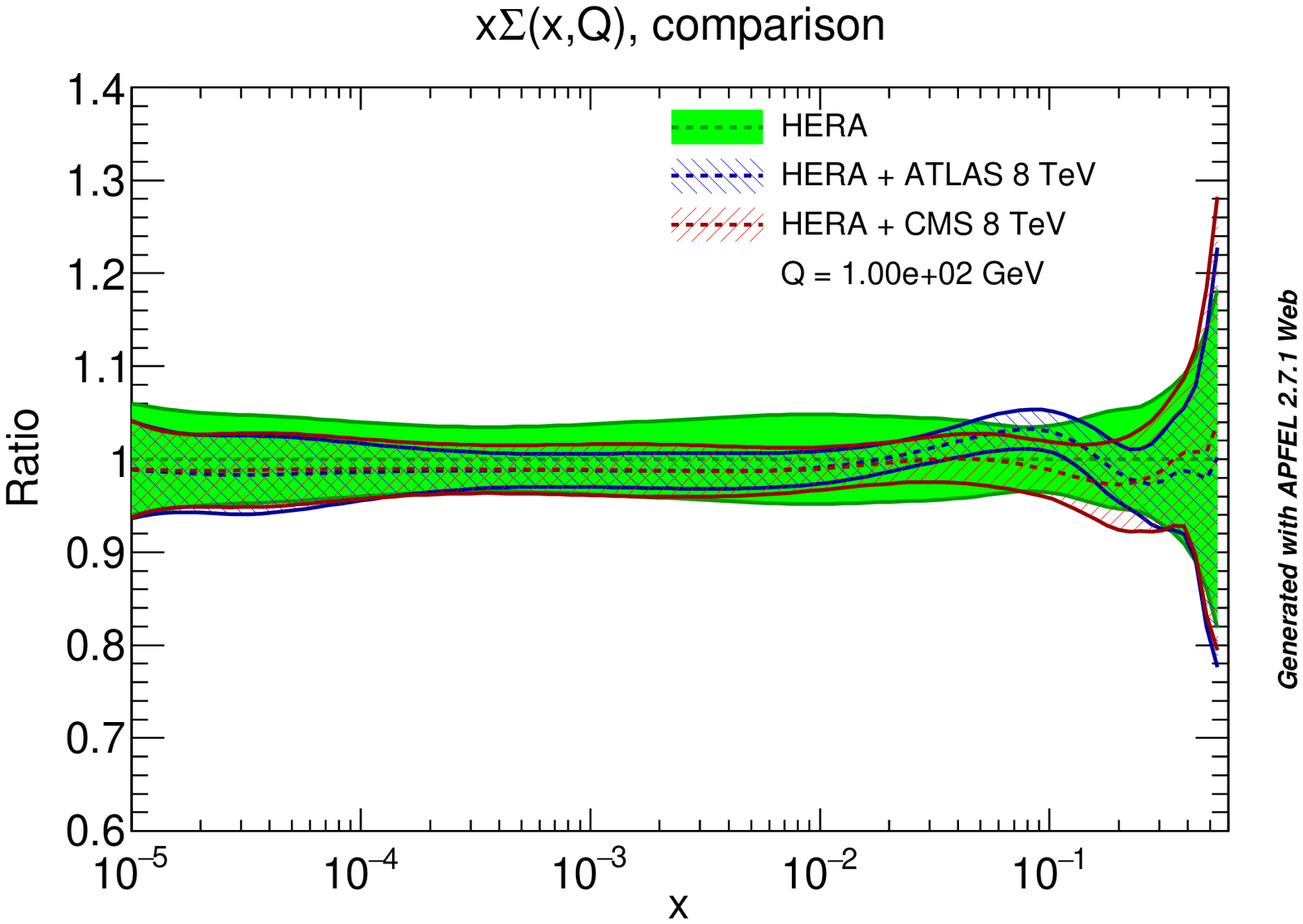}
        \includegraphics[width=0.45\textwidth]{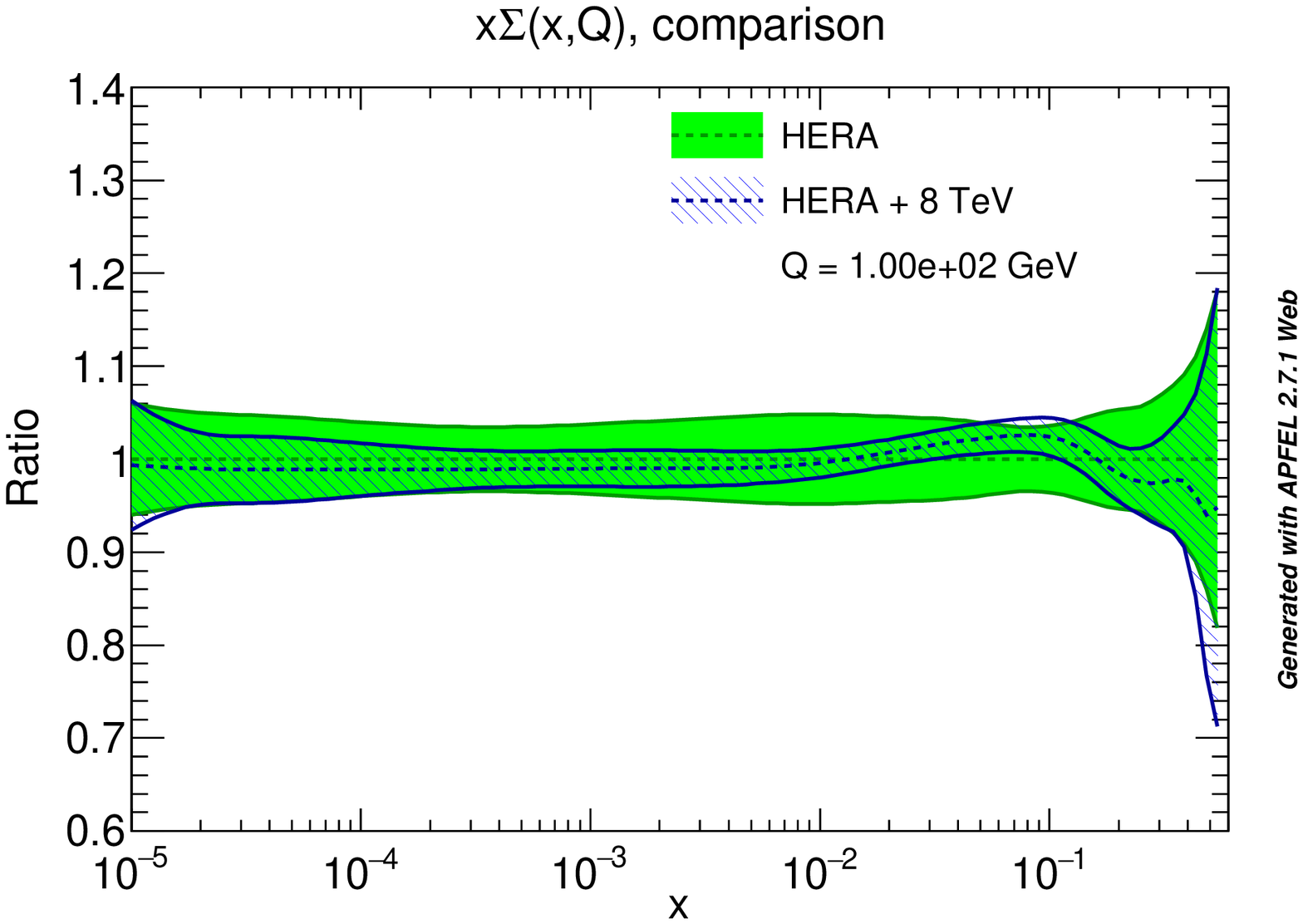}\\
        \caption{\label{fig:1p:one}
Impact of the inclusion of 8 TeV $\pt$ data with $\Delta=1\%$ on the gluon and the singlet PDFs in a HERA-only fit.}
\end{figure}

We now study the implications of these fits for the PDF sets. All plots have been done by using
the on-line interface of APFEL~\cite{Bertone:2013vaa}.  We consider the gluon and the singlet-quark combination.  
To avoid too large a proliferation of plots we focus on the $\Delta=1\%$ and $\Delta=0\%$ cases. 
In Fig.~\ref{fig:1p:one} we display the impact of the inclusion of these data on the gluon and singlet-quark PDFs 
by adding them with an additional uncertainty $\Delta=1\%$.  As can be seen from the upper left panel of 
of Fig.~\ref{fig:1p:one}, including either the ATLAS 8 TeV and CMS 8 TeV data sets leads to a gluon consistent with the HERA 
result but with a slightly smaller uncertainty. The upper right panel shows that HERA+8 TeV gives a gluon similar to HERA-only 
but with a significantly smaller uncertainty for $x>10^{-3}$.

The situation for the singlet-quark distribution is similar.  However the ATLAS and CMS data seem to pull in
slightly different directions, the former preferring a harder singlet in the $x=10^{-1}$ region, as it can be observed
in the lower-left panel.  The lower-right panel shows that the ATLAS data have a stronger pull in the fit and that the
simultaneous inclusion of the ATLAS and CMS data at 8 TeV leads to a significantly reduced uncertainty.

The effects of the $\Delta=1\%$ fits on the down-quark and up-quark distributions
is similar to the effect on the singlet and thus is not shown here: the PDF errors when HERA 
and the 8 TeV data sets are simultaneously fit decreases significantly for both the up and down distributions.

\begin{figure}[tbp]
        \centering
        \includegraphics[width=0.45\textwidth]{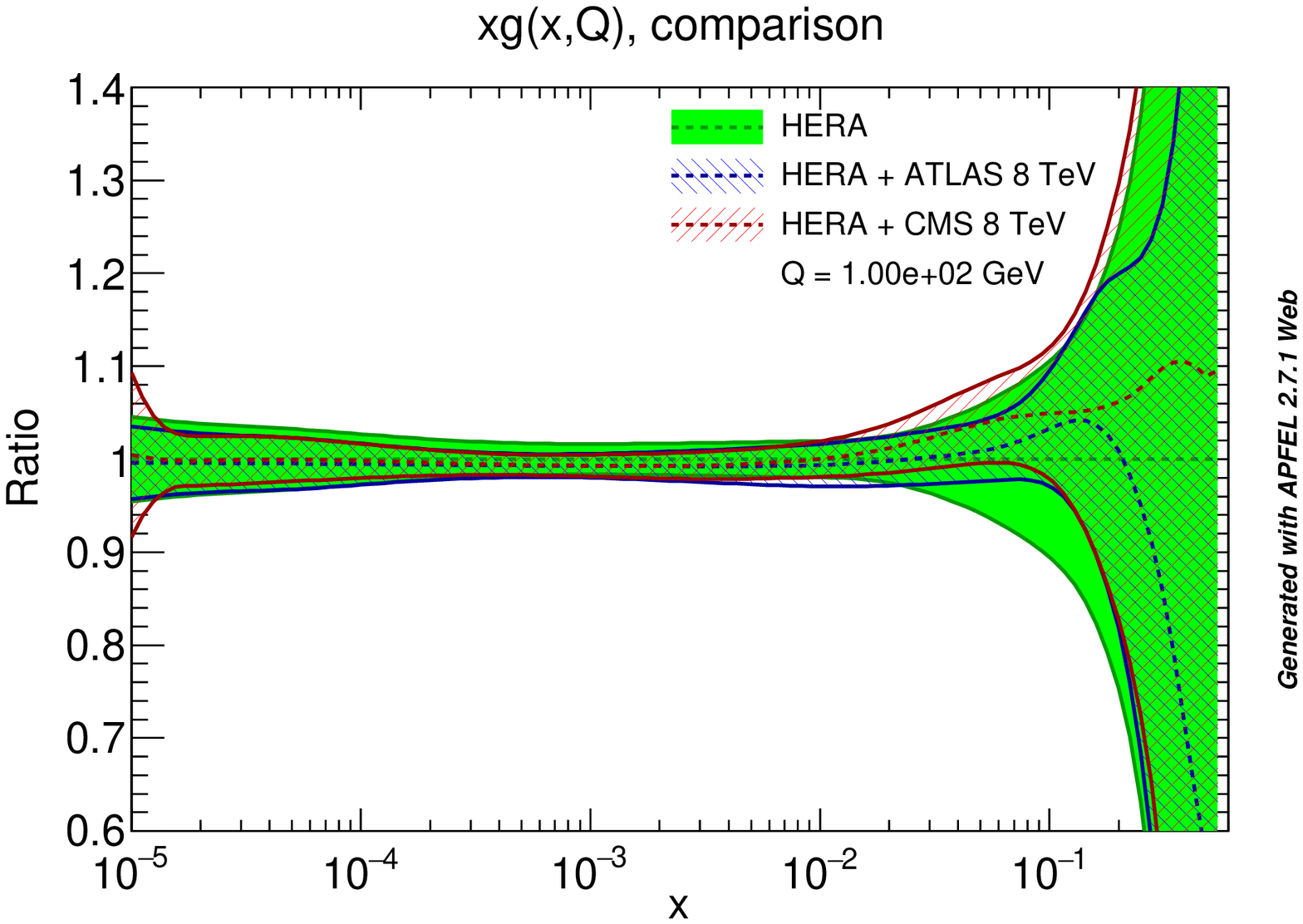}
        \includegraphics[width=0.45\textwidth]{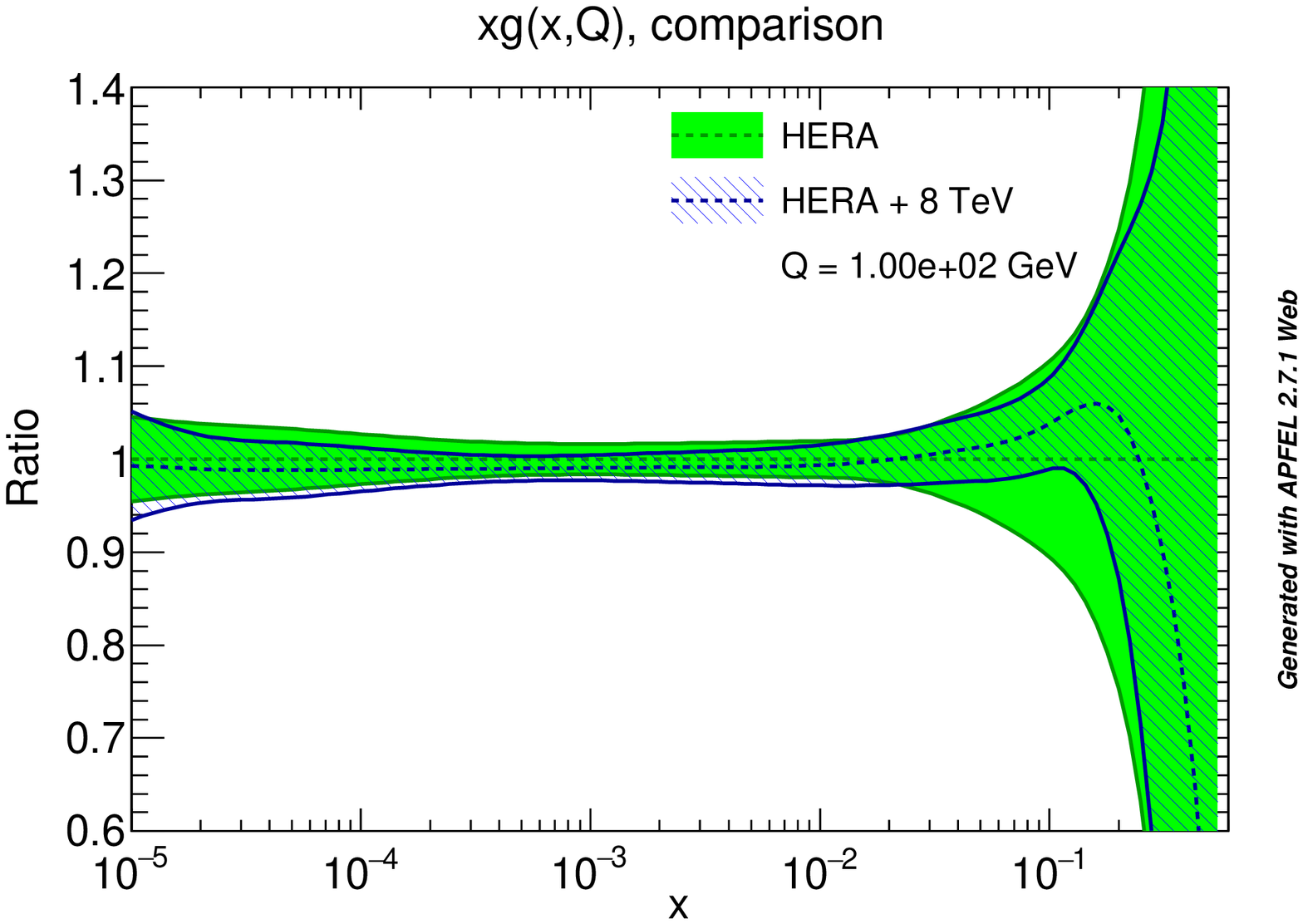}\\
        \includegraphics[width=0.45\textwidth]{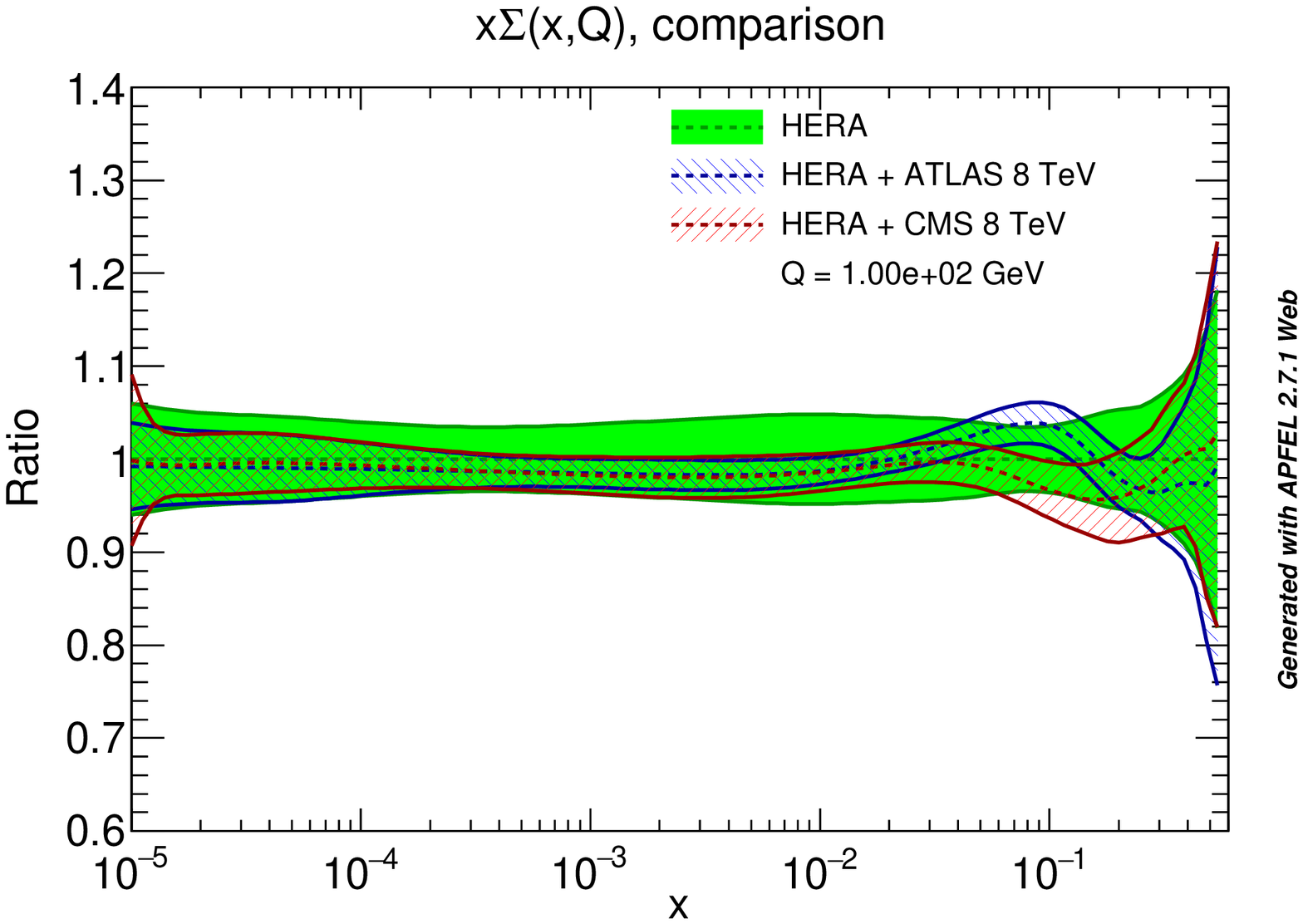}
        \includegraphics[width=0.45\textwidth]{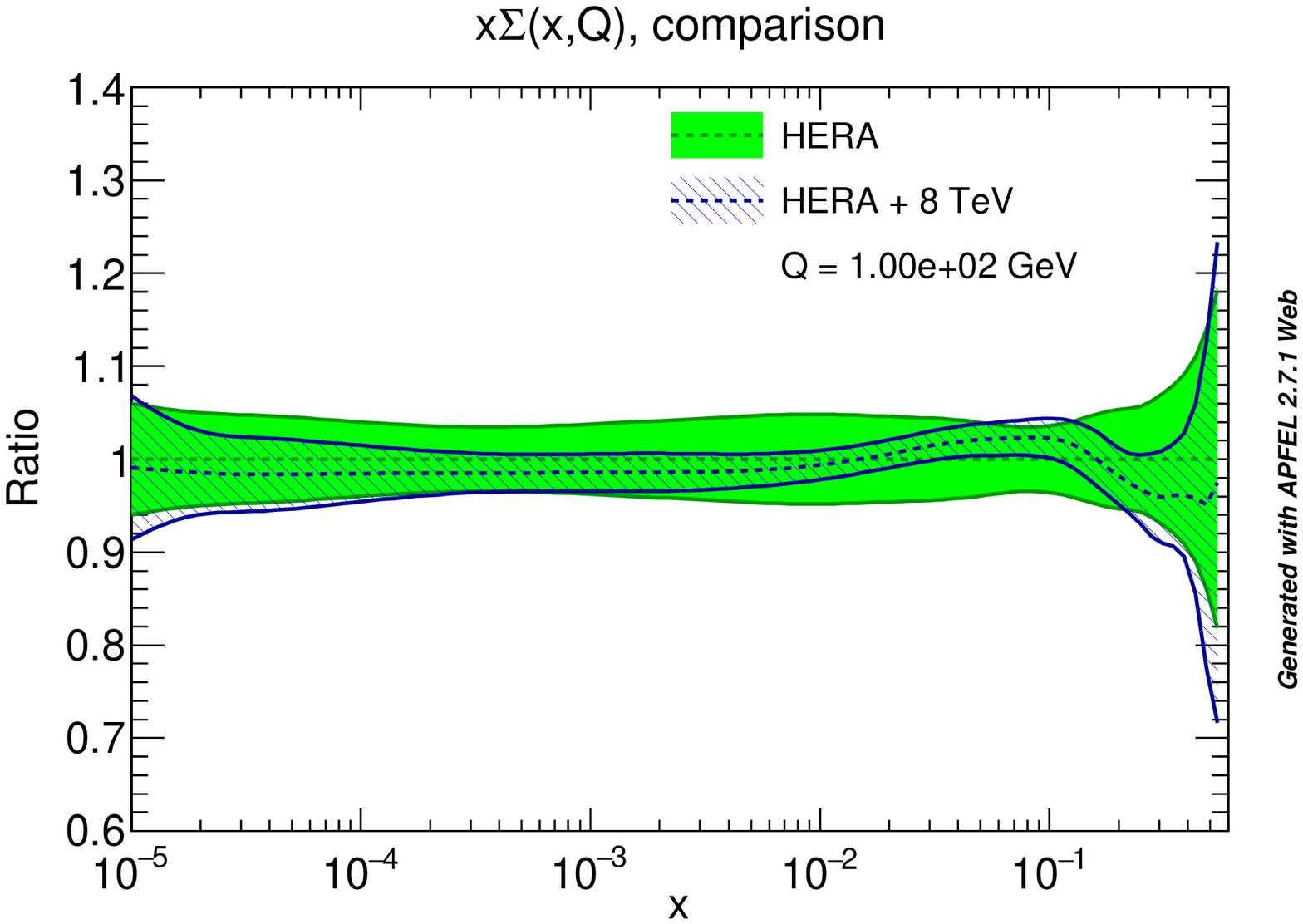}\\
        \caption{Impact of the inclusion of 8 TeV Z pT data with $\Delta= 0\%$ error on the gluon in a HERA-only fit\label{fig:0p:one}}
\end{figure}

In Fig.~\ref{fig:0p:one} we show the results for the PDFs assuming no additional uncertainty, $\Delta=0\%$.  
The observed patterns of PDF shifts when 8 TeV data sets are included is very similar to those seen for $\Delta=1\%$, with only 
small differences in the estimated PDF errors in certain $x$ regions.  

\subsection{Normalized versus unnormalized distributions}
\label{sec:norm}
In this Section we focus on the inclusion of the normalized ATLAS 7
TeV data and give details on the tension we observe with the 8 TeV data. 
We consider a NNLO fit, applying the following cuts
\begin{align*}
& \pt\, > 30\,{\rm GeV}\\
& \pt\, < 500\,{\rm GeV},
\end{align*}
where the latter is motivated by the fact that in the last $\pt$ bin the EW corrections are larger than the sum in quadrature of the 
statistical and uncorrelated systematic uncertainties of the data.  We are left then with 39 data points for the ATLAS 7 TeV distribution.

\begin{table}[ht]
 \centering
 \caption{Overview of fits run with HERA-only as a baseline including
   the normalized ATLAS 7 TeV along with the other data sets.  For each fit, we indicate
 which measurements from ATLAS and CMS has been included, whether an
 uncorrelated uncertainty has been added to the $\chi^2$ (in
 brackets unless it is set to 0).}
 \label{tab:descrfitsHERA7}
 \begin{small}
 \begin{tabular}{c|c|cc|cc|cc}
 \hline
       & (a) & (k) & (l) & (m) & (n) & (o) & (p) \\
 \hline
 HERA            & y & y          &y          &y           &y           & y &y \\
 ATLAS7TEV  & n & y(1\%)  & y(1\%) &y(0.5\%) &y(0.5\%)&y  &y \\
 ATLAS8TEV  & n & n          & y(1\%) &n           &y(0.5\%)&n  &y \\
 CMS8TEV     & n & n          & y(1\%) &n            &y(0.5\%)&n &y\\
 \hline
 \end{tabular}
 \end{small}
 \end{table}
We summarize the fits in Table~\ref{tab:descrfitsHERA7}.  
These are labelled (k)-(p). The baseline is the same as the one presented in the previous section.
Fits (k), (m) and (o) add individually the ATLAS 7 TeV data by adding an uncorrelated uncertainty of 1\%, 0.5\% and none 
respectively. Fits (l), (n) and (p) add them along with the unnormalized ATLAS and CMS data at 8 TeV with 
an extra uncorrelated uncertainty of 1\%, 0.5\% and none 
respectively.

The results of fits (k)-(p) are summarized in Table~\ref{tab:overview7}.
For each fit the $\chi^2$ per degree of freedom ($\chi^2_{{\rm d.o.f.}}$) of the experiments included in the fit, 
and of the prediction for the observables not included in the fit (in brackets), are
displayed.  The additional uncorrelated uncertainty added to the fit is denoted by $\Delta$.  
Again, we have repeated the baseline HERA-only fit (a) at the beginning of each Table section for ease of comparison.
\begin{table}[ht]
\centering
\caption{Fully correlated $\chi^2_{{\rm d.o.f.}}$ for the fits described in Table~\ref{tab:descrfitsHERA7}.
The numbers in brackets correspond to the
  $\chi^2$ for experiments which are not fitted.  The total $\chi^2$ is
  computed over all baseline HERA data the included $\pt$ distributions.}
\label{tab:overview7}
\begin{small}
\begin{tabular}{c|c|ccccc}
\hline
fit id & extra $\Delta$ & $\chi^2_{\rm ATLAS7tev}$ & $\chi^2_{\rm ATLAS8tev,m}$
  &$\chi^2_{\rm ATLAS8tev,y}$ & $\chi^2_{\rm CMS8tev}$ & $\chi^2_{\rm tot}$\\
\hline
(a) & 1\% & (21.8) & (1.00) & (1.56) & (1.55) & 1.168 \\  
\hline
(k) & 1\% &  1.39 & (1.39) & (2.04) & (1.41)   & 1.176 \\ 
(l) & 1\%  &  1.64 &  1.05   & 1.17   & 1.27   & 1.171\\ 
\hline
(a) & 0.5\% & (27.6) & (1.10) & (2.83) & (2.46) & 1.168 \\
\hline
(m) & 0.5\% &  1.58  & (1.54)  & (3.36) & (2.11)  & 1.186 \\ 
(n) & 0.5\% &  2.13  &  1.18   &  1.98  &  2.21  & 1.253\\ 
\hline
(a) & no  & (30.6) & (1.15) & (4.65) & (3.46)  & 1.168 \\
\hline
(o)   & no& 1.74   & (1.69) & (4.79) & (3.06)  & 1.185 \\ 
(p)  & no & 2.35   & 1.24   & 2.81   & 3.19    &1.301 \\ 
\hline
\end{tabular}
\end{small}
\end{table}
  A few things are apparent from the table.  
\begin{itemize}

\item The ATLAS 7 TeV data is inconsistent with the HERA-only fit, with a $\chi^2_{{\rm d.o.f.}}$ over 20 regardless of the $\Delta$ chosen.
A primary reason for this is that the ATLAS 7 TeV data is normalized to the fiducial cross section in each rapidity bin, while the 8 TeV data sets 
are unnormalized.  The normalization performed for the ATLAS 7 TeV data introduces correlations between the low-$\pt$ bins and the 
$\pt>30$ GeV region to which we must restrict our fit due to the theoretical considerations discussed earlier. 
Due to this cut on the data the covariance matrix provided by the experiments for the whole data set cannot be used to consistently include
the 7 TeV data in the fit. 
It would be interesting to revisit this issue if the unnormalized data became available.  

\item Studying fits (l), (n) and (p) shows that it is hard to simultaneously fit the ATLAS 7 TeV data with the 8 TeV data sets.  
In table \ref{tab:overview8} we observed that fitting the 8 TeV data leads to a $\chi^2_{{\rm d.o.f.}}$ of 18 for the ATLAS 7 TeV data in fit (d).
In table \ref{tab:overview7} we see that the $\chi^2_{{\rm d.o.f.}}$ of the 8 TeV data deteriorates when we attempt to include the 7 TeV too.

\end{itemize}

\begin{figure}[htbp]
        \centering
        \includegraphics[width=0.45\textwidth]{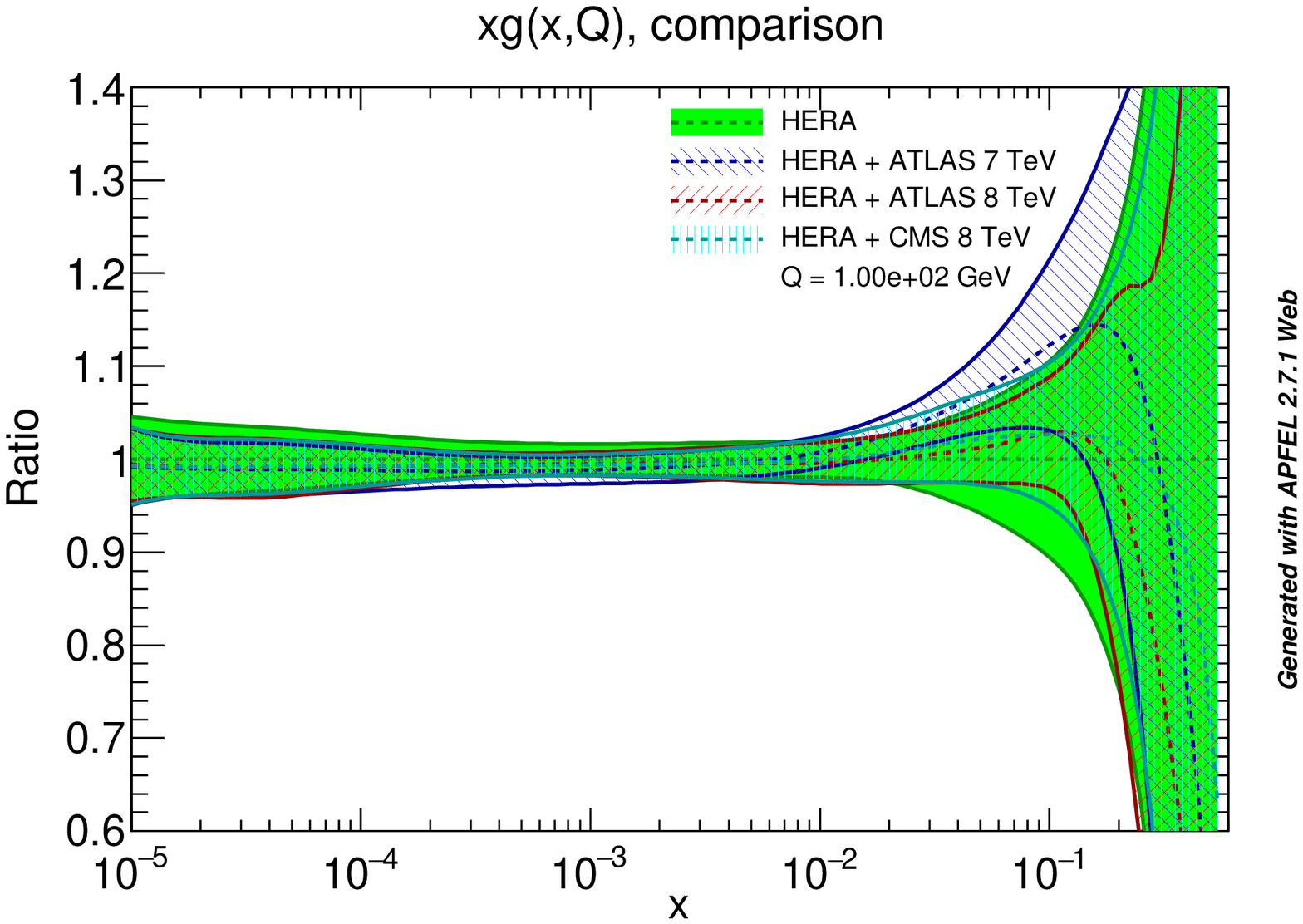}
        \includegraphics[width=0.45\textwidth]{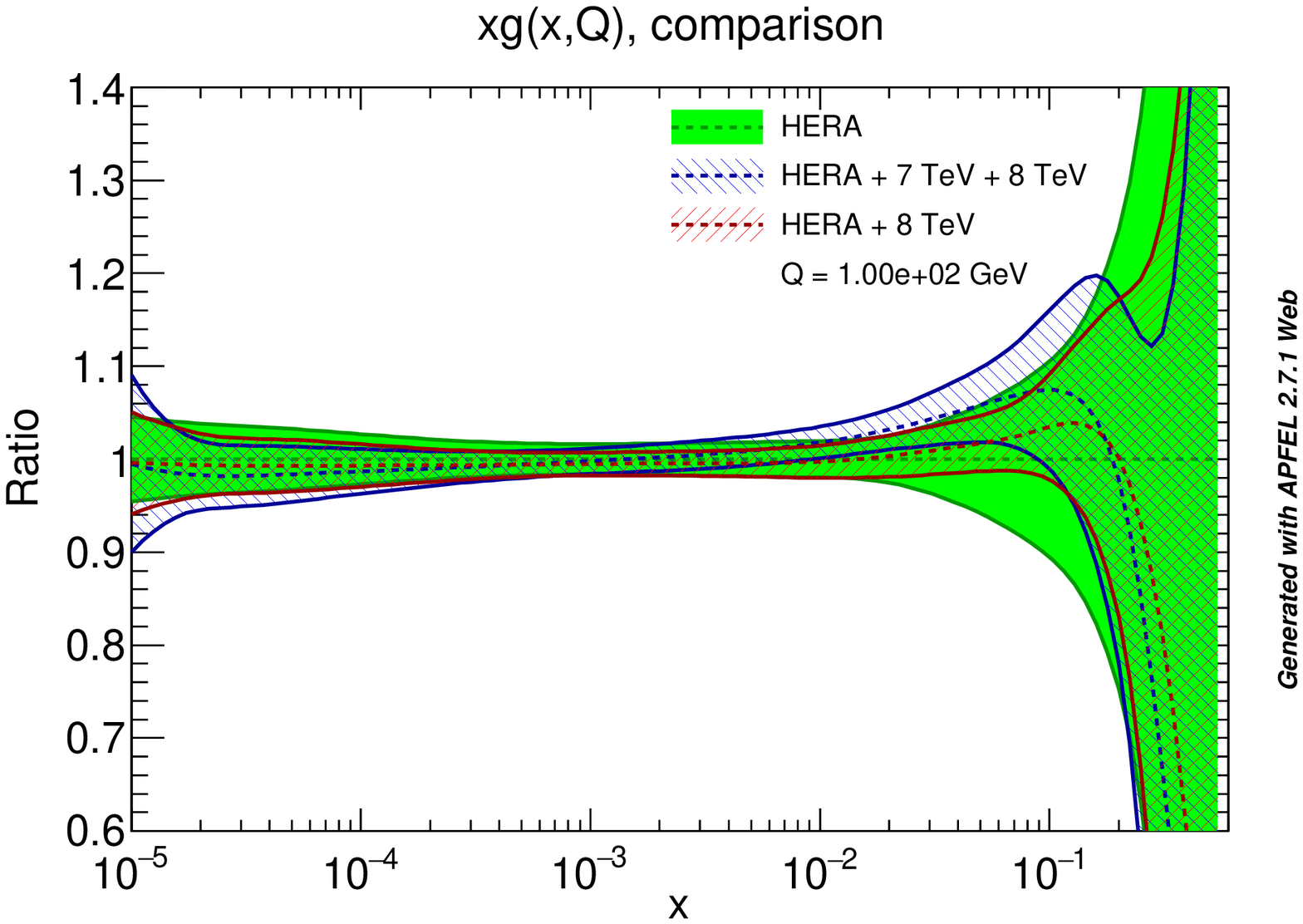}\\
       \includegraphics[width=0.45\textwidth]{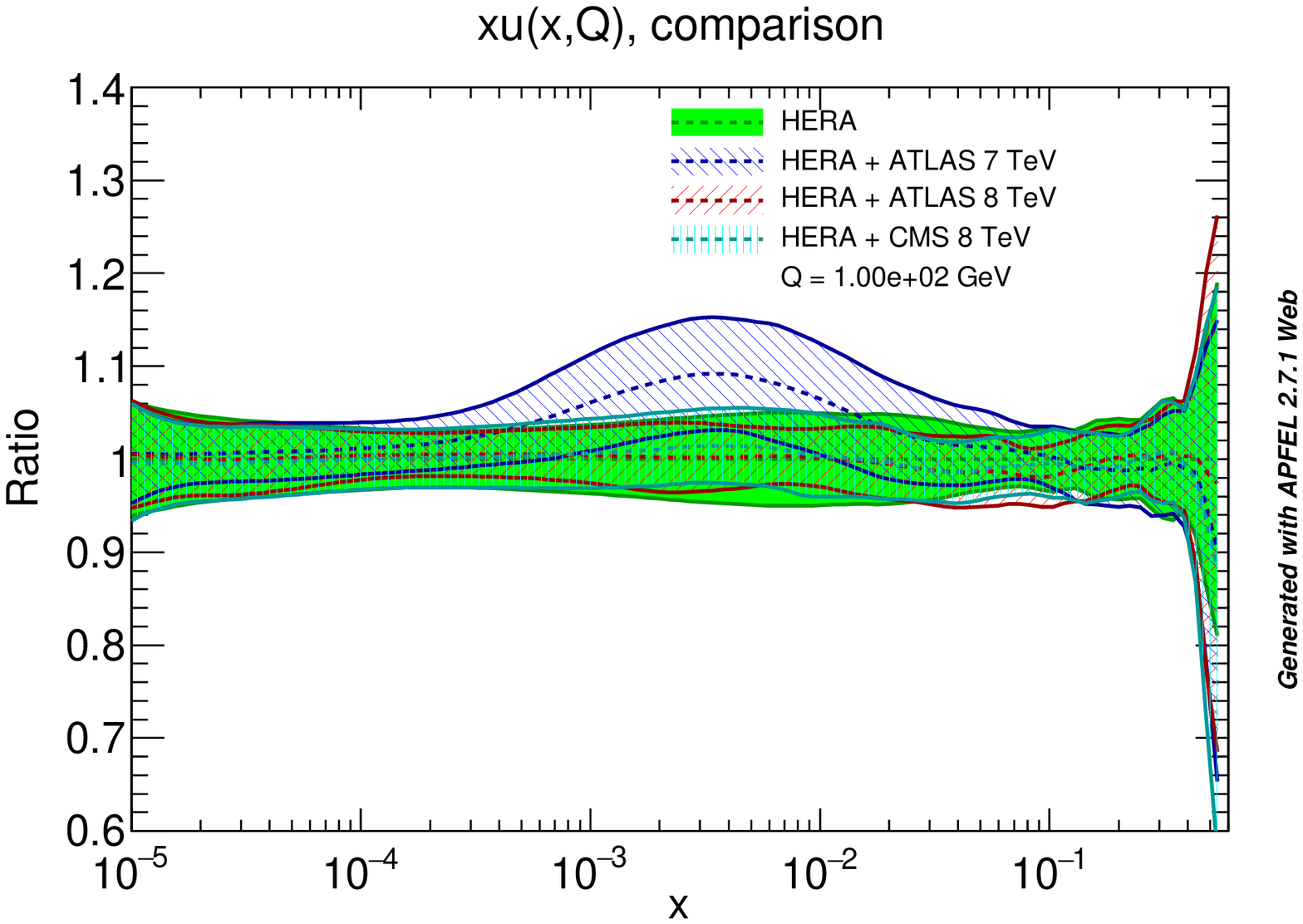}
        \includegraphics[width=0.45\textwidth]{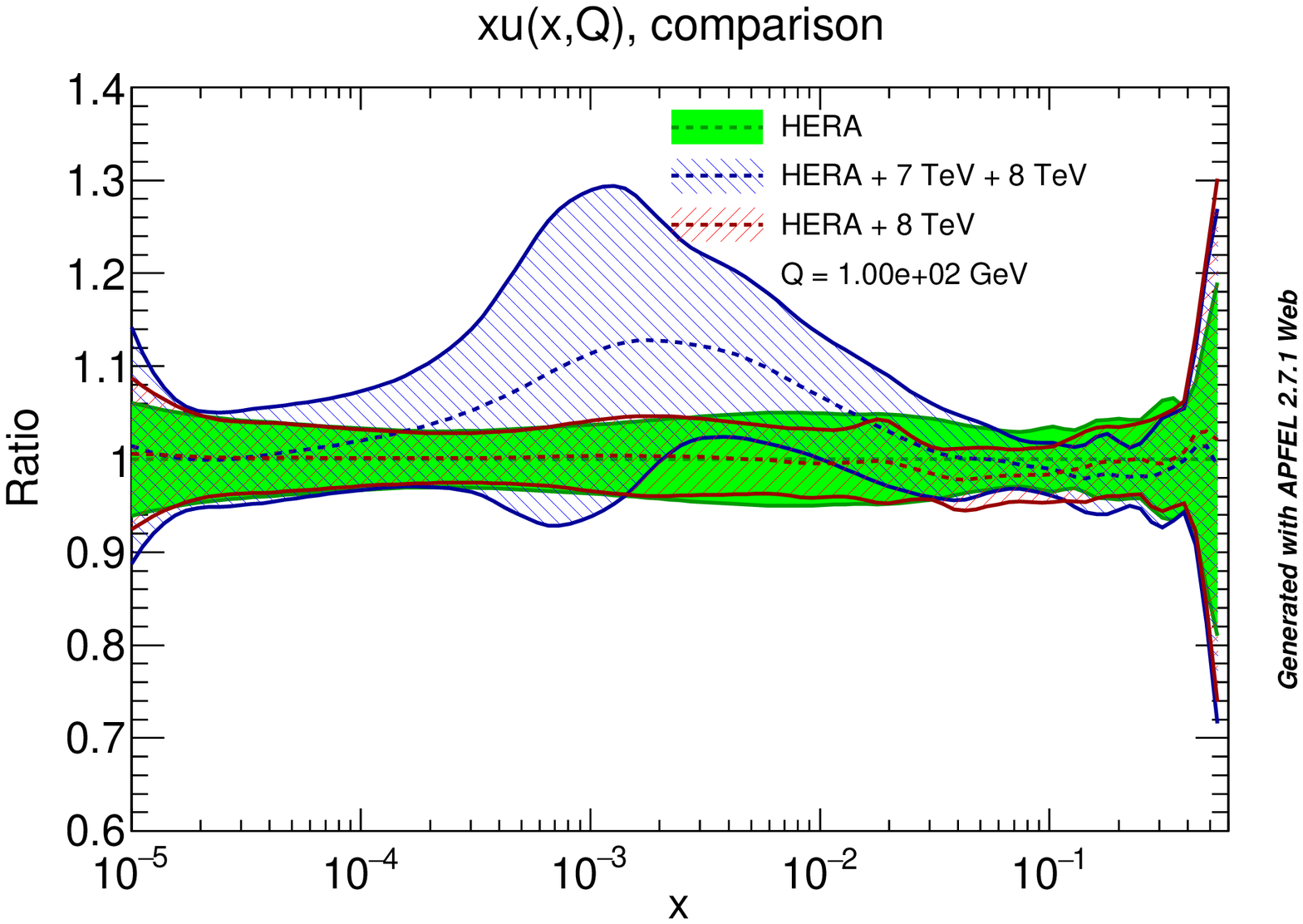}\\
        \includegraphics[width=0.45\textwidth]{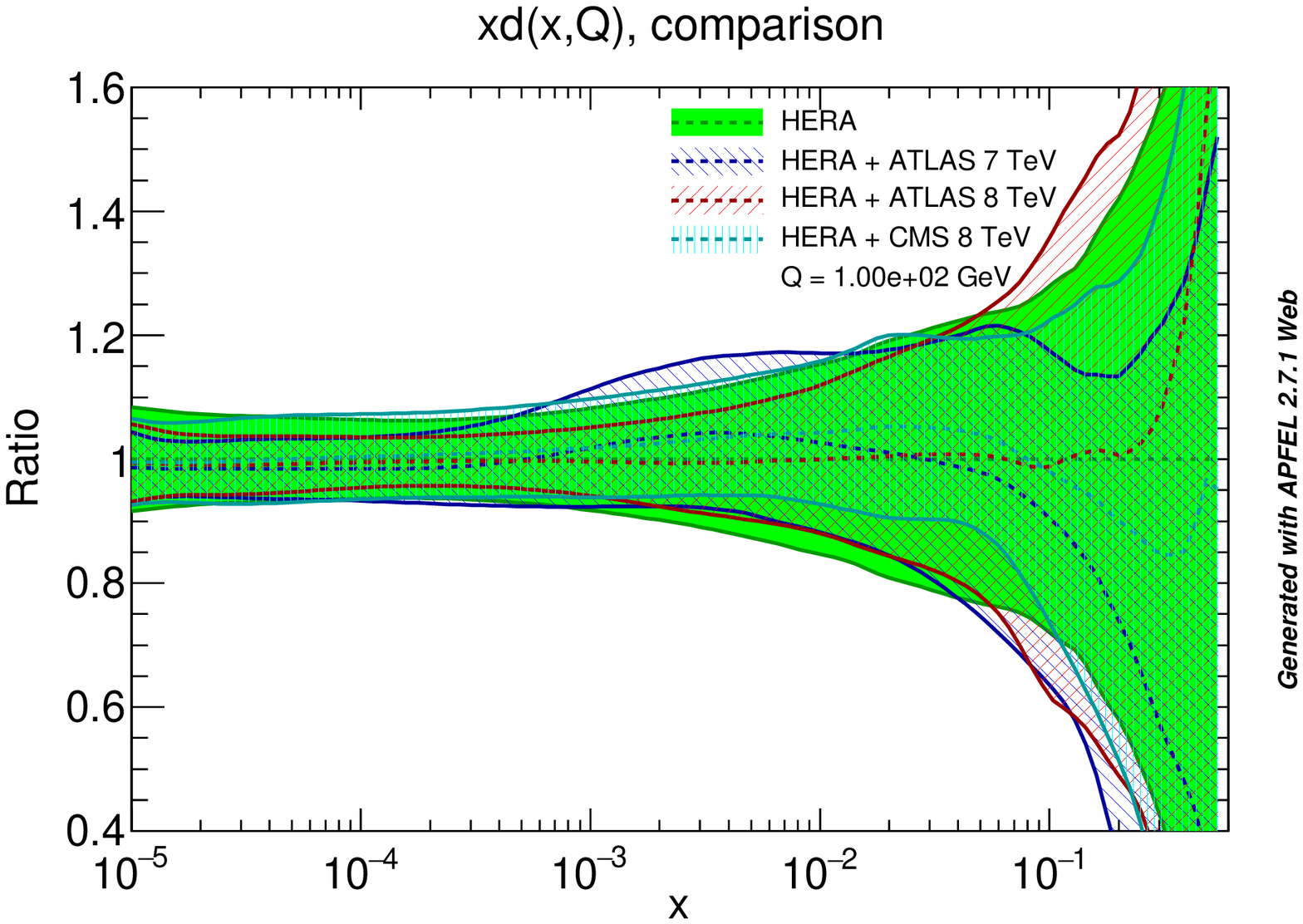}
        \includegraphics[width=0.45\textwidth]{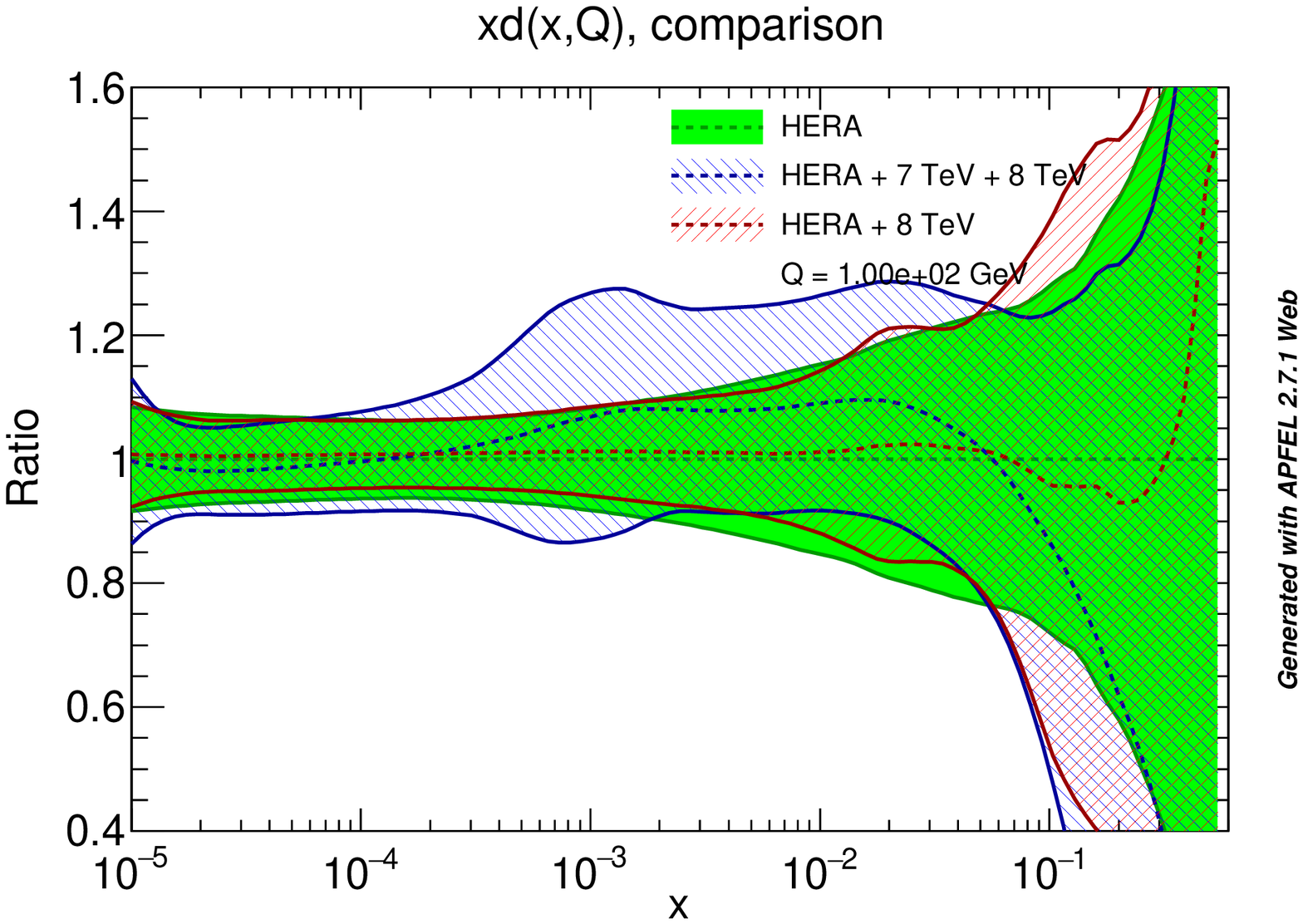}
        \caption{Impact of the inclusion of 7 TeV $\pt$ data with 1\% error on the gluon(top row), up (middle row) and down (bottom row)
in a HERA-only fit.\label{fig:1p:two}}
\end{figure}

We now study the implications of these fits for the PDF sets. 
We consider the gluon, up-quark and down-quark distributions and focus on the $\Delta=1\%$
case only, as we have seen that PDFs remain basically unchanged upon a reduction of  $\Delta$.
In Fig.~\ref{fig:1p:two} we display the impact of the inclusion of these data on the gluon, up and down quark PDFs 
by adding them with an additional uncertainty $\Delta=1\%$.  
An important feature of these plots is the difference between the impact of the ATLAS 7 TeV data 
on the gluon, compared to the impact of the 8 TeV data sets. As can be seen from the upper left panel  
of Fig.~\ref{fig:1p:two}, including either the ATLAS 8 TeV and CMS 8 TeV data sets leads to a gluon consistent 
with the HERA result but with a slightly smaller uncertainty.  Adding the ATLAS 7 TeV data leads to an increased 
gluon distribution for $x>5\cdot 10^{-3}$.  The upper right panel shows that HERA+8 TeV gives a gluon similar 
to HERA-only but with a significantly smaller uncertainty for $x>10^{-2}$.  Attempting to fit both 7 TeV and ATLAS 8 TeV 
data leads to an increased uncertainty, which is barely visible.  
The tension present between the ATLAS 7 TeV data, and the combined HERA+8 TeV data observed 
for the gluon PDF is also observed for the up and down distributions.  The middle right panel shows 
that the error on the up-quark PDF is greatly increased for $x \approx 10^{-3}$ when we attempt to 
simultaneously fit all data.  The reason for this can be seen from the left middle panel.  The ATLAS 7 TeV data 
prefers a peak in the up-quark distribution at this value.  In contrast, the upper right panel shows a decrease in the PDF error when 
HERA and the 8 TeV data sets are simultaneously fit.  A similar pattern is observed for the down-quark distribution, as is shown in 
the lower two panels of Fig.~\ref{fig:1p:two}.

\begin{figure}[htbp]
        \centering
       \includegraphics[width=0.45\textwidth]{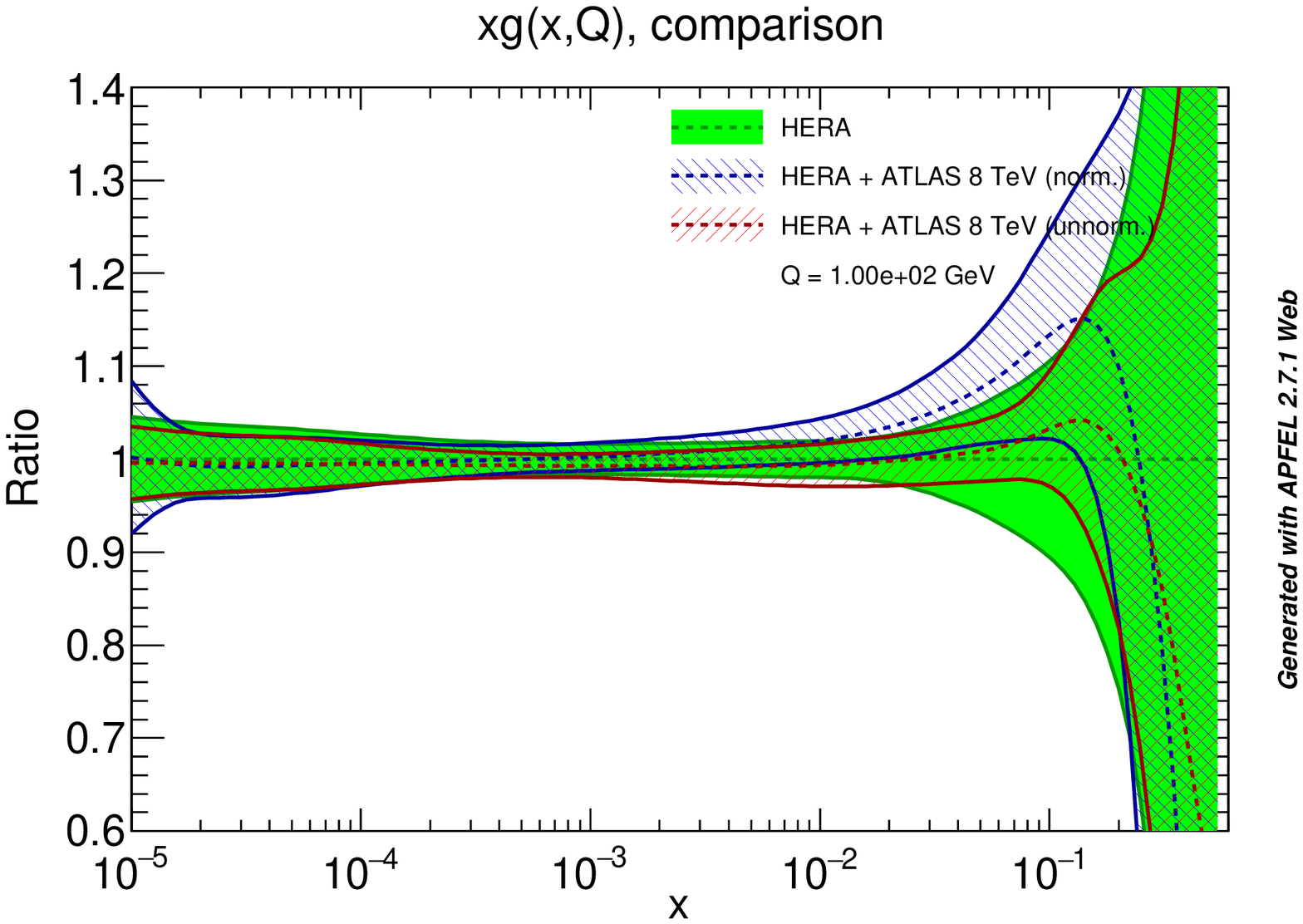}
       \includegraphics[width=0.45\textwidth]{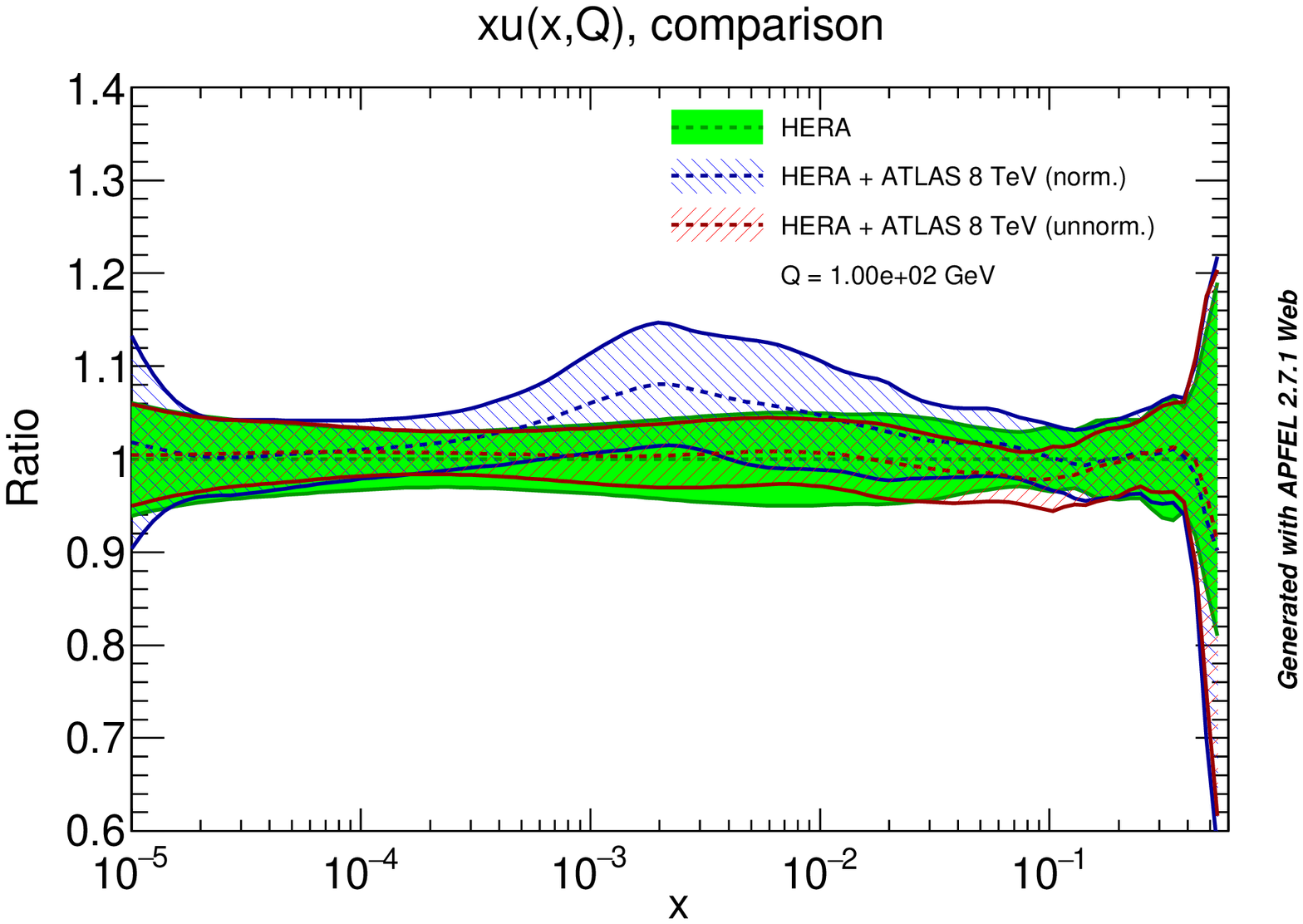}
        \caption{Impact of the inclusion of normalized versus unnormalized 8 TeV  $\pt$ data 
PDFs added to the HERA-only fit. \label{fig:norm}}
\end{figure}

In order to confirm that the origin of the anomalous behaviour of PDFs
 upon the inclusion of the 7 TeV data is due to the fact that they are normalized, 
we perform an additional fit including the precise ATLAS 8 TeV normalized data in the $Z$-peak region in a HERA-only fit.
As far as the quality of the fit is concerned, we observe that these data are harder to fit than the 7 TeV ones, as the  $\chi^2_{\rm d.o.f.}$
ranges from 9 (for a fit with $\Delta=0$\%) to 2.1 (for a fit with $\Delta=1$\%). As far as PDFs are concerned,
in Fig.~\ref{fig:norm} it is apparent that, while the inclusion of the on-peak ATLAS 8 TeV unnormalized data
reduces the uncertainty of the gluon and up-quark distributions, the inclusion of the on-peak ATLAS 8 TeV normalized data
inflates their uncertainties, thus pointing to their inconsistency with respect to the baseline.

\subsection{Impact of NLO EW corrections}

\begin{figure}[htbp]
        \centering
        \includegraphics[width=0.45\textwidth]{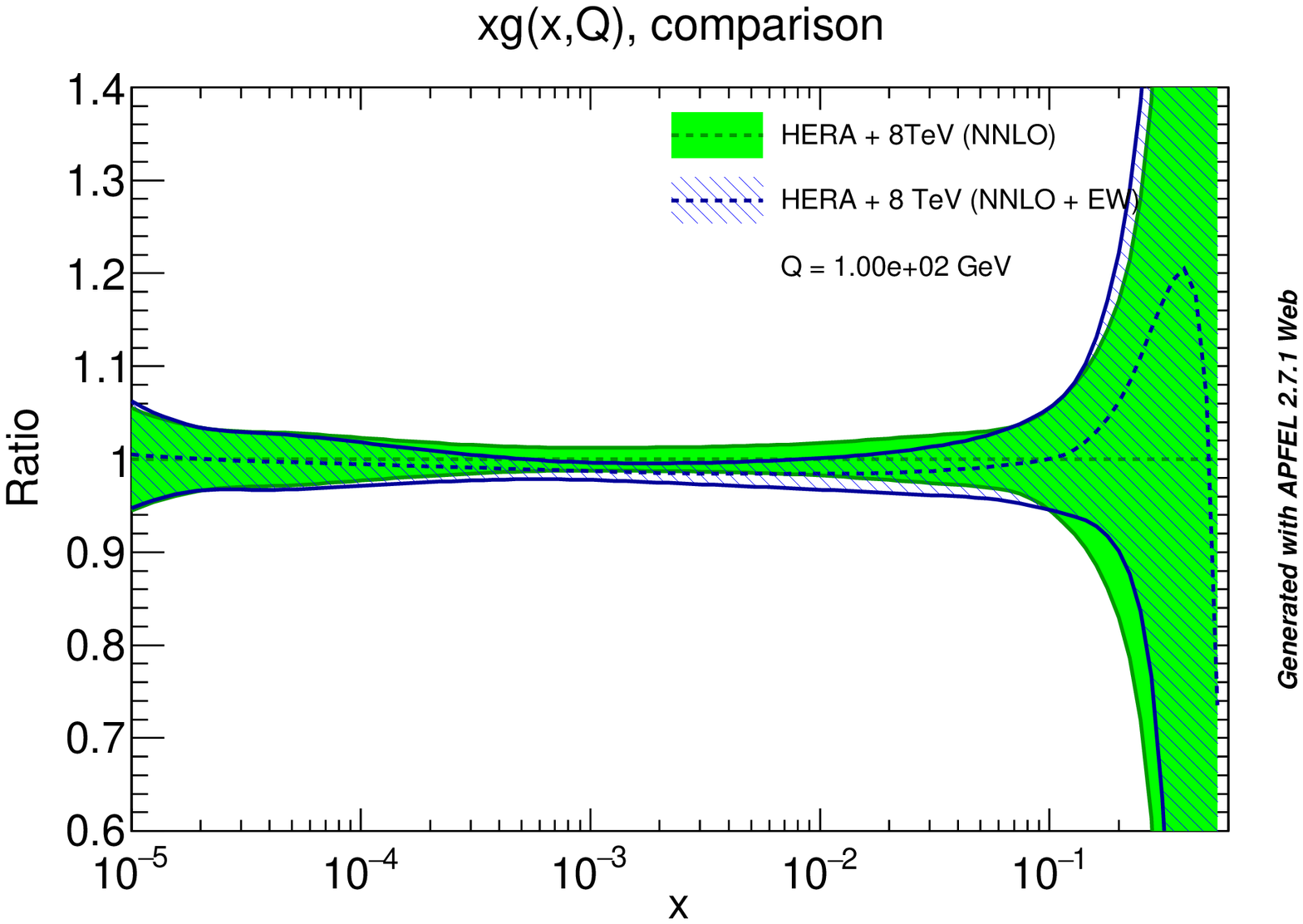}
        \includegraphics[width=0.45\textwidth]{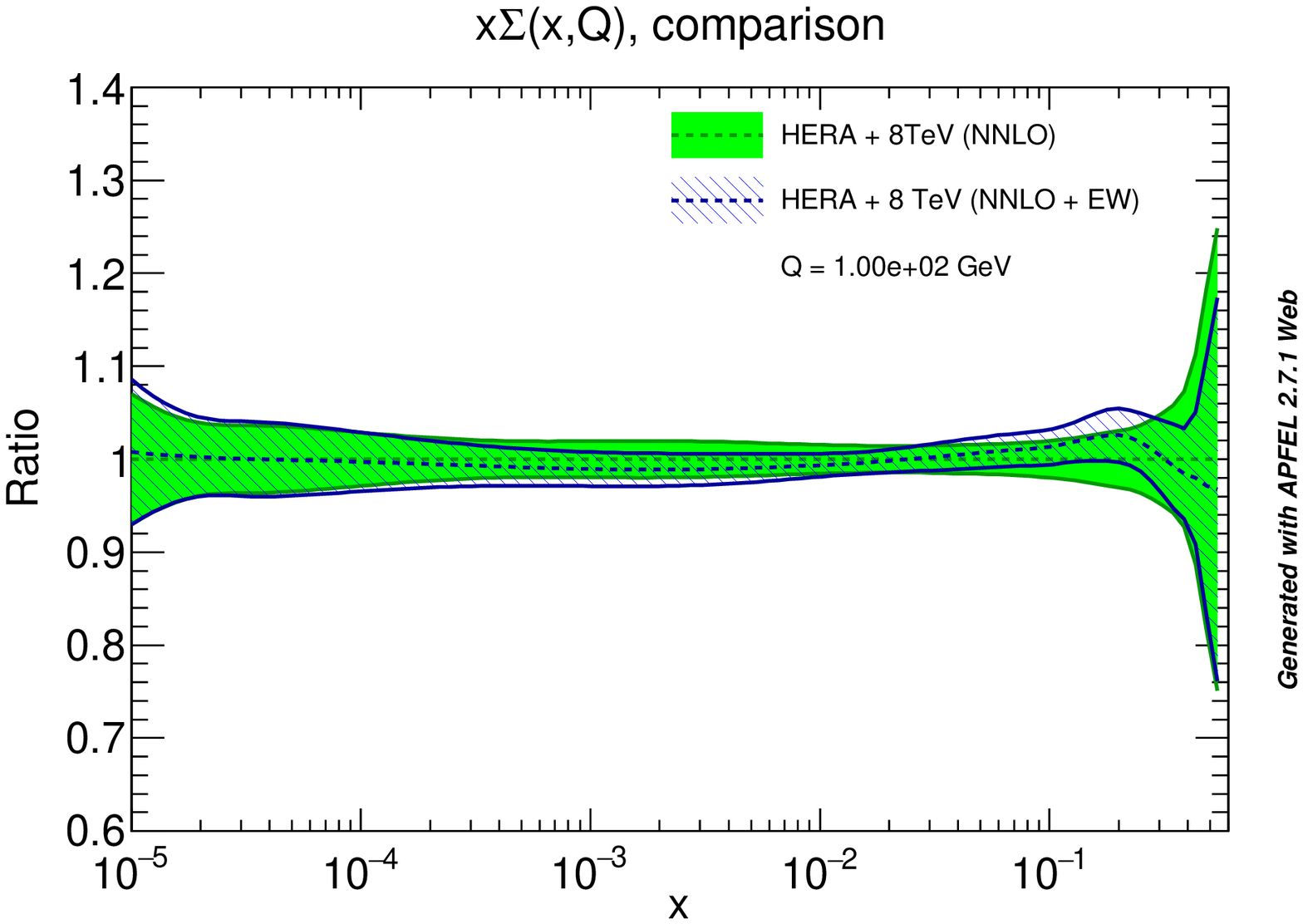}
        \includegraphics[width=0.45\textwidth]{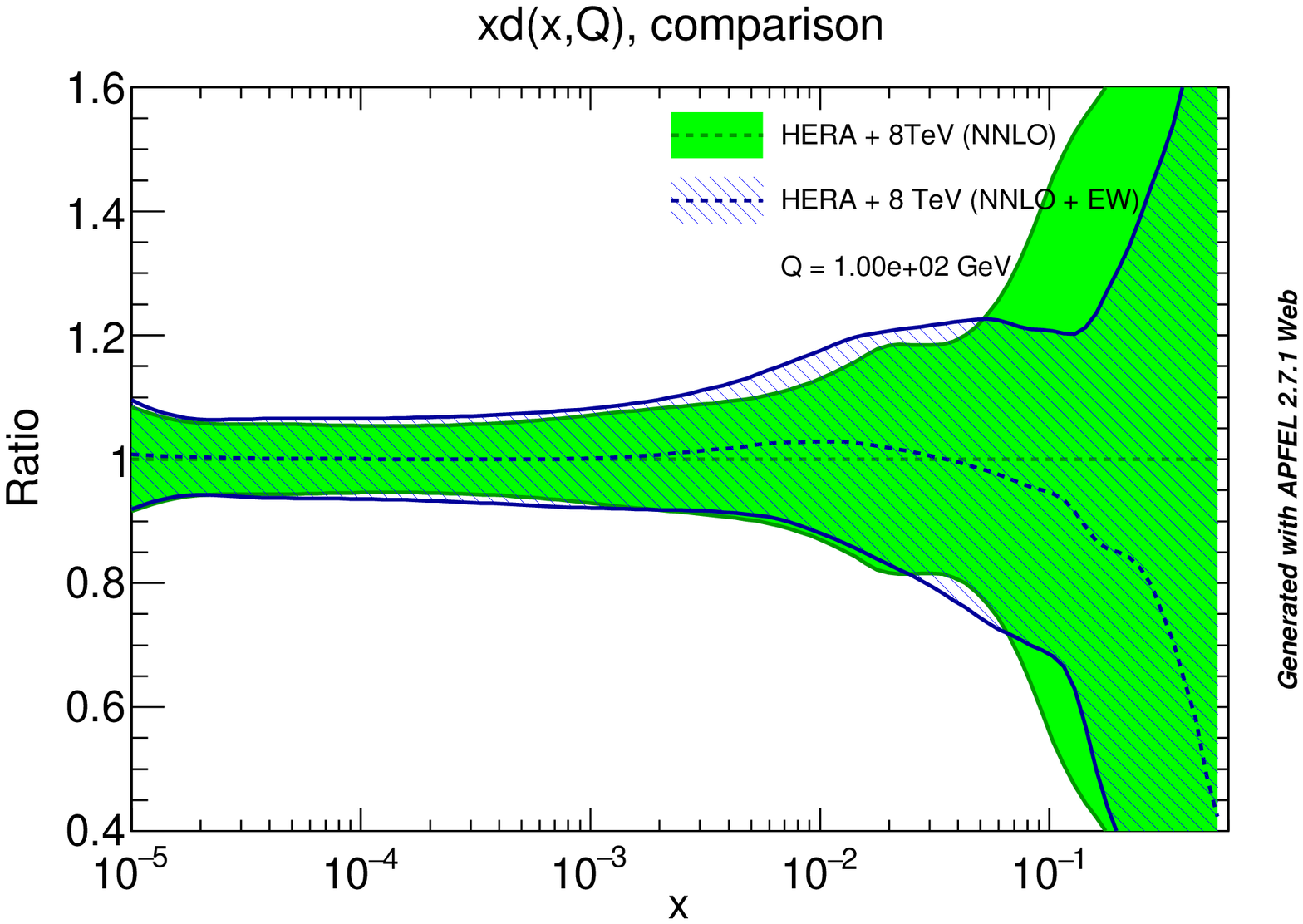}
        \includegraphics[width=0.45\textwidth]{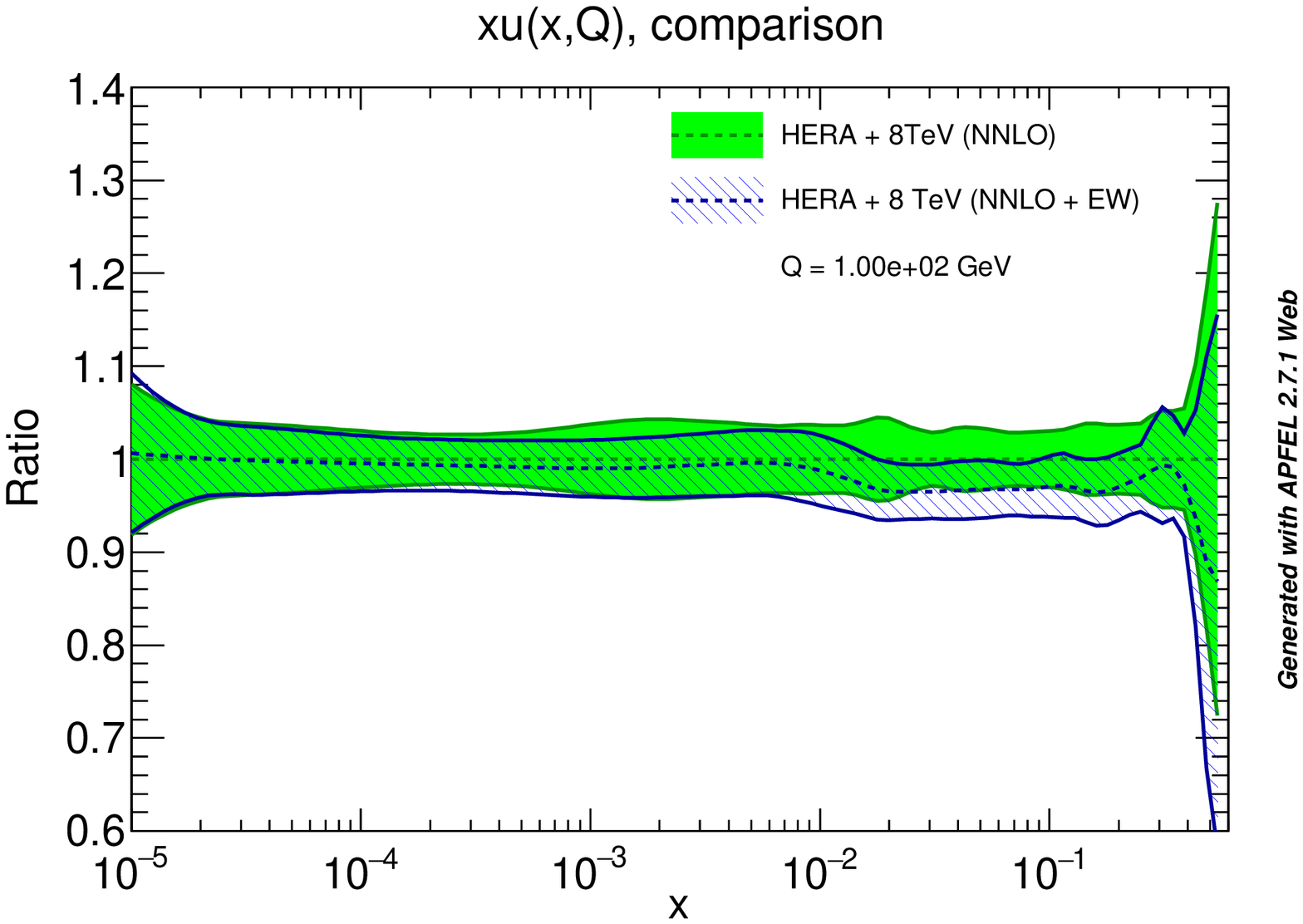}
        \caption{Impact of the inclusion of 8 TeV $\pt$ data with $\Delta=1\%$ PDFs using NNLO or NNLO+EW theory. \label{fig:ewk}}
\end{figure}

Another interesting aspect that we can investigate is the impact of electroweak corrections on the obtained PDFs.  To probe this 
we perform fits to the HERA and 8 TeV data sets, with NNLO QCD corrections and both with and without EW corrections.  
We recall that in the pure NNLO QCD fit we remove bins where the EW corrections are larger than the combined uncorrelated 
uncertainty, as explained previously.  We first display the gluon, singlet, down-quark and up-quark distributions with and without 
EW corrections in Fig.~\ref{fig:ewk}.  The EW corrections have a small but noticeable effect on the PDFs, lowering both the gluon 
and singlet distributions in the intermediate-$x$ regions.  The $\chi^2_{\rm d.o.f.}$ is shown in Table~\ref{tab:ewk}.  
The quality of the fit deteriorates slightly upon including EW corrections.  This results primarily not because EW corrections worsen 
the agreement between theory and data, but because with EW corrections included we are able to include additional high-$\pt$ bins 
in the fit that were excluded in the pure NNLO QCD fit, and these bins are slightly more discrepant than the lower-$\pt$ ones.  
The agreement with the 7 TeV data is marginally improved upon including EW corrections, although it is still inconsistent with the 
HERA+8 TeV combined fit.
\begin{table}[htb]
\centering
\caption{Fully correlated $\chi^2$ for the experiments in the HERA + $\pt$ 8 TeV fit.}
\label{tab:ewk}
\begin{small}
\begin{tabular}{c|c|c|ccccc}
\hline
fit id & extra $\Delta$ & Theory  & $\chi^2_{\rm ATLAS7tev}$ & $\chi^2_{\rm ATLAS8tev,m}$
  &$\chi^2_{\rm ATLAS8tev,y}$ & $\chi^2_{\rm CMS8tev}$ & $\chi^2_{\rm tot}$\\
\hline
(e) & 1\%  & NNLO    & (18) & 0.90   & 0.77  &  1.42   & 1.156\\ 
(q) & 1\%  & NNLO+EW & (16) & 1.00   & 0.87   & 1.72   & 1.182\\ 
\hline
\end{tabular}
\end{small}
\end{table}

\subsection{Impact of the $\pt$ data on a global fit}

Having investigated the impact of the LHC $\pt$ data in a fit consisting of only HERA data, 
which allowed us to consider several aspects of this new data in detail, we turn to their 
inclusion in a global fit of the available measurements.  We follow the NNPDF3.0 analysis with 
the modifications explained in Section~\ref{sec:fitsettings}. 
We set the additional uncorrelated error to $\Delta=1\%$, and, having established that we cannot consistently
include the normalized 7 TeV data in a PDF fit, we only add the unnormalized 8 TeV data to the global baseline. 
The results for the $\chi^2$ per degree of freedom of each fit is shown in Table~\ref{tab:overviewGLOB}. 
The $\chi^2_{\rm d.o.f}$ of the fitted $\pt$ distributions reveals a mild tension 
between the CMS and ATLAS data sets, with $\chi^2_{\rm d.o.f}$ of the CMS set reaching 1.32, while the 
ATLAS 8 TeV sets give a $\chi^2_{\rm d.o.f}$ below one. 
We notice that when including the 8 TeV data the $\chi^2_{\rm d.o.f}$ of the (not-fitted) ATLAS 7 TeV
data deteriorates. 
\begin{table}[htbp]
\centering
\caption{Fully-correlated $\chi^2$ per degree of freedom when the $\pt$ data is added to the global
  fits. 
The numbers in brackets correspond to the
  $\chi^2$ for experiments which are not fitted.  The total $\chi^2$ is
  computed over all data in the baseline fit and the included $\pt$
distributions.  We have labeled our slightly-modified NNPDF3.0 global baseline as NN30red in the table below.}
\label{tab:overviewGLOB}
\begin{small}
\begin{tabular}{c|ccccc}
\hline
fit & $\chi^2_{\rm ATLAS7tev}$ & $\chi^2_{\rm ATLAS8tev,mdist}$
  &$\chi^2_{\rm ATLAS8tev,ydist}$ & $\chi^2_{\rm CMS8tev}$ & $\chi^2_{\rm tot}$\\
\hline
NN30red & (6.93) & (0.98) & (1.06) & (1.41) & 1.17677 \\ 
\hline
NN30red + 8 TeV & (7.87) &  0.96  & 0.88  &  1.32 & 1.17690 \\ 
\hline
\end{tabular}
\end{small}
\end{table}

\begin{figure}[tbp]
        \centering
        \includegraphics[width=0.32\textwidth]{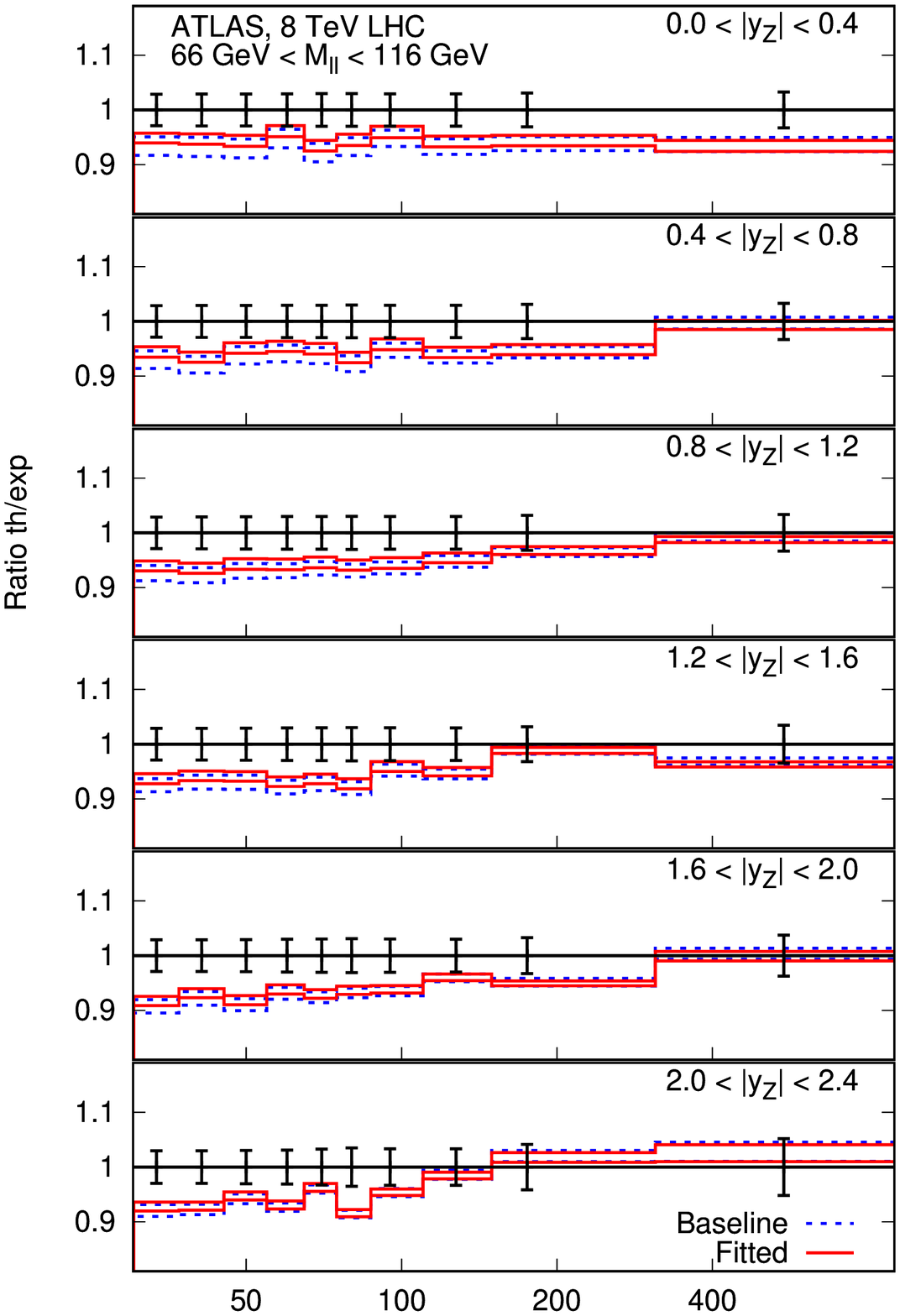}
        \includegraphics[width=0.32\textwidth]{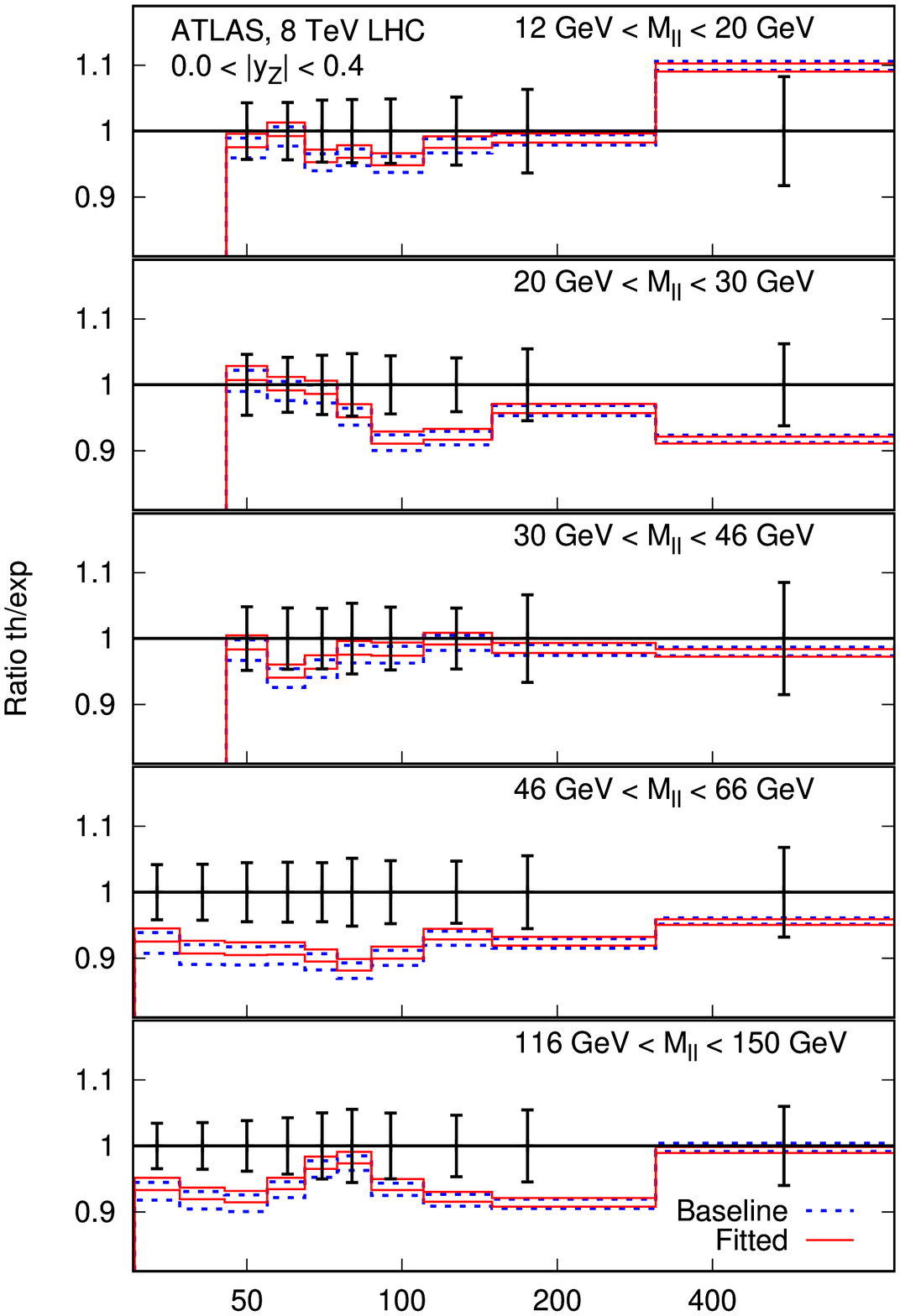}
        \includegraphics[width=0.32\textwidth]{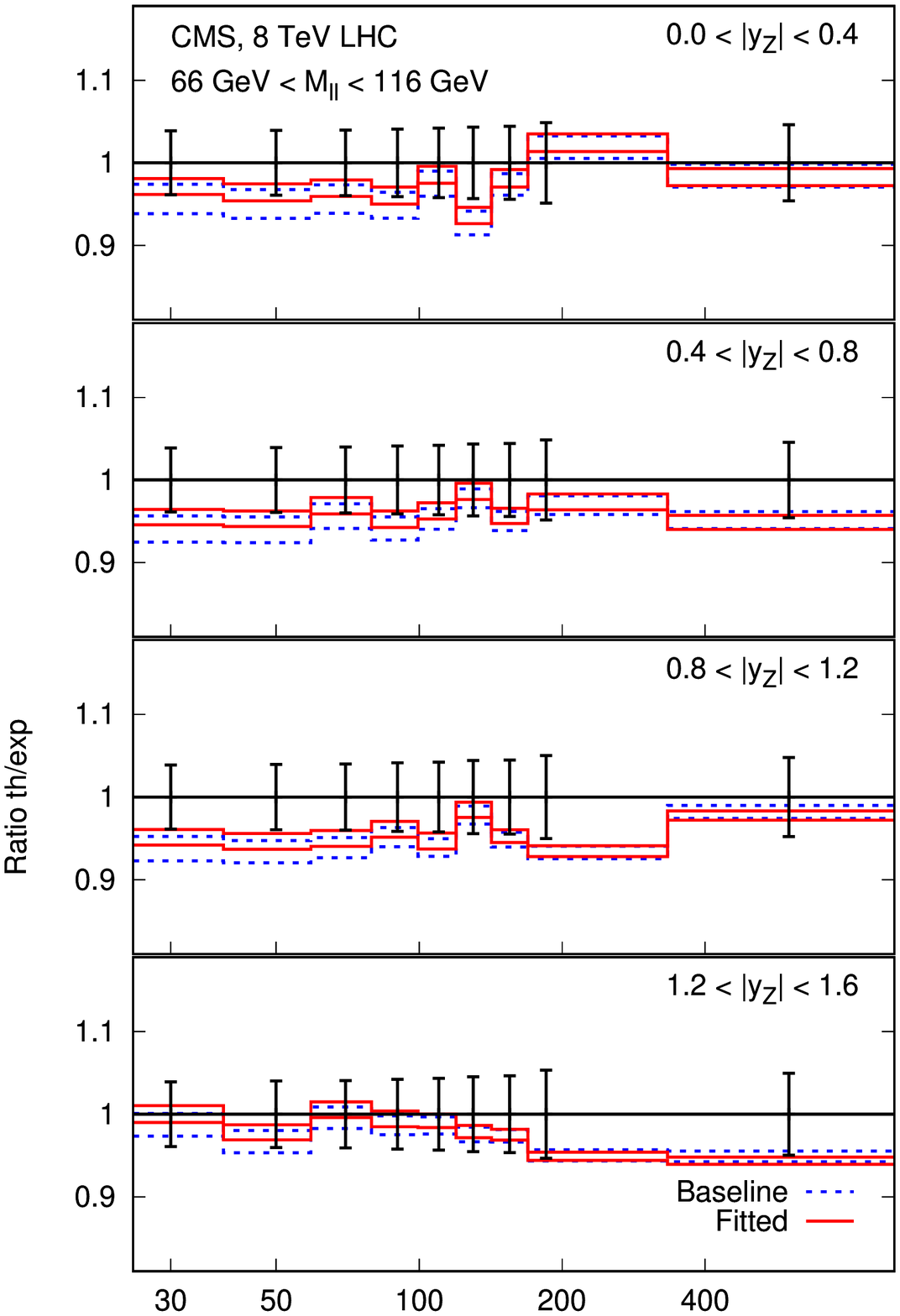}
        \caption{$\pt$ observables computed at NNLO with input PDFs before and after the addition of
the $\pt$ data in the global baseline.\label{fig:obs}}
\end{figure}

In Fig.~\ref{fig:obs} we display the agreement of the NNLO predictions and the data before and after
the fit. We observe that the agreement improves and uncertainties shrink.

\begin{figure}[tbp]
        \centering
        \includegraphics[width=0.45\textwidth]{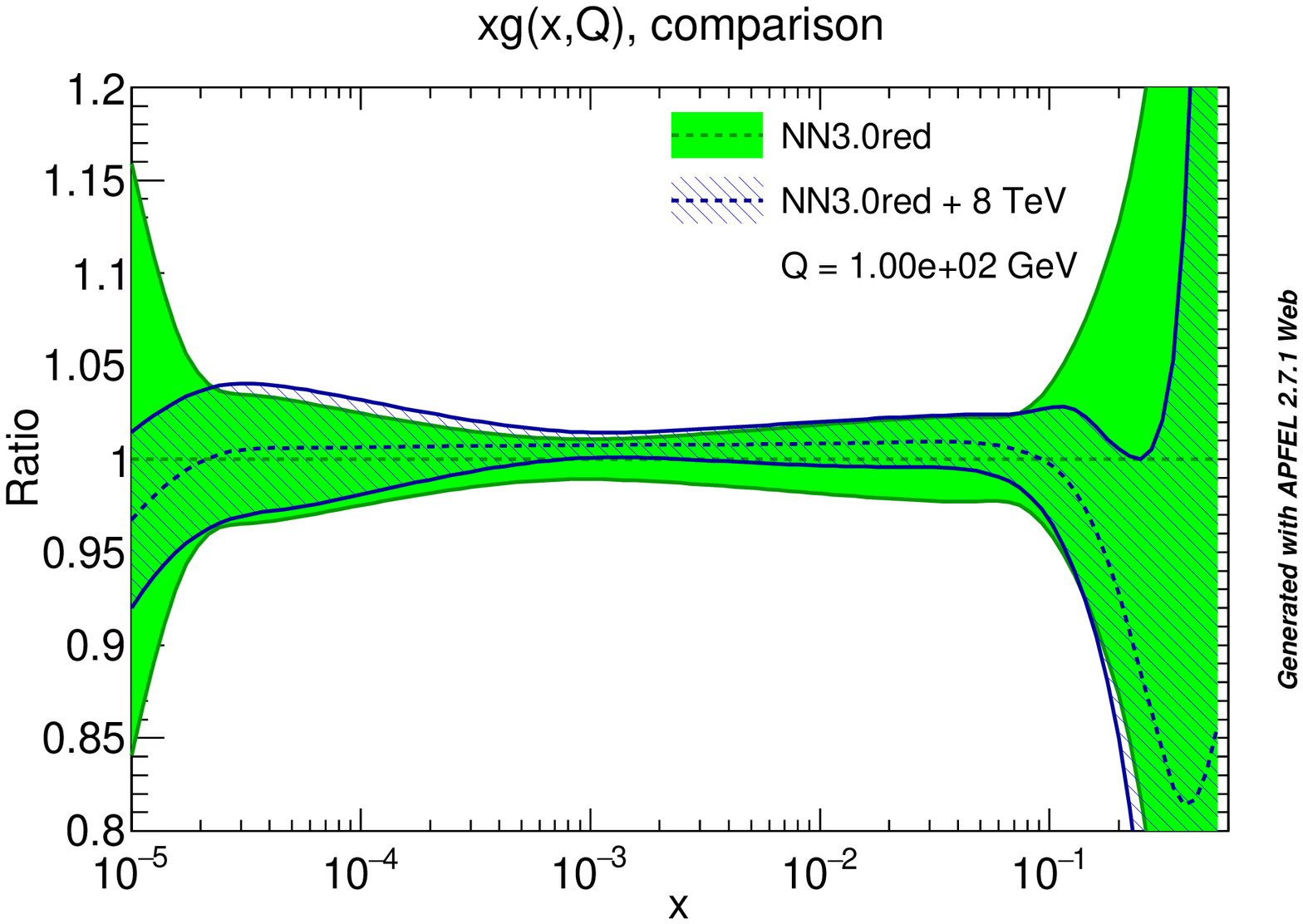}
        \includegraphics[width=0.45\textwidth]{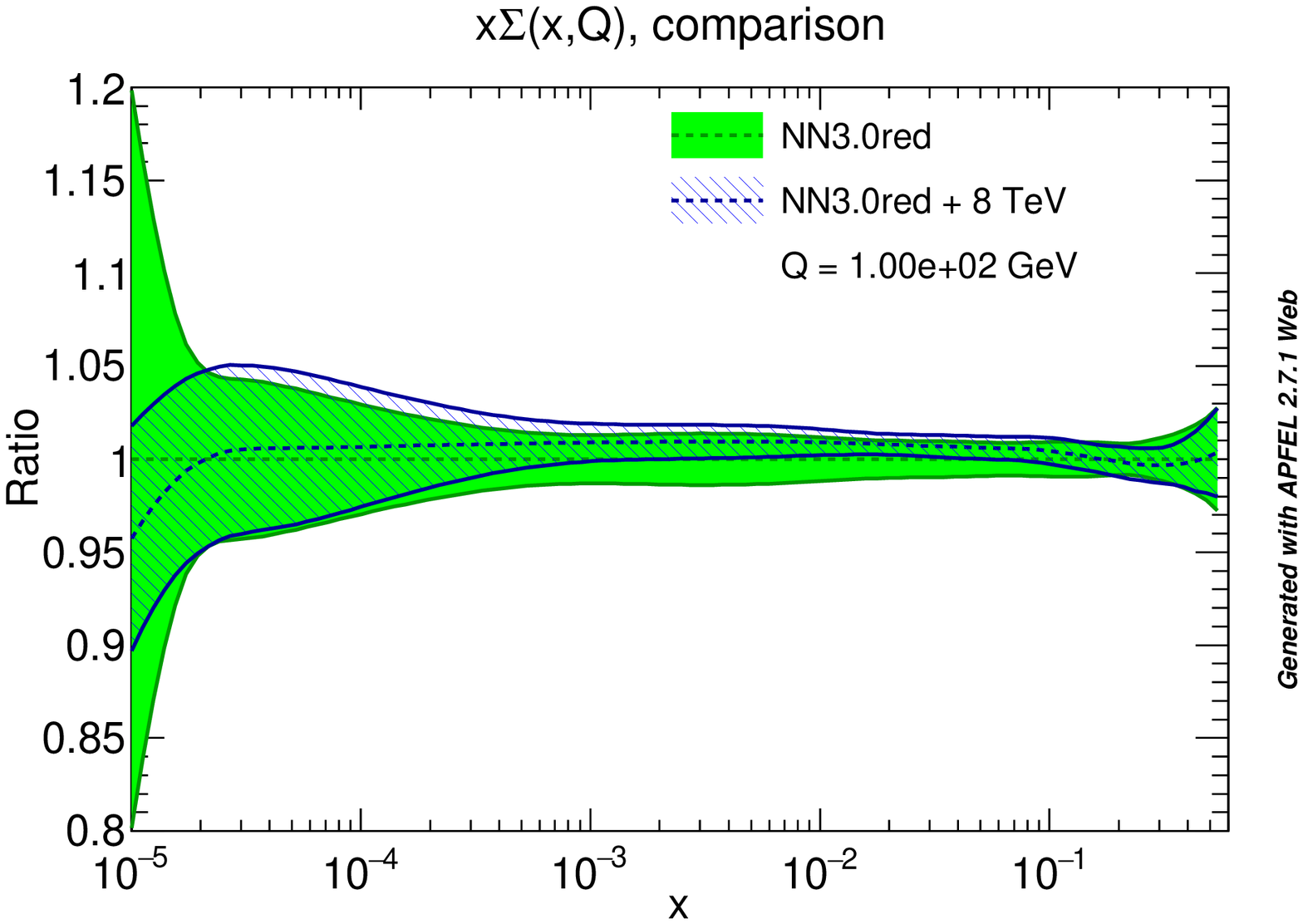}
        \caption{Impact of the inclusion of the 8 TeV $\pt$ data on the global gluon and singlet-quark distributions.\label{fig:global1}}
\end{figure}

\begin{figure}[tbp]
        \centering
        \includegraphics[width=0.45\textwidth]{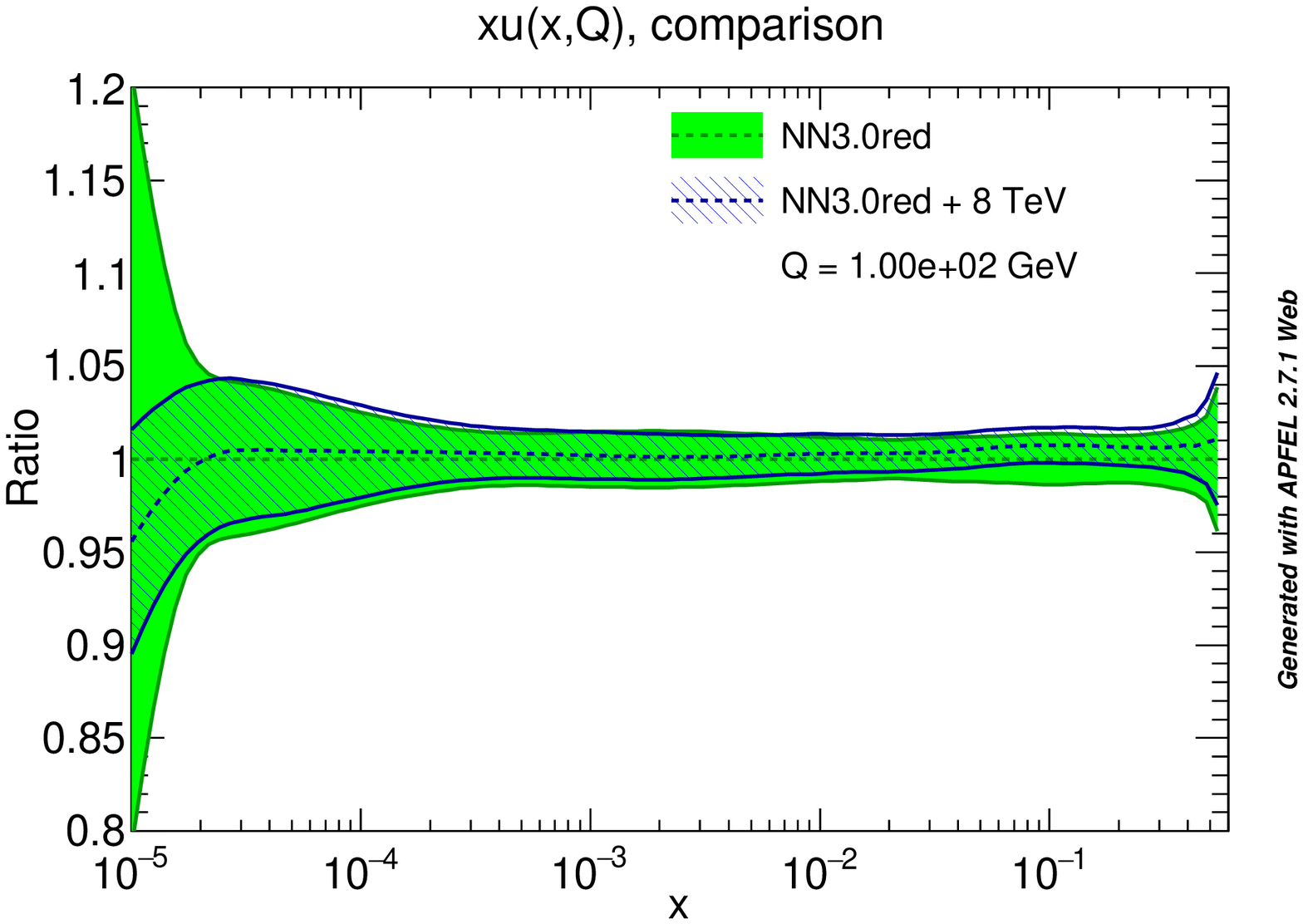}
        \includegraphics[width=0.45\textwidth]{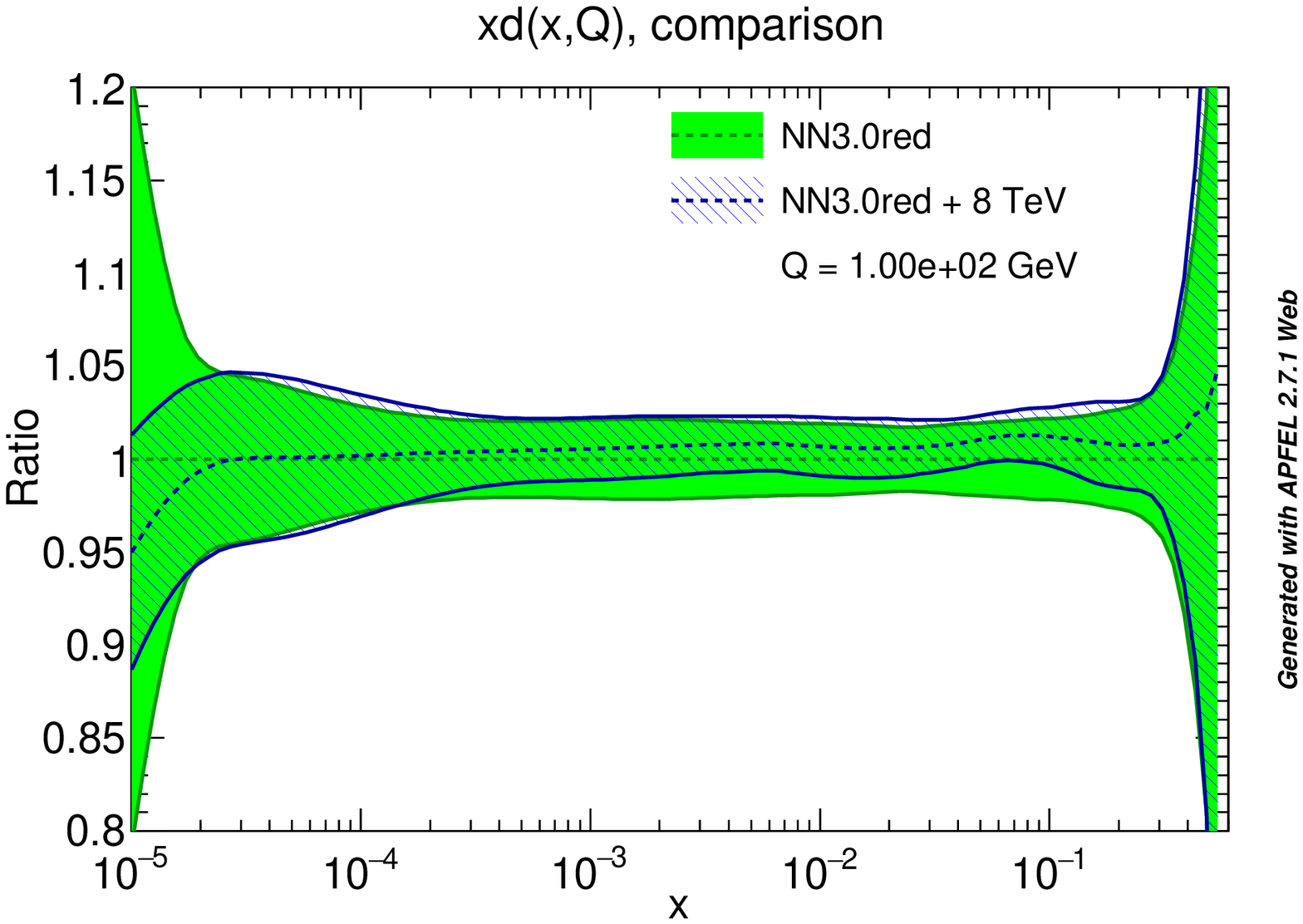}
        \caption{Impact of the inclusion of the 8 TeV $\pt$ data on the global up-quark and down-quark distributions.\label{fig:global2}}
\end{figure}

In Fig.~\ref{fig:global1} and \ref{fig:global2} we show the impact of the precise 8 TeV $\pt$ data on 
the various PDFs determined from the global fit of the available data. 
The observed shifts of the PDFs are similar to those seen in the HERA-only fit.  The reduction
of the uncertainty is milder but still significant.  
The new PDFs obtained after including the 8 TeV $\pt$ data are consistent 
with those found in the baseline.  

It is interesting to compare our results with those presented in~\cite{Czakon:2016olj}, in which a similar baseline 
was used and the impact of including top-pair production differential distributions in PDF fits was studied in detail 
for the first time.  
The gluon is pulled in the same direction by both data sets, thus displaying a perfect compatibility between these 
two complementary measurements.  The inclusion of the $\pt$ data decreases the uncertainties on the gluon PDF 
more than the top-pair data in the intermediate-$x$ region between $10^{-3}$ and $10^{-2}$.  The impact of the 
top-pair data is much stronger for $x>10^{-2}$.  This result follows the correlation patterns presented in Section 5.1 
for $\pt$ and in~\cite{Czakon:2016olj} for top-quark differential distributions, from which it is clear that the latter are 
strongly correlated with the gluon in the large-$x$ region, while the former are mostly correlated with the gluon (and 
slightly less with the light-quark distributions) in the intermediate-$x$ region.  Given that these two observables provide 
such strong and complementary constraints, we expect that their impact in a joint fit will be stronger than the impact of 
the jet data, which were traditionally thought to be the best probe of the gluon in the intermediate and large-$x$ regions.

To conclude, we explore the stability of our results upon increasing the $\pt$ cut from 30 GeV to 50 GeV.
Both the gluon and singlet central values are very stable, with uncertainties that are larger when a larger $\pt$ cut is used.
We note that the number of $\pt$ data points in the fit decreases from 48 to 40 for the ATLAS 8 TeV on-peak data, from 
44 to 36 for the ATLAS 8 TeV off-peak data and from 28 to 24 for the CMS 8 TeV on-peak data. Thus an increase in 
the PDF uncertainty when the cut is raised is expected. Everything else is consistent with expectations.
\begin{figure}[tbp]
        \centering
        \includegraphics[width=0.45\textwidth]{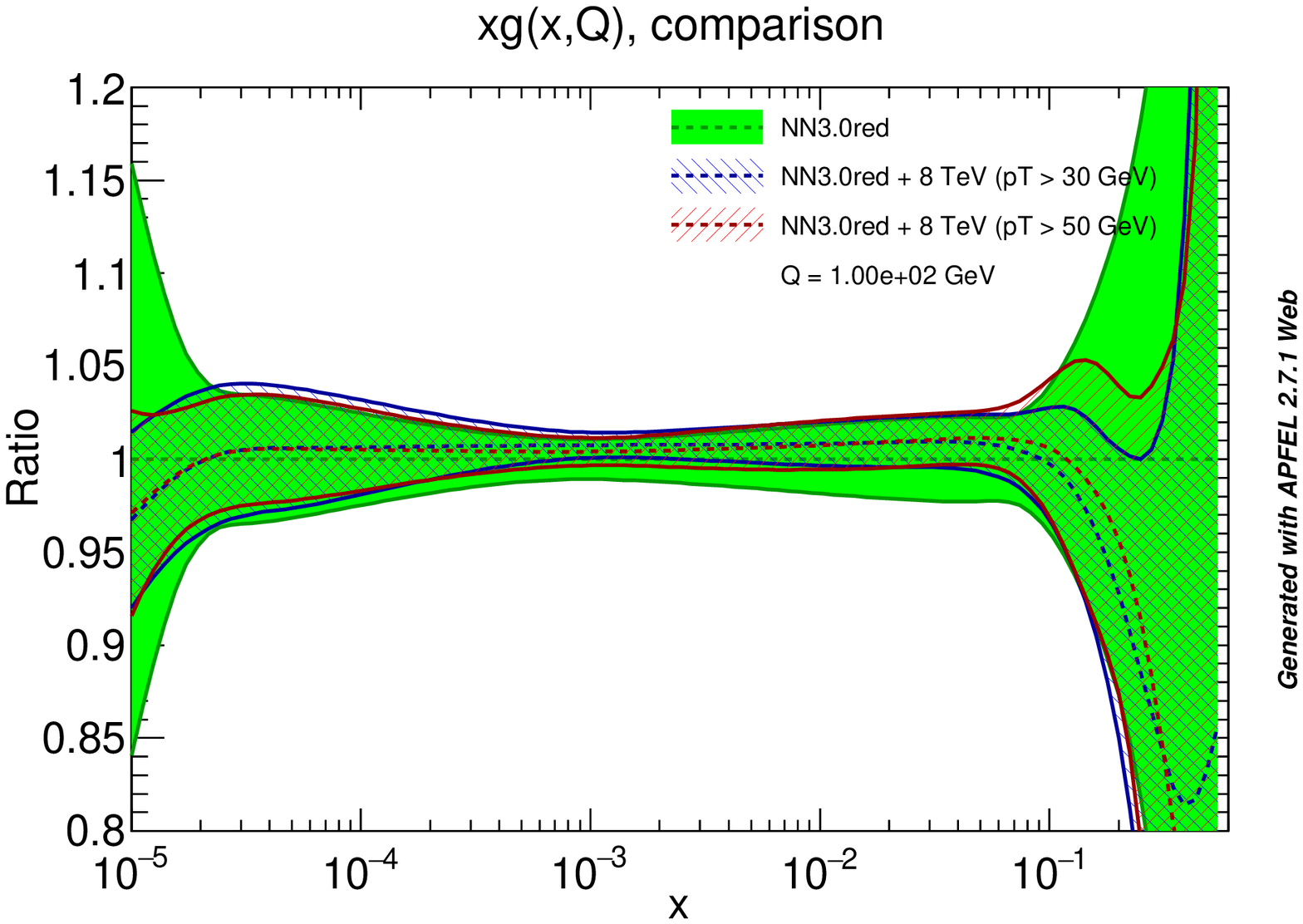}
        \includegraphics[width=0.45\textwidth]{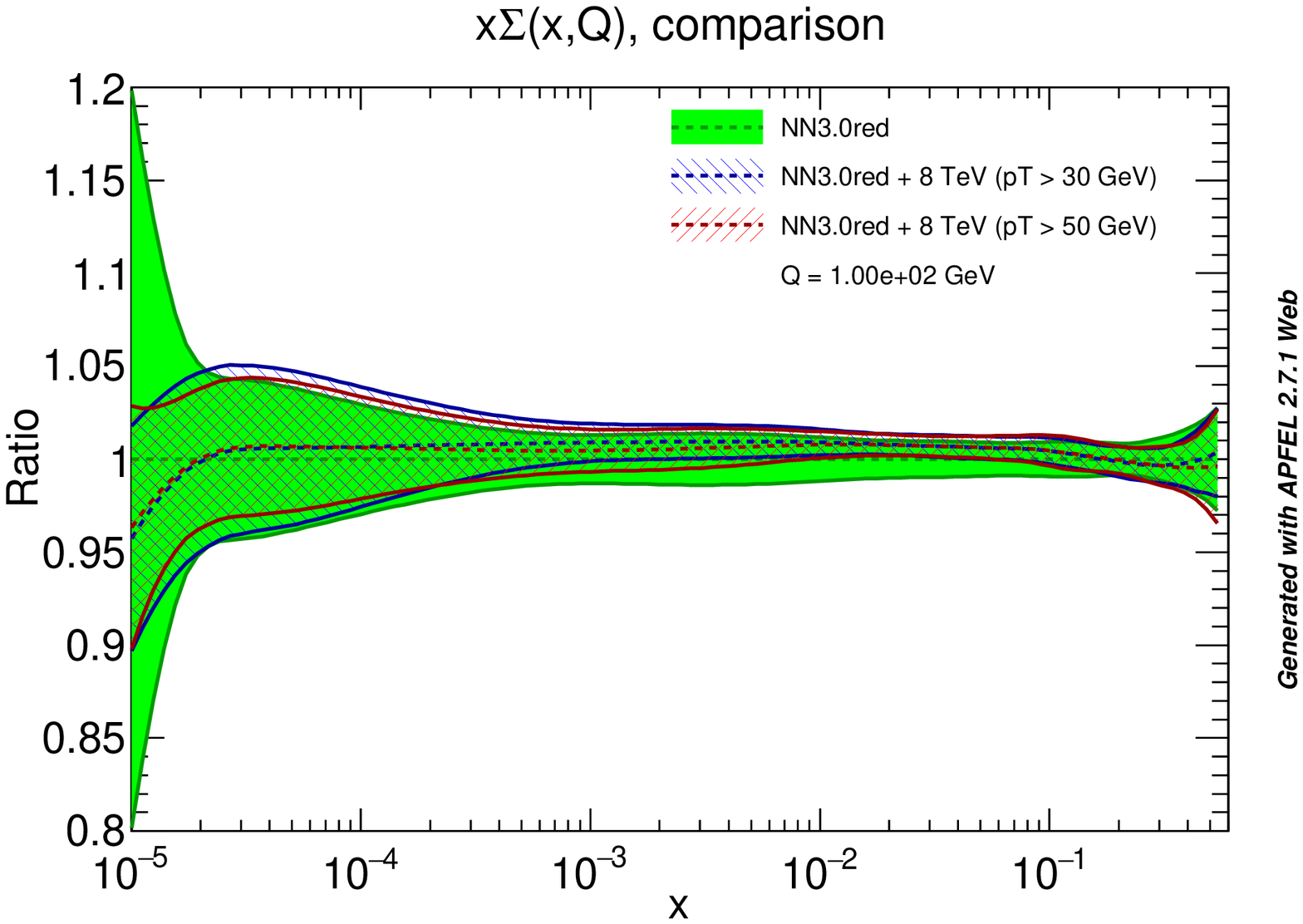}
        \caption{Impact of the choice of $\pt$ cut on the gluon and
          singlet-quark distributions. \label{fig:cut}}
\end{figure}

\section{Phenomenological implications}
\label{sec:pheno}

Having derived a new global fit of PDFs with the 8 TeV $\pt$ data included, it is interesting to investigate 
the impact of these new measurements on quantities of phenomenological interest. 

\begin{figure}[tbp]
        \centering
        \includegraphics[width=0.32\textwidth]{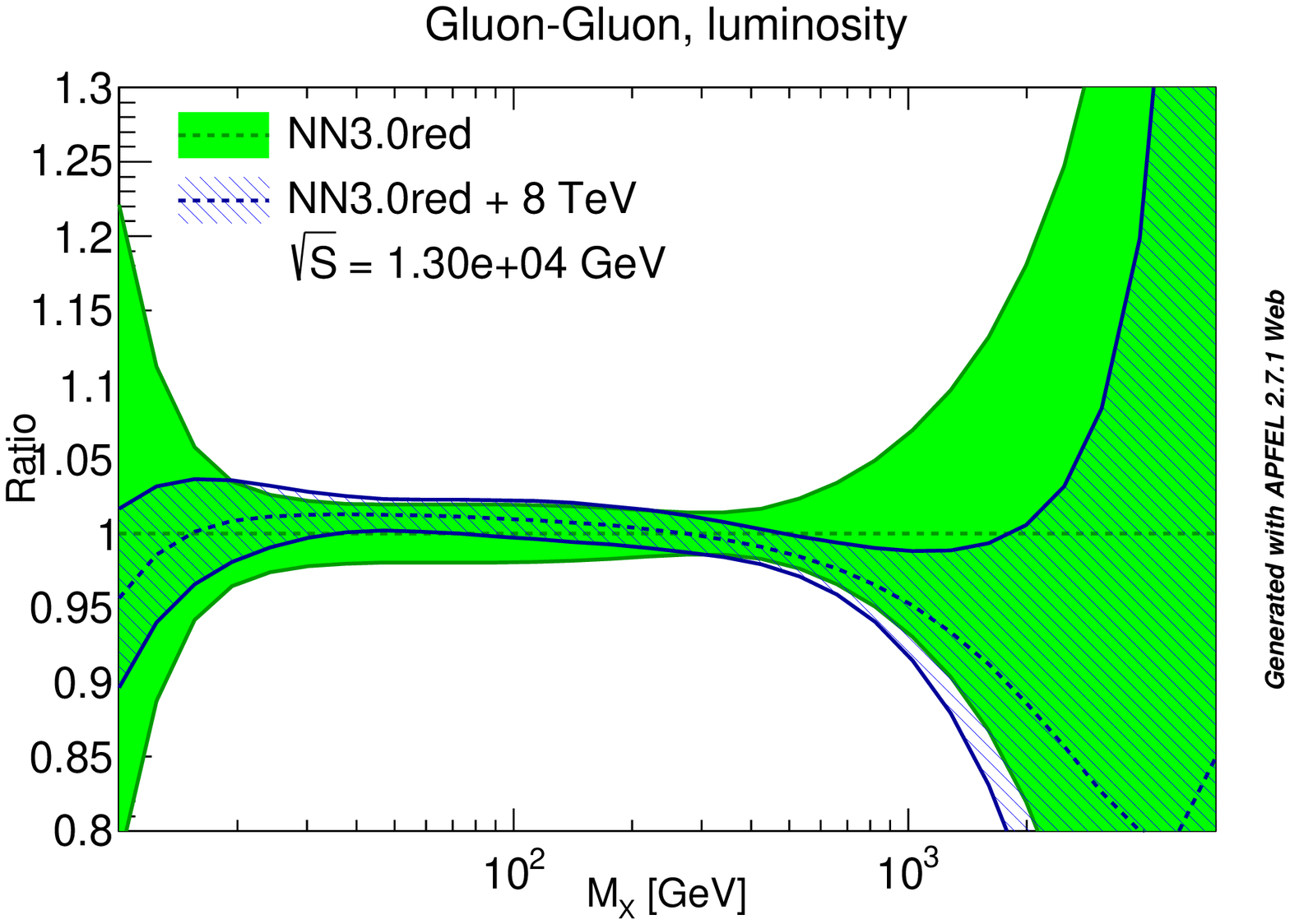}
        \includegraphics[width=0.32\textwidth]{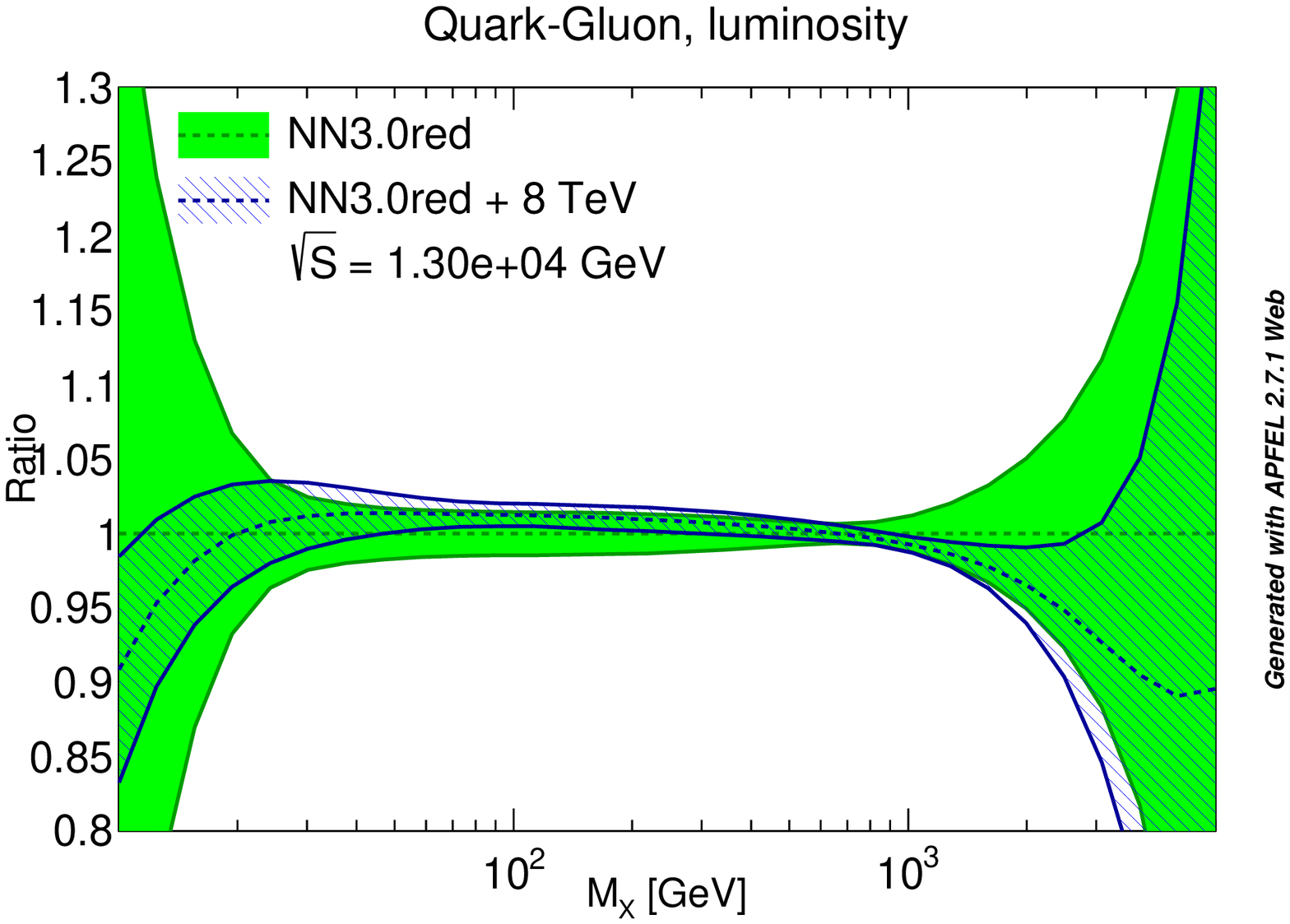} 
        \includegraphics[width=0.32\textwidth]{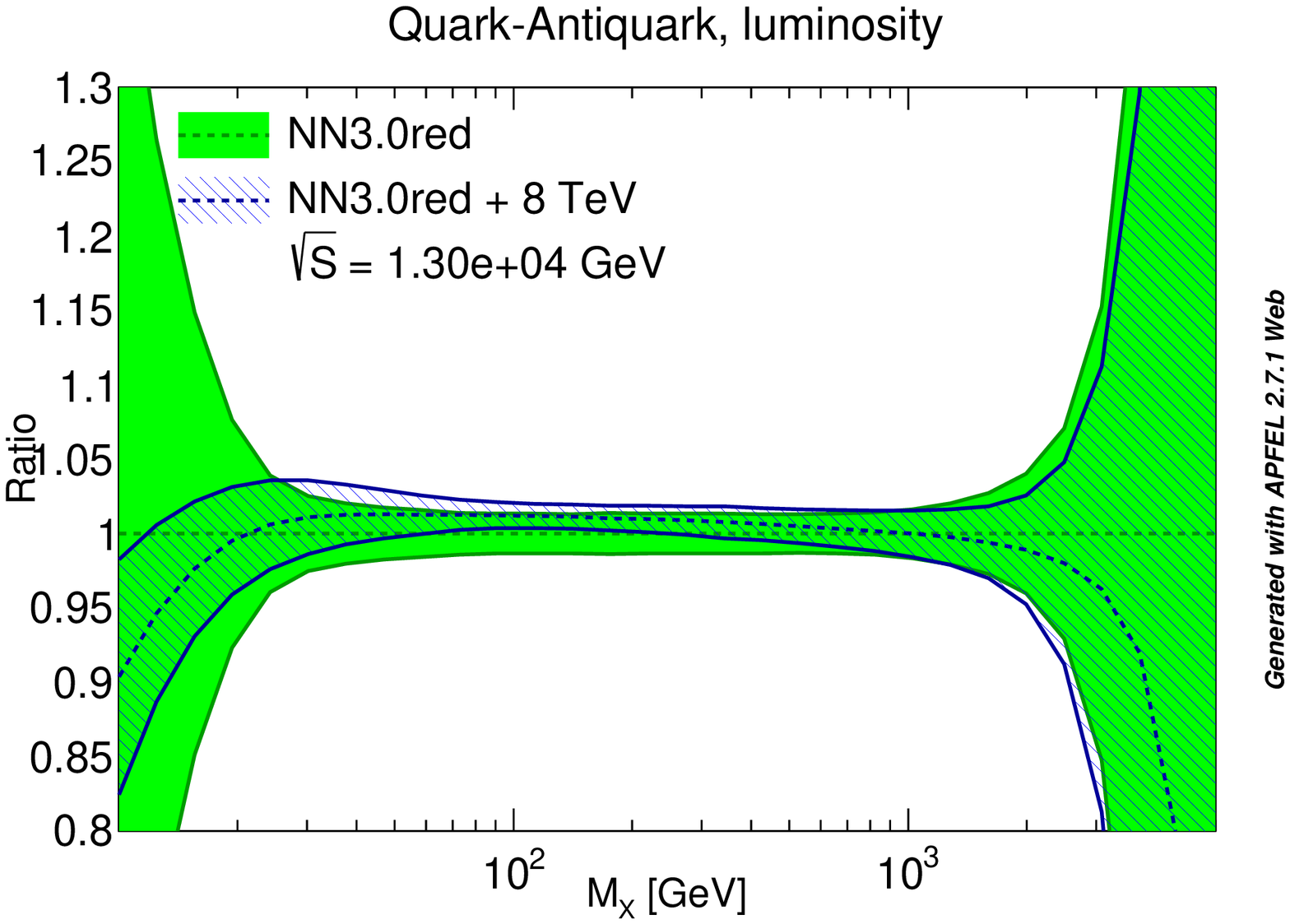} 
        \caption{Impact of the inclusion of $\pt$ data taken at 8 TeV on
various parton-parton luminosities at LHC 13 TeV.\label{fig:lumi}}
\end{figure}

Parton luminosities directly show the impact of the inclusion of a given data set on 
the computation of processes.
A comparison of the 13 TeV parton-parton luminosities before the $\pt$ data, and after including the unnormalized 8 TeV data, 
is presented in Fig.~\ref{fig:lumi}.  The uncertainties  significantly decrease  in all three luminosities, 
while their central values remain nearly the same as before.

Furthermore, we present below the 13 TeV predictions for both the gluon-fusion Higgs production 
cross section and the VBF Higgs production cross section before and after the inclusion of the 
$\pt$ data in our global baseline fit.  For the gluon-fusion production cross section we set 
$m_H=125$ GeV and $\mu_R=\mu_F = m_H/2$ and use the code {\tt ggHiggs v3.5}~\cite{Bonvini:2016frm} 
to compute the result through N$^3$LO in QCD perturbation theory~\cite{Anastasiou:2016cez}.  
The result below includes no charm or bottom quarks running in the loop, and no quark mass effects beyond leading order.
The impact on the Higgs production cross section uncertainties is significant.  The error on the gluon-fusion production cross section 
is reduced by 30\%, following the corresponding improvement in the gluon-gluon-luminosity observed in Fig.~\ref{fig:lumi}.  
The central value is increased by only 1\%, indicating consistency with the cross section obtained using the previous global fit.
For Higgs production in Vector Boson Fusion we compute the total cross section to  N$^3$LO in QCD using the {\tt proVBFH-inclusive}
code~\cite{provbfh} based on the computation presented in~\cite{Cacciari:2015jma,Dreyer:2016oyx}.

\begin{table}[htbp]
\centering
\caption{Predictions for the Higgs cross sections in 13 TeV $pp$ collisions before and after inclusion of the $\pt$ data in the global fits.  
The indicated errors are the PDF errors computed according to the NNPDF prescription.}
\label{tab:Higgscr}
\begin{small}
\begin{tabular}{c|c|c}
\hline
& Before $\pt$ data & After $\pt$ data \\
\hline
$\sigma_{gg \to H}$ [pb] & $48.22 \pm 0.89$ (1.8\%)  & $48.61 \pm 0.61$ (1.3\%) \\
\hline
$\sigma_{\rm VBF}$ [pb] & $3.92 \pm 0.06$ (1.5\%) &  $3.96 \pm 0.04$ (1.0\%)\\  
\hline
\end{tabular}
\end{small}
\end{table}

\section{Conclusions}
\label{sec:conc}

In this manuscript we have included for the first time the precision $\pt$ measurements from the LHC into a global fit of parton distribution
functions to next-to-next-to-leading order in QCD.  This result is made possible by the recent theoretical predictions of this process to the
necessary order.  We have performed a detailed study of the impact of various perturbative corrections, including higher-order QCD and
electroweak corrections, on the agreement between theory and data.  To asses in detail the impact of these new data we have tested the 
effect of adding them to several baseline fits, including a DIS HERA-only PDF determination and a global fit with settings closely following 
those of NNPDF3.0.

The major findings of our study are summarized below.  In their current form the normalized ATLAS 7 TeV data cannot be fit simultaneously 
with the 8 TeV $\pt$ data. It also cannot be fit together with HERA data, nor in a global fit.  The normalization performed on the 7 TeV data 
ties together the low and high $\pt$ regions. When we perform the fit on the high$-\pt$ region needed for a stable fixed-order QCD prediction, 
thus on a region in $\pt$ which is different from the one used to normalise the data, the correlations between the bins are lost.  
The inclusion of this data requires either the experimental covariance matrix for the $\pt > 30$ GeV range only, the unnormalized data, 
or the inclusion of low-$\pt$ resummation in the theoretical prediction.  
This last option would introduce an additional theoretical uncertainty into the fit.  

The extreme precision of the 8 TeV $\pt$ data binned in rapidity, with uncertainties at the few-per-mille level for the majority of bins, 
necessitates the introduction of an additional uncorrelated uncertainty for a fit with a low $\chi^2$ per degree of freedom.  This additional 
parameter is meant to cover the residual theoretical uncertainty and the Monte-Carlo integration uncertainty on the theoretical prediction, 
as well as possible under-reported experimental errors.  While the introduction of this extra uncertainty improves the $\chi^2$ per degree of
 freedom of the fit, we have varied the chosen value of this parameter to check that it has little impact on the actual PDFs obtained from the fit. 

Including the 8 TeV $\pt$ data into a global fit based on the NNPDF3.0 settings results in a significant reduction of the 13 TeV gluon-gluon, 
quark-gluon and quark-antiquark luminosity errors.  To quantify this we have computed the gluon-fusion Higgs production using our NNPDF3.0
baseline, before and after including the $\pt$ data in the fit.  We find that the PDF uncertainty on the Higgs cross section decreases by 30\%, 
while the central value of the prediction increases by 1\%, within the previously-estimated uncertainty.  We caution that this quantitative 
estimate of uncertainty reduction holds upon including only the $\pt$ data into the NNPDF3.0 baseline fit.  If additional data sets are included 
as well, these numbers will change.  However, given the power of the $\pt$ data found in our study, we expect that future global fits using this 
data will observe similar results.

\section*{Acknowledgements}
We thank Jan Kretzschmar, Maarten Boonekamp and Stefano Camarda for insightful discussions
on the ATLAS experimental data.  We thank Gavin Salam for discussions on possible non-perturbative corrections to the $\pt$ distribution, 
and Alexander Mueck for sending results for the NLO electroweak
corrections. We thank Alex Mitov for useful discussions on the
comparison with the inclusion of top data.
We acknowledge extensive discussions and cross checks of results and PDF fits with the members of the NNPDF Collaboration, 
in particular Zahari Dim, Luca Rottoli, Stefano Carrazza, Nathan Hartland and Juan Rojo.
We thank the Kavli Institute for Theoretical Physics in Santa Barbara for hosting the authors during the completion of this manuscript. 
This research was supported in part by the National Science Foundation under Grant No. NSF PHY11-25915 to the Kavli Institute of 
Theoretical Physics in Santa Barbara. 
R.~B. is supported by the DOE contract DE-AC02-06CH11357. F.~P. is supported by the DOE grants DE-FG02- 91ER40684 and 
DE-AC02-06CH11357. M.~U. is supported by a Royal Society Dorothy Hodgkin Research Fellowship and partially supported by the 
STFC grant ST/L000385/1. A.~G. is supported by the European Union�s Horizon 2020 research and innovation programme under the 
Marie Sk\l odowska-Curie grant agreement No 659128 - NEXTGENPDF. This research used resources of the Argonne Leadership 
Computing Facility, which is a DOE Office of Science User Facility supported under Contract DE-AC02-06CH11357.

\renewcommand{\em}{}
\bibliographystyle{UTPstyle}
\input{ZpT.bbl}

\end{document}

%% file: ZpT.bbl
\providecommand{\href}[2]{#2}\begingroup\raggedright\endgroup

%% file: ZpT.bbl
\begin{thebibliography}{10}

\bibitem{Hamberg:1990np}
R.~Hamberg, W.~L. van Neerven, and T.~Matsuura, {\it {A complete calculation of
  the order $\alpha-s^{2}$ correction to the Drell-Yan $K$ factor}},  {\em
  Nucl. Phys.} {\bf B359} (1991) 343--405. [Erratum: Nucl.
  Phys.B644,403(2002)].

\bibitem{Anastasiou:2003ds}
C.~Anastasiou, L.~J. Dixon, K.~Melnikov, and F.~Petriello, {\it {High precision
  QCD at hadron colliders: Electroweak gauge boson rapidity distributions at
  NNLO}},  {\em Phys. Rev.} {\bf D69} (2004) 094008,
  [\href{http://xxx.lanl.gov/abs/hep-ph/0312266}{{\tt hep-ph/0312266}}].

\bibitem{Melnikov:2006kv}
K.~Melnikov and F.~Petriello, {\it {Electroweak gauge boson production at
  hadron colliders through O(alpha(s)**2)}},  {\em Phys. Rev.} {\bf D74} (2006)
  114017, [\href{http://xxx.lanl.gov/abs/hep-ph/0609070}{{\tt
  hep-ph/0609070}}].

\bibitem{Catani:2009sm}
S.~Catani, L.~Cieri, G.~Ferrera, D.~de~Florian, and M.~Grazzini, {\it {Vector
  boson production at hadron colliders: a fully exclusive QCD calculation at
  NNLO}},  {\em Phys. Rev. Lett.} {\bf 103} (2009) 082001,
  [\href{http://xxx.lanl.gov/abs/0903.2120}{{\tt arXiv:0903.2120}}].

\bibitem{Gavin:2010az}
R.~Gavin, Y.~Li, F.~Petriello, and S.~Quackenbush, {\it {FEWZ 2.0: A code for
  hadronic Z production at next-to-next-to-leading order}},  {\em Comput. Phys.
  Commun.} {\bf 182} (2011) 2388--2403,
  [\href{http://xxx.lanl.gov/abs/1011.3540}{{\tt arXiv:1011.3540}}].

\bibitem{Malik:2013kba}
S.~A. Malik and G.~Watt, {\it {Ratios of W and Z Cross Sections at Large Boson
  $p_T$ as a Constraint on PDFs and Background to New Physics}},  {\em JHEP}
  {\bf 02} (2014) 025, [\href{http://xxx.lanl.gov/abs/1304.2424}{{\tt
  arXiv:1304.2424}}].

\bibitem{Rolph:2015hoa}
A.~Rolph and M.~Ubiali, {\it {PDFs and LHC data: current and future
  constraints}},  {\em EPJ Web Conf.} {\bf 90} (2015) 07001.

\bibitem{Chatterjee:2016sxt}
R.~M. Chatterjee, M.~Guchait, and R.~Placakyte, {\it {QCD analysis of the CMS
  inclusive differential $Z$ production data at $\sqrt {s} =$ 8??TeV}},  {\em
  Phys. Rev.} {\bf D94} (2016), no.~3 034035,
  [\href{http://xxx.lanl.gov/abs/1603.09619}{{\tt arXiv:1603.09619}}].

\bibitem{Currie:2016bfm}
J.~Currie, E.~W.~N. Glover, and J.~Pires, {\it {NNLO QCD predictions for single
  jet inclusive production at the LHC}},  {\em Phys. Rev. Lett.} {\bf 118}
  (2017), no.~7 072002, [\href{http://xxx.lanl.gov/abs/1611.01460}{{\tt
  arXiv:1611.01460}}].

\bibitem{Beneke:2012wb}
M.~Beneke, P.~Falgari, S.~Klein, J.~Piclum, C.~Schwinn, M.~Ubiali, and F.~Yan,
  {\it {Inclusive Top-Pair Production Phenomenology with TOPIXS}},  {\em JHEP}
  {\bf 07} (2012) 194, [\href{http://xxx.lanl.gov/abs/1206.2454}{{\tt
  arXiv:1206.2454}}].

\bibitem{Czakon:2013tha}
M.~Czakon, M.~L. Mangano, A.~Mitov, and J.~Rojo, {\it {Constraints on the gluon
  PDF from top quark pair production at hadron colliders}},  {\em JHEP} {\bf
  07} (2013) 167, [\href{http://xxx.lanl.gov/abs/1303.7215}{{\tt
  arXiv:1303.7215}}].

\bibitem{Czakon:2016olj}
M.~Czakon, N.~P. Hartland, A.~Mitov, E.~R. Nocera, and J.~Rojo, {\it {Pinning
  down the large-x gluon with NNLO top-quark pair differential distributions}},
   {\em JHEP} {\bf 04} (2017) 044,
  [\href{http://xxx.lanl.gov/abs/1611.08609}{{\tt arXiv:1611.08609}}].

\bibitem{Ball:2014uwa}
{\bf NNPDF} Collaboration, R.~D. Ball et~al., {\it {Parton distributions for
  the LHC Run II}},  {\em JHEP} {\bf 04} (2015) 040,
  [\href{http://xxx.lanl.gov/abs/1410.8849}{{\tt arXiv:1410.8849}}].

\bibitem{Aad:2014xaa}
{\bf ATLAS} Collaboration, G.~Aad et~al., {\it {Measurement of the $Z/\gamma^*$
  boson transverse momentum distribution in $pp$ collisions at $\sqrt{s}$ = 7
  TeV with the ATLAS detector}},  {\em JHEP} {\bf 09} (2014) 145,
  [\href{http://xxx.lanl.gov/abs/1406.3660}{{\tt arXiv:1406.3660}}].

\bibitem{Aad:2015auj}
{\bf ATLAS} Collaboration, G.~Aad et~al., {\it {Measurement of the transverse
  momentum and $\phi ^*_{\eta }$ distributions of Drell-Yan lepton pairs in
  proto-proton collisions at $\sqrt{s}=8$ TeV with the ATLAS detector}},  {\em
  Eur. Phys. J.} {\bf C76} (2016), no.~5 291,
  [\href{http://xxx.lanl.gov/abs/1512.02192}{{\tt arXiv:1512.02192}}].

\bibitem{Khachatryan:2015oaa}
{\bf CMS} Collaboration, V.~Khachatryan et~al., {\it {Measurement of the Z
  boson differential cross section in transverse momentum and rapidity in
  proton-proton collisions at 8 TeV}},  {\em Phys. Lett.} {\bf B749} (2015)
  187--209, [\href{http://xxx.lanl.gov/abs/1504.03511}{{\tt
  arXiv:1504.03511}}].

\bibitem{Boughezal:2015ded}
R.~Boughezal, J.~M. Campbell, R.~K. Ellis, C.~Focke, W.~T. Giele, X.~Liu, and
  F.~Petriello, {\it {Z-boson production in association with a jet at
  next-to-next-to-leading order in perturbative QCD}},  {\em Phys. Rev. Lett.}
  {\bf 116} (2016), no.~15 152001,
  [\href{http://xxx.lanl.gov/abs/1512.01291}{{\tt arXiv:1512.01291}}].

\bibitem{Ridder:2016nkl}
A.~Gehrmann-De~Ridder, T.~Gehrmann, E.~W.~N. Glover, A.~Huss, and T.~A. Morgan,
  {\it {The NNLO QCD corrections to Z boson production at large transverse
  momentum}},  {\em JHEP} {\bf 07} (2016) 133,
  [\href{http://xxx.lanl.gov/abs/1605.04295}{{\tt arXiv:1605.04295}}].

\bibitem{Gehrmann-DeRidder:2016jns}
A.~Gehrmann-De~Ridder, T.~Gehrmann, E.~W.~N. Glover, A.~Huss, and T.~A. Morgan,
  {\it {NNLO QCD corrections for Drell-Yan $p_T^Z$ and $\phi^*$ observables at
  the LHC}},  {\em JHEP} {\bf 11} (2016) 094,
  [\href{http://xxx.lanl.gov/abs/1610.01843}{{\tt arXiv:1610.01843}}].

\bibitem{Bertone:2013vaa}
V.~Bertone, S.~Carrazza, and J.~Rojo, {\it {APFEL: A PDF Evolution Library with
  QED corrections}},  {\em Comput. Phys. Commun.} {\bf 185} (2014) 1647--1668,
  [\href{http://xxx.lanl.gov/abs/1310.1394}{{\tt arXiv:1310.1394}}].

\bibitem{Carrazza:2014gfa}
S.~Carrazza, A.~Ferrara, D.~Palazzo, and J.~Rojo, {\it {APFEL Web}},  {\em J.
  Phys.} {\bf G42} (2015), no.~5 057001,
  [\href{http://xxx.lanl.gov/abs/1410.5456}{{\tt arXiv:1410.5456}}].

\bibitem{Bertone:2016lga}
V.~Bertone, S.~Carrazza, and N.~P. Hartland, {\it {APFELgrid: a high
  performance tool for parton density determinations}},
  \href{http://xxx.lanl.gov/abs/1605.02070}{{\tt arXiv:1605.02070}}.

\bibitem{Forte:2010ta}
S.~Forte, E.~Laenen, P.~Nason, and J.~Rojo, {\it {Heavy quarks in
  deep-inelastic scattering}},  {\em Nucl. Phys.} {\bf B834} (2010) 116--162,
  [\href{http://xxx.lanl.gov/abs/1001.2312}{{\tt arXiv:1001.2312}}].

\bibitem{Agashe:2014kda}
{\bf Particle Data Group} Collaboration, K.~A. Olive et~al., {\it {Review of
  Particle Physics}},  {\em Chin. Phys.} {\bf C38} (2014) 090001.

\bibitem{Dittmaier:2011ti}
{\bf LHC Higgs Cross Section Working Group} Collaboration, S.~Dittmaier et~al.,
  {\it {Handbook of LHC Higgs Cross Sections: 1. Inclusive Observables}},
  \href{http://xxx.lanl.gov/abs/1101.0593}{{\tt arXiv:1101.0593}}.

\bibitem{nnpdf31}
{\bf NNPDF} Collaboration, R.~D. Ball et~al., {\it {Parton distributions from
  high-precision collider data}},  {\em In preparation}.

\bibitem{Aaron:2012qi}
{\bf H1} Collaboration, F.~D. Aaron et~al., {\it {Inclusive Deep Inelastic
  Scattering at High $Q^2$ with Longitudinally Polarised Lepton Beams at
  HERA}},  {\em JHEP} {\bf 09} (2012) 061,
  [\href{http://xxx.lanl.gov/abs/1206.7007}{{\tt arXiv:1206.7007}}].

\bibitem{Collaboration:2010ry}
{\bf H1} Collaboration, F.~Aaron et~al., {\it {Measurement of the Inclusive
  $e^{\pm}p$ Scattering Cross Section at High Inelasticity y and of the
  Structure Function $F_L$}},  {\em Eur.Phys.J.} {\bf C71} (2011) 1579,
  [\href{http://xxx.lanl.gov/abs/1012.4355}{{\tt arXiv:1012.4355}}].

\bibitem{Abramowicz:2012bx}
{\bf ZEUS} Collaboration, H.~Abramowicz et~al., {\it {Measurement of high-Q2
  neutral current deep inelastic $e^+p$ scattering cross sections with a
  longitudinally polarized positron beam at HERA}},  {\em Phys. Rev.} {\bf D87}
  (2013), no.~5 052014, [\href{http://xxx.lanl.gov/abs/1208.6138}{{\tt
  arXiv:1208.6138}}].

\bibitem{Abramowicz:2015mha}
{\bf ZEUS, H1} Collaboration, H.~Abramowicz et~al., {\it {Combination of
  measurements of inclusive deep inelastic ${e^{\pm }p}$ scattering cross
  sections and QCD analysis of HERA data}},  {\em Eur. Phys. J.} {\bf C75}
  (2015), no.~12 580, [\href{http://xxx.lanl.gov/abs/1506.06042}{{\tt
  arXiv:1506.06042}}].

\bibitem{Abramowicz:1900rp}
{\bf ZEUS, H1} Collaboration, H.~Abramowicz et~al., {\it {Combination and QCD
  Analysis of Charm Production Cross Section Measurements in Deep-Inelastic ep
  Scattering at HERA}},  {\em Eur. Phys. J.} {\bf C73} (2013), no.~2 2311,
  [\href{http://xxx.lanl.gov/abs/1211.1182}{{\tt arXiv:1211.1182}}].

\bibitem{Aaron:2009af}
{\bf H1} Collaboration, F.~D. Aaron et~al., {\it {Measurement of the Charm and
  Beauty Structure Functions using the H1 Vertex Detector at HERA}},  {\em Eur.
  Phys. J.} {\bf C65} (2010) 89--109,
  [\href{http://xxx.lanl.gov/abs/0907.2643}{{\tt arXiv:0907.2643}}].

\bibitem{Abramowicz:2014zub}
{\bf ZEUS} Collaboration, H.~Abramowicz et~al., {\it {Measurement of beauty and
  charm production in deep inelastic scattering at HERA and measurement of the
  beauty-quark mass}},  {\em JHEP} {\bf 09} (2014) 127,
  [\href{http://xxx.lanl.gov/abs/1405.6915}{{\tt arXiv:1405.6915}}].

\bibitem{Arneodo:1996kd}
{\bf New Muon} Collaboration, M.~Arneodo et~al., {\it {Accurate measurement of
  F2(d) / F2(p) and R**d - R**p}},  {\em Nucl. Phys.} {\bf B487} (1997) 3--26,
  [\href{http://xxx.lanl.gov/abs/hep-ex/9611022}{{\tt hep-ex/9611022}}].

\bibitem{Arneodo:1996qe}
{\bf New Muon} Collaboration, M.~Arneodo et~al., {\it {Measurement of the
  proton and deuteron structure functions, F2(p) and F2(d), and of the ratio
  sigma-L / sigma-T}},  {\em Nucl. Phys.} {\bf B483} (1997) 3--43,
  [\href{http://xxx.lanl.gov/abs/hep-ph/9610231}{{\tt hep-ph/9610231}}].

\bibitem{bcdms1}
{\bf BCDMS} Collaboration, A.~C. Benvenuti et~al., {\it A high statistics
  measurement of the proton structure functions $f_2 (x, q^2)$ and $r$ from
  deep inelastic muon scattering at high $q^2$},  {\em Phys. Lett.} {\bf B223}
  (1989) 485.

\bibitem{bcdms2}
{\bf BCDMS} Collaboration, A.~C. Benvenuti et~al., {\it A high statistics
  measurement of the deuteron structure functions $f_2 (x, q^2)$ and $r$ from
  deep inelastic muon scattering at high $q^2$},  {\em Phys. Lett.} {\bf B237}
  (1990) 592.

\bibitem{Whitlow:1991uw}
L.~W. Whitlow, E.~M. Riordan, S.~Dasu, S.~Rock, and A.~Bodek, {\it {Precise
  measurements of the proton and deuteron structure functions from a global
  analysis of the SLAC deep inelastic electron scattering cross-sections}},
  {\em Phys. Lett.} {\bf B282} (1992) 475--482.

\bibitem{Onengut:2005kv}
{\bf CHORUS} Collaboration, G.~Onengut et~al., {\it {Measurement of nucleon
  structure functions in neutrino scattering}},  {\em Phys. Lett.} {\bf B632}
  (2006) 65--75.

\bibitem{Goncharov:2001qe}
{\bf NuTeV} Collaboration, M.~Goncharov et~al., {\it {Precise measurement of
  dimuon production cross-sections in muon neutrino Fe and muon anti-neutrino
  Fe deep inelastic scattering at the Tevatron}},  {\em Phys. Rev.} {\bf D64}
  (2001) 112006, [\href{http://xxx.lanl.gov/abs/hep-ex/0102049}{{\tt
  hep-ex/0102049}}].

\bibitem{MasonPhD}
D.~A. Mason, {\it {Measurement of the strange - antistrange asymmetry at NLO in
  QCD from NuTeV dimuon data}}, . FERMILAB-THESIS-2006-01.

\bibitem{Moreno:1990sf}
G.~Moreno et~al., {\it {Dimuon production in proton - copper collisions at
  $\sqrt{s}$ = 38.8-GeV}},  {\em Phys. Rev.} {\bf D43} (1991) 2815--2836.

\bibitem{Webb:2003ps}
{\bf NuSea} Collaboration, J.~C. Webb et~al., {\it {Absolute Drell-Yan dimuon
  cross sections in 800-GeV/c p p and p d collisions}},
  \href{http://xxx.lanl.gov/abs/hep-ex/0302019}{{\tt hep-ex/0302019}}.

\bibitem{Webb:2003bj}
J.~C. Webb, {\it {Measurement of continuum dimuon production in 800-GeV/c
  proton nucleon collisions}},
  \href{http://xxx.lanl.gov/abs/hep-ex/0301031}{{\tt hep-ex/0301031}}.

\bibitem{Towell:2001nh}
{\bf FNAL E866/NuSea} Collaboration, R.~S. Towell et~al., {\it {Improved
  measurement of the anti-d/anti-u asymmetry in the nucleon sea}},  {\em Phys.
  Rev.} {\bf D64} (2001) 052002,
  [\href{http://xxx.lanl.gov/abs/hep-ex/0103030}{{\tt hep-ex/0103030}}].

\bibitem{Aaltonen:2010zza}
{\bf CDF} Collaboration, T.~A. Aaltonen et~al., {\it {Measurement of
  $d\sigma/dy$ of Drell-Yan $e^+e^-$ pairs in the $Z$ Mass Region from
  $p\bar{p}$ Collisions at $\sqrt{s}=1.96$ TeV}},  {\em Phys. Lett.} {\bf B692}
  (2010) 232--239, [\href{http://xxx.lanl.gov/abs/0908.3914}{{\tt
  arXiv:0908.3914}}].

\bibitem{Abazov:2007jy}
{\bf D0} Collaboration, V.~M. Abazov et~al., {\it {Measurement of the shape of
  the boson rapidity distribution for $p \bar{p} \to Z/gamma^* \to e^{+} e^{-}$
  + $X$ events produced at $\sqrt{s}$ of 1.96-TeV}},  {\em Phys. Rev.} {\bf
  D76} (2007) 012003, [\href{http://xxx.lanl.gov/abs/hep-ex/0702025}{{\tt
  hep-ex/0702025}}].

\bibitem{Aad:2011dm}
{\bf ATLAS} Collaboration, G.~Aad et~al., {\it {Measurement of the inclusive
  $W^\pm$ and Z/gamma cross sections in the electron and muon decay channels in
  $pp$ collisions at $\sqrt{s}=7$ TeV with the ATLAS detector}},  {\em Phys.
  Rev.} {\bf D85} (2012) 072004, [\href{http://xxx.lanl.gov/abs/1109.5141}{{\tt
  arXiv:1109.5141}}].

\bibitem{Aad:2013iua}
{\bf ATLAS} Collaboration, G.~Aad et~al., {\it {Measurement of the high-mass
  Drell--Yan differential cross-section in pp collisions at sqrt(s)=7 TeV with
  the ATLAS detector}},  {\em Phys. Lett.} {\bf B725} (2013) 223--242,
  [\href{http://xxx.lanl.gov/abs/1305.4192}{{\tt arXiv:1305.4192}}].

\bibitem{Aad:2011fp}
{\bf ATLAS} Collaboration, G.~Aad et~al., {\it {Measurement of the Transverse
  Momentum Distribution of $W$ Bosons in $pp$ Collisions at $\sqrt{s}=7$ TeV
  with the ATLAS Detector}},  {\em Phys. Rev.} {\bf D85} (2012) 012005,
  [\href{http://xxx.lanl.gov/abs/1108.6308}{{\tt arXiv:1108.6308}}].

\bibitem{Chatrchyan:2012xt}
{\bf CMS} Collaboration, S.~Chatrchyan et~al., {\it {Measurement of the
  electron charge asymmetry in inclusive $W$ production in $pp$ collisions at
  $\sqrt{s}=7$ TeV}},  {\em Phys. Rev. Lett.} {\bf 109} (2012) 111806,
  [\href{http://xxx.lanl.gov/abs/1206.2598}{{\tt arXiv:1206.2598}}].

\bibitem{Chatrchyan:2013mza}
{\bf CMS} Collaboration, S.~Chatrchyan et~al., {\it {Measurement of the muon
  charge asymmetry in inclusive $pp \to W+X$ production at $\sqrt s =$ 7 TeV
  and an improved determination of light parton distribution functions}},  {\em
  Phys. Rev.} {\bf D90} (2014), no.~3 032004,
  [\href{http://xxx.lanl.gov/abs/1312.6283}{{\tt arXiv:1312.6283}}].

\bibitem{Chatrchyan:2013uja}
{\bf CMS} Collaboration, S.~Chatrchyan et~al., {\it {Measurement of associated
  W + charm production in pp collisions at $\sqrt{s}$ = 7 TeV}},  {\em JHEP}
  {\bf 02} (2014) 013, [\href{http://xxx.lanl.gov/abs/1310.1138}{{\tt
  arXiv:1310.1138}}].

\bibitem{CMSDY}
{\bf CMS} Collaboration, S.~Chatrchyan et~al., {\it {Measurement of the
  differential and double-differential Drell-Yan cross sections in
  proton-proton collisions at $\sqrt{s} =$ 7 TeV}},  {\em JHEP} {\bf 1312}
  (2013) 030, [\href{http://xxx.lanl.gov/abs/1310.7291}{{\tt
  arXiv:1310.7291}}].

\bibitem{Aaij:2012vn}
{\bf LHCb} Collaboration, R.~Aaij et~al., {\it {Inclusive $W$ and $Z$
  production in the forward region at $\sqrt{s} = 7$ TeV}},  {\em JHEP} {\bf
  06} (2012) 058, [\href{http://xxx.lanl.gov/abs/1204.1620}{{\tt
  arXiv:1204.1620}}].

\bibitem{Aaij:2012mda}
{\bf LHCb} Collaboration, R.~Aaij et~al., {\it {Measurement of the
  cross-section for $Z \to e^+e^-$ production in $pp$ collisions at
  $\sqrt{s}=7$ TeV}},  {\em JHEP} {\bf 02} (2013) 106,
  [\href{http://xxx.lanl.gov/abs/1212.4620}{{\tt arXiv:1212.4620}}].

\bibitem{Abulencia:2007ez}
{\bf CDF} Collaboration, A.~Abulencia et~al., {\it {Measurement of the
  Inclusive Jet Cross Section using the {\boldmath $k_{\rm T}$}
  algorithmin{\boldmath $p\overline{p}$} Collisions at{\boldmath $\sqrt{s}$} =
  1.96 TeV with the CDF II Detector}},  {\em Phys. Rev.} {\bf D75} (2007)
  092006, [\href{http://xxx.lanl.gov/abs/hep-ex/0701051}{{\tt
  hep-ex/0701051}}]. [Erratum: Phys. Rev.D75,119901(2007)].

\bibitem{Aad:2011fc}
{\bf ATLAS} Collaboration, G.~Aad et~al., {\it {Measurement of inclusive jet
  and dijet production in $pp$ collisions at $\sqrt{s}=7$ TeV using the ATLAS
  detector}},  {\em Phys. Rev.} {\bf D86} (2012) 014022,
  [\href{http://xxx.lanl.gov/abs/1112.6297}{{\tt arXiv:1112.6297}}].

\bibitem{Chatrchyan:2012bja}
{\bf CMS} Collaboration, S.~Chatrchyan et~al., {\it {Measurements of
  differential jet cross sections in proton-proton collisions at $\sqrt{s}=7$
  TeV with the CMS detector}},  {\em Phys. Rev.} {\bf D87} (2013), no.~11
  112002, [\href{http://xxx.lanl.gov/abs/1212.6660}{{\tt arXiv:1212.6660}}].
  [Erratum: Phys. Rev.D87,no.11,119902(2013)].

\bibitem{Aad:2013lpa}
{\bf ATLAS} Collaboration, G.~Aad et~al., {\it {Measurement of the inclusive
  jet cross section in pp collisions at sqrt(s)=2.76 TeV and comparison to the
  inclusive jet cross section at sqrt(s)=7 TeV using the ATLAS detector}},
  {\em Eur. Phys. J.} {\bf C73} (2013), no.~8 2509,
  [\href{http://xxx.lanl.gov/abs/1304.4739}{{\tt arXiv:1304.4739}}].

\bibitem{Boughezal:2016isb}
R.~Boughezal, X.~Liu, and F.~Petriello, {\it {Phenomenology of the Z-boson plus
  jet process at NNLO}},  {\em Phys. Rev.} {\bf D94} (2016), no.~7 074015,
  [\href{http://xxx.lanl.gov/abs/1602.08140}{{\tt arXiv:1602.08140}}].

\bibitem{Boughezal:2015dva}
R.~Boughezal, C.~Focke, X.~Liu, and F.~Petriello, {\it {$W$-boson production in
  association with a jet at next-to-next-to-leading order in perturbative
  QCD}},  {\em Phys. Rev. Lett.} {\bf 115} (2015), no.~6 062002,
  [\href{http://xxx.lanl.gov/abs/1504.02131}{{\tt arXiv:1504.02131}}].

\bibitem{Boughezal:2015eha}
R.~Boughezal, X.~Liu, and F.~Petriello, {\it {$N$-jettiness soft function at
  next-to-next-to-leading order}},  {\em Phys. Rev.} {\bf D91} (2015), no.~9
  094035, [\href{http://xxx.lanl.gov/abs/1504.02540}{{\tt arXiv:1504.02540}}].

\bibitem{Gaunt:2015pea}
J.~Gaunt, M.~Stahlhofen, F.~J. Tackmann, and J.~R. Walsh, {\it {N-jettiness
  Subtractions for NNLO QCD Calculations}},  {\em JHEP} {\bf 09} (2015) 058,
  [\href{http://xxx.lanl.gov/abs/1505.04794}{{\tt arXiv:1505.04794}}].

\bibitem{Boughezal:2016wmq}
R.~Boughezal, J.~M. Campbell, R.~K. Ellis, C.~Focke, W.~Giele, X.~Liu,
  F.~Petriello, and C.~Williams, {\it {Color Singlet Production at NNLO in
  MCFM}},  \href{http://xxx.lanl.gov/abs/1605.08011}{{\tt arXiv:1605.08011}}.

\bibitem{Denner:2011vu}
A.~Denner, S.~Dittmaier, T.~Kasprzik, and A.~Muck, {\it {Electroweak
  corrections to dilepton + jet production at hadron colliders}},  {\em JHEP}
  {\bf 06} (2011) 069, [\href{http://xxx.lanl.gov/abs/1103.0914}{{\tt
  arXiv:1103.0914}}].

\bibitem{Hollik:2015pja}
W.~Hollik, B.~A. Kniehl, E.~S. Scherbakova, and O.~L. Veretin, {\it
  {Electroweak corrections to $Z$-boson hadroproduction at finite transverse
  momentum}},  {\em Nucl. Phys.} {\bf B900} (2015) 576--602,
  [\href{http://xxx.lanl.gov/abs/1504.07574}{{\tt arXiv:1504.07574}}].

\bibitem{Kuhn:2004em}
J.~H. Kuhn, A.~Kulesza, S.~Pozzorini, and M.~Schulze, {\it {Logarithmic
  electroweak corrections to hadronic Z+1 jet production at large transverse
  momentum}},  {\em Phys. Lett.} {\bf B609} (2005) 277--285,
  [\href{http://xxx.lanl.gov/abs/hep-ph/0408308}{{\tt hep-ph/0408308}}].

\bibitem{Kuhn:2005az}
J.~H. Kuhn, A.~Kulesza, S.~Pozzorini, and M.~Schulze, {\it {One-loop weak
  corrections to hadronic production of Z bosons at large transverse momenta}},
   {\em Nucl. Phys.} {\bf B727} (2005) 368--394,
  [\href{http://xxx.lanl.gov/abs/hep-ph/0507178}{{\tt hep-ph/0507178}}].

\bibitem{Li:2012wna}
Y.~Li and F.~Petriello, {\it {Combining QCD and electroweak corrections to
  dilepton production in FEWZ}},  {\em Phys. Rev.} {\bf D86} (2012) 094034,
  [\href{http://xxx.lanl.gov/abs/1208.5967}{{\tt arXiv:1208.5967}}].

\bibitem{Dittmaier:2015rxo}
S.~Dittmaier, A.~Huss, and C.~Schwinn, {\it {Dominant mixed QCD-electroweak
  O($\alpha$$_s$$\alpha$) corrections to Drell�Yan processes in the resonance
  region}},  {\em Nucl. Phys.} {\bf B904} (2016) 216--252,
  [\href{http://xxx.lanl.gov/abs/1511.08016}{{\tt arXiv:1511.08016}}].

\bibitem{Gavintalk}
G.~Salam, ``Talk presented at the kitp workshop lhc run 2 and the precision
  frontier.'' \url{http://online.kitp.ucsb.edu/online/lhc16/salam/}, 2016.

\bibitem{Alekhin:2017kpj}
S.~Alekhin, J.~Bl�mlein, S.~Moch, and R.~Placakyte, {\it {Parton Distribution
  Functions, $\alpha_s$ and Heavy-Quark Masses for LHC Run II}},
  \href{http://xxx.lanl.gov/abs/1701.05838}{{\tt arXiv:1701.05838}}.

\bibitem{Dulat:2015mca}
S.~Dulat, T.-J. Hou, J.~Gao, M.~Guzzi, J.~Huston, P.~Nadolsky, J.~Pumplin,
  C.~Schmidt, D.~Stump, and C.~P. Yuan, {\it {New parton distribution functions
  from a global analysis of quantum chromodynamics}},  {\em Phys. Rev.} {\bf
  D93} (2016), no.~3 033006, [\href{http://xxx.lanl.gov/abs/1506.07443}{{\tt
  arXiv:1506.07443}}].

\bibitem{Harland-Lang:2014zoa}
L.~A. Harland-Lang, A.~D. Martin, P.~Motylinski, and R.~S. Thorne, {\it {Parton
  distributions in the LHC era: MMHT 2014 PDFs}},  {\em Eur. Phys. J.} {\bf
  C75} (2015), no.~5 204, [\href{http://xxx.lanl.gov/abs/1412.3989}{{\tt
  arXiv:1412.3989}}].

\bibitem{Carrazza:2016htc}
S.~Carrazza, S.~Forte, Z.~Kassabov, and J.~Rojo, {\it {Specialized minimal PDFs
  for optimized LHC calculations}},  {\em Eur. Phys. J.} {\bf C76} (2016),
  no.~4 205, [\href{http://xxx.lanl.gov/abs/1602.00005}{{\tt
  arXiv:1602.00005}}].

\bibitem{Carrazza:2017bjw}
S.~Carrazza, {\it {Modeling NNLO jet corrections with neural networks}},  in
  {\em {23rd Cracow Epiphany Conference on Particle Theory Meets the First Data
  from LHC Run 2 Cracow, Poland, January 9-12, 2017}}, 2017.
\newblock \href{http://xxx.lanl.gov/abs/1704.00471}{{\tt arXiv:1704.00471}}.

\bibitem{Bonvini:2016frm}
M.~Bonvini, S.~Marzani, C.~Muselli, and L.~Rottoli, {\it {On the Higgs cross
  section at N$^{3}$LO+N$^{3}$LL and its uncertainty}},  {\em JHEP} {\bf 08}
  (2016) 105, [\href{http://xxx.lanl.gov/abs/1603.08000}{{\tt
  arXiv:1603.08000}}].

\bibitem{Anastasiou:2016cez}
C.~Anastasiou, C.~Duhr, F.~Dulat, E.~Furlan, T.~Gehrmann, F.~Herzog,
  A.~Lazopoulos, and B.~Mistlberger, {\it {High precision determination of the
  gluon fusion Higgs boson cross-section at the LHC}},  {\em JHEP} {\bf 05}
  (2016) 058, [\href{http://xxx.lanl.gov/abs/1602.00695}{{\tt
  arXiv:1602.00695}}].

\bibitem{provbfh}
F.~A. Dreyer and A.~Karlberg. \url{http://provbfh.hepforge.org/}.

\bibitem{Cacciari:2015jma}
M.~Cacciari, F.~A. Dreyer, A.~Karlberg, G.~P. Salam, and G.~Zanderighi, {\it
  {Fully Differential Vector-Boson-Fusion Higgs Production at
  Next-to-Next-to-Leading Order}},  {\em Phys. Rev. Lett.} {\bf 115} (2015),
  no.~8 082002, [\href{http://xxx.lanl.gov/abs/1506.02660}{{\tt
  arXiv:1506.02660}}].

\bibitem{Dreyer:2016oyx}
F.~A. Dreyer and A.~Karlberg, {\it {Vector-Boson Fusion Higgs Production at
  Three Loops in QCD}},  {\em Phys. Rev. Lett.} {\bf 117} (2016), no.~7 072001,
  [\href{http://xxx.lanl.gov/abs/1606.00840}{{\tt arXiv:1606.00840}}].

\end{thebibliography}
